\documentclass{elsart}

\usepackage[dvips
]{graphicx}
\usepackage[centerlast,small]{caption2}

\usepackage{latexsym} 
\usepackage{amssymb}
\usepackage{amsmath}  

\usepackage{bm} 
\newcommand{\underdot}[1]{\underset{\scriptstyle\dot{}}{#1}}

\newcommand{\Real}{\mathop{\mathrm{Re}}}
\newcommand{\Imag}{\mathop{\mathrm{Im}}}
\newcommand{\Exp}{\mathop{\mathrm{e}}\nolimits}
\newcommand{\tr}{\mathop{\mathrm{tr}}}

\newcommand{\ie}{i.e.}
\newcommand{\Ie}{I.e.}
\newcommand{\eg}{e.g.}
\newcommand{\Eg}{E.g.}
\newcommand{\vs}{vs.}
\newcommand{\ibid}{ibid}

\begin{document}

\begin{frontmatter}

\title{Sub-collision hyperfine structure \\
       of nonlinear-optical resonance \\
       with field scanning}

\author[IAE]{K. I. Gus'kov\corauthref{corr}\thanksref{RFBR}},
\ead{guskiv@narod.ru} 
\author[IAE]{V. I. Kovalevsky},
\author[MIPT]{A. G. Rudavets}, 
and \author[NSU]{\'{E}. G. Saprykin\thanksref{RFBR}}

\address[IAE]{Institute of Automation and Electrometry SB
RAS, Novosibirsk, 630090, Russia}
\address[MIPT]{Department of Applied Physics, Moscow Institute
of Physics and Technology, Dolgoprudny, 141700, Russia}
\address[NSU]{Novosibirsk State University, Novosibirsk,
630090, Russia}

\corauth[corr]{Corresponding author.}

\thanks[RFBR]{Partially supported by the Russian Foundation
for Basic Researches.} 

\begin{abstract}
Some experimental evidences for methane are produced that the
simple transition from frequency scanning of nonlinear-optical
resonances to magnetic one may  be  accompanied  with  transition
from sub-Doppler collisionally broadened structure  to
sub-collision hyperfine  one. It is conditioned by nonlinearity of
splitting of hyperfine sublevel for molecules in the adiabatically
varied magnetic field and respectively breaking the analogy of
magnetic and frequency scannings. The exact calculation of the
resonance structure is considered for molecules with only one spin
subsystem. The approximately spin-additive calculation of the
structure is given for sufficiently fast rotating molecules with
greater number of spin subsystems. Within the same approximation
an example of hyperfine doubling in the magnetic and electric
spectra of nonlinear-optical resonance is considered for
fluoromethane.

\end{abstract}

\end{frontmatter}

\section{Introduction}

Nonlinear-optical resonance (NOR) is considered here in the
intensity of laser radiation passed through molecular gas cell
with low pressure ($\gtrsim 1 \,\mathrm{mtorr}$). The laser
frequency $\omega$ is tuned in a resonance with the frequency of
spectral line $\omega_{mn}$, corresponding to investigated
vibration-rotation transition ${[}^m_n$. The requirement to its
fixing is practically absent and frequency detuning $\varOmega
=\omega -\omega_{mn}$ can be somewhere in limits of Doppler width
of this line.

The NOR (or NOR spectrum) can be scanned with laser frequency
(NOR/Fr, \ie\ NOR with frequency scanning). However we will take
an interest in an alternative method of NOR scanning by means of
varying external field while the laser frequency is invariable. It
is supposed, that the resonancely absorbing gas is located in
(spatially homogeneous and slowly\footnote{It is exacter,
adiabatically\label{adiabat} slowly \cite[Chap.\,XVII
\S\,14]{Mes62}.} varied) magnetic (/M) or electric (/E) field; the
latter is meaningful to use only if there is a sublevel
degeneration on parity. In both of the cases, there are two
factors, namely, the crossings of field $M$-sublevels and the
processes of their anti-crossings [in other words, processes of
their repulsion, caused by hyperfine interactions in the initial
(diagonalizing only field interactions) basis of wave functions]
take place, respectively, with unequal and equal field projections
$M$ of full angular momentum $F_j$ of a molecule in the hyperfine
multiplets extracted by light, $(J_j,I^{\bm{\cdot}})$ with $j
=m,n$ [in the parentheses, respectively, rotation angular momentum
of the molecule and spins\footnote{The superscript
${}^{\bm{\cdot}}$ will be defined on p.~\pageref{bmcdot_set}.} of
all its nuclei are designated] \cite{PS72}. As a matter of fact
and it will be shown in the given paper, just these two factors
(and without collisional complexities) determine all major
characteristics of a field spectrum of NOR. Let us underline the
heterogeneity of the factors: the first is connected to diagonal
elements of the interactions and the second to off-diagonal ones.
Under transition in final (diagonalizing the sum of field and
hyperfine interactions) wave function basis, the account of
repulsing interactions of sublevels brings to that the amplitudes
of hyperfine components of optical (electro-dipole) transition
between these multiplets become field-dependent \cite[Chap.\,2
\S\,13]{AChKh93}. Owing to the dependence, there are structures in
field spectra of NOR, which we shall name ``ballast'' ones,
underlining the energetic aspect of known process of anti-crossing
of magnetic sublevels \cite{EFW63,Eck67}. With respect to Raman
structures connected with crossings of magnetic sublevels, ballast
ones have always opposite sign. They are present only in
nonlinear-optical correction to a transmission of laser radiation
and are absent in its birefringence. The ability of rotational
(or, more precisely, vibration-rotational) subsystem to absorb a
laser radiation will increase with hyperfine coupling of ballast
spin subsystems. These ballast subsystems, if they are uncoupled,
do not naturally interact with light themselves (from here the
term ``ballast''). Coupling (or uncoupling) implies a varying of
hyperfine constants, however we reach that by the smooth varying
of the external field, imposed on molecule gas, absorbing
radiation. In this case, disbalancing of nuclear spin and rotation
subsystems takes place, that can be effectively considered as a
rupture of hyperfine couplings, see below (\ref {hfz:c}). The
disbalancing is possible because of usual distinction of Zeeman
frequencies of a precession, \ie, when spin $g$-factor of any
homogeneous nuclear subsystem of a molecule differs from rotation
$g$-factor of the molecule. This preliminary qualitative picture
will be verified in the following sections of our paper, and also
we shall slightly touch upon linear Stark effect with its
analogous electric spectra.

The NOR with field scanning (NOR/Fi, where option /Fi is /M or /E;
one of advantages for given ranking of the letters consists in
that the variable part of the abbreviation
%
%
appears in the end) can be observed with various orientations of
(amplitudely varied) external field $\bm {B}^{(0)}$ with respect
to direction of wave vector $\bm{k}$ of laser radiation. The
parallel or perpendicular orientation is used and respectively
designated by subscript, $\parallel$ or $\perp$ after
/Fi.\label{par_per} The shape of NOR/Fi spectrum essentially
depends from polarization of laser radiation.

From the very beginning of our research, the term ``anomalous'',
appearing below in description of an observed structure
NOR/M$_\parallel$ for methane,  implied its sub-collision
property, \ie\ a disposition inside collision contour of the
resonance. It is necessary to take into account, that the
constants of hyperfine interactions (HFI) in methane did not
exceed its collision constant $\nu$ for our working pressure
($\gtrsim 1\,\mathrm{mtorr}$). The adequate theoretical model of
NOR/M with this unusual property had find out only after the
computer calculations, carried out for methane with the exact
account of all its HFI.

NOR/Fr allows to look inside of inhomogeneous Dop\-p\-ler contour
of a line. NOR/Fi, as we shall show below, allows to look inside of
homogeneous collision contour of the same line, and we can observe
a sub-collision hyperfine structure (HFS) [see below the
respective equations (\ref{rambound}), (\ref{anglebound}), and
(\ref{balbound})].

Sometimes the field spectroscopy can be the real alternative to
the frequency one. It is important, which type of field structures
is selected. In the meantime it appears that the ballast HFS of
NOR/Fi is more convenient and informative than Raman one. In this
case the field spectroscopy can be viewed as a spectroscopy of
intramolecular tops. We consider its variants for molecular levels
without doubling and with doubling on parity. Respectively, our
examples will be the molecules of methane and fluoromethane. For
the first molecule all hyperfine (spin-rotation) constants can be
spectrally determined, but for the second one they can be only
partially determined from the spectra. Here to the aid there come
researches of spin conversion in molecules
\cite{Gus95,IB98,Gus99}. These additional researches open a
possibility of the determination of those hyperfine constants,
which do not usually completely affect the spectra. The combined
solution of these problems allows to receive the complete set of
hyperfine spin-rotation constants for molecules of fluoromethane
symmetry.

The purpose of this work is the theoretical research of influence
hyperfine (spin-rotation) interactions on NOR/Fi in molecules of
methane and fluoromethane symmetry. For it we are going:

\textbullet{} to analyze the probable reasons of appearance for
``anomalous'' structures of NOR/M in methane. It is necessary to
make a choice between two models, one of theirs takes into account
a HFS, other --- a collision structure (it is exacter, a
collision-hyperfine one).

\textbullet{} to apply the mathematical means that are adequate to
the problem, \ie\  permitting to simplify the analysis of NOR/Fi
caused by a multiplet structure of resonance levels.

\textbullet{} to receive an approximation (\ie\ the first summands
of expansion on a small parameter $J^{-2}$) for NOR/Fi in a case,
when researched multispin molecule is in fast rotation states, and
to apply the approximation to methane and fluoromethane. In the
latter case we take into account hyperfine parity doubling for
rotation $JK$-levels with $K=1$.

\section{Some experimental evidences for a sub-collision structure
\newline of NOR/M}

When HFS of levels is absent or can be neglect, in a polarizing
nonlinear sub-Dopppler spectroscopy there is a useful analogy
between magnetic and frequency scanning of NOR.  Let us begin from
its brief description, see
\cite
{RSh91}. Let we have molecules in a gas cell and are capable
optically to initiate electro-dipole transitions ${[}^m_n$ between
two molecular vibration-rotation levels labelled with $j =m,n$ and
degenerated on magnetic projections $M_{J_j}$ of rotation angular
momentum with $J_j =\max(M_{J_j})$. We shall connect collision
relaxation constants (level half-width $\nu_j$ and transition one
$\nu$) by means of an equation $\nu_m +\nu_n =2\nu$; in all our
examples it will completely be $\nu_j =\nu$. The gas cell is
located in a varied magnetic field $\bm{B}^{(0)} =B^{(0)}
\bm{u}_z$ and through it along the field a
bichromatic\footnote{\Eg, see \cite{GA79}.} light wave is
propagated. Its electrical component
\begin{equation}
\bm{E}(r_k,t)=\Real \sum_{q=\pm1} E^{(\omega)}_{\underdot{q}}(r_k)
\Exp^{\mathrm{i}(kr_k -\omega_qt)} \bm{u}_{\dot{q}}.
\label{bichrom}
\end{equation}
The frequency $\omega_q \equiv\omega +q\omega_\varDelta \simeq
\omega$, \ie\  frequency detuning\footnote{We sometimes designate
the negative sign (of an index especially) as overbar.}
$\omega_{1,\bar{1}} \equiv \omega_1 -\omega_{\bar{1}}
=2\omega_\varDelta$ is small, comparable with $\nu_j$. In rarefied
gas the amplitudes $E^{(\omega)}_{\underdot{q}}(r_k)$ of the
extracted components weakly depend on projection $r_k
=\bm{r}\cdot\bm{k} /k$. The wave vector $\bm{k}$ is parallel to
field $\bm{B}^{(0)}$, namely, $\bm{k} =k\bm{u}_z$. With this
longitudinal orientation of fields the NOR/M/Fr in transmission is
described by expression
\begin{subequations}
\begin{equation}
 \varDelta s^{(3)} \propto\sum_{q=\pm1}
  |e_{\underdot{q}}e_{\underdot{\bar{q}}}|^2
     (\bm{\vee}^m_{nq} +\bm{\wedge}^m_{nq})
 =2|e_{\underdot{1}}e_{\underdot{\bar{1}}}|^2
  \Real(\bm{\vee}^m_{n1} +\bm{\wedge}^m_{n1}).
\end{equation}
On the cell input the unit vector of wave polarization is
\begin{gather}
 \bm{e} \equiv\bm{E}(0,0)/E(0,0)
=\sum_{q} e_{\underdot{q}{}} \bm{u}_{\dot{q}{}}. \label{e_polar}
\\ \intertext{The sum}
 \bm{\vee}^m_{nq} +\bm{\wedge}^m_{nq}
=\frac{a^2_{m2}\,\nu^2/\nu_n}{\nu_m -\mathrm{i}2q(\varDelta_J
+\omega_\varDelta)} +(m\leftrightarrow n). \label{without_hfs}
\end{gather}
\label{fm_analog}
\end{subequations}
Here the second summand turns out from the first with mentioned
permutation of indices. Numerical factor $a^2_{j\varkappa}$ is
expressed by means of $6J$-symbol, see
\cite
{RSh91}. If $J_m =J_n -1$ then $a^2_{m2} < a^2_{n2}$, and their
levelling goes with increase of $J_j$. Zeeman
frequency\footnote{Everywhere we use angular frequencies
\cite[Chap.\,2 \S\,1]{Sten84} and it is convenient to measure them
by means of angular Hertz, $\protect\mathrm{Hza}
=2\pi\,\protect\mathrm{s}^{-1}$.\label{Hz-angular} The
abbreviation is adopted from the software
\emph{MathCad}.\label{Hz}} $\varDelta_J =\gamma_J B^{(0)}$, where
$\gamma_J$ --- rotational gyromagnetic ratio (for methane
$2\gamma_J \simeq0.477 \,\mathrm{kHza} \,\mathrm{Gs}^{-1}$). One
can see from (\ref{without_hfs}), that $\varDelta_J$ is analogue
of $\omega_\varDelta$ --- a frequency detuning of circular
components from their average value $\omega$.

With Raman scattering of light, when there are two contrarily
polarized photons in the combinative transitions, the considered
multilevel system is a set of three-level subsystems $\bm{\vee}$-
and $\bm{\wedge}$-types. These two-photon transitions go between
various magnetic $M_J$-sublevels of the same level $j$ with
difference $\varDelta M_J =\pm2$. Generally speaking, the
interaction includes both dipole transitions between levels and
the multipole transitions between sublevels. Therefore their
relaxational constants $\nu$ and $\nu_j$ affect NOR together.
However for us its three components [being responsible for effects
of population changing (Bennett's holes), nonlinear interference
(NIF) and field splitting] are not separated and are contained in
both summands $\bm{\vee}^m_{nq}$ and $\bm{\wedge}^m_{nq}$ in
proportion $2:1:1$ (see
%
\cite
{RSh91}). Thus, this classification is
not useful in all our following considerations and does not reveal
itself in any way.

In the beginning of our research it seemed, that the simple
reduced expression (\ref{without_hfs}) is quite enough for
description of methane NOR/M/Fr in conditions, when the collision
constants $\nu_j$ ($\simeq\nu$) are greater than all its hyperfine
constants $C^{\varsigma}_{\mathrm{a}}$, and it is possible to
explain all expected deviations by means of disregarded collision
features \cite{RSh91}.

\begin{figure}\centering
\includegraphics[width=0.414\textwidth]{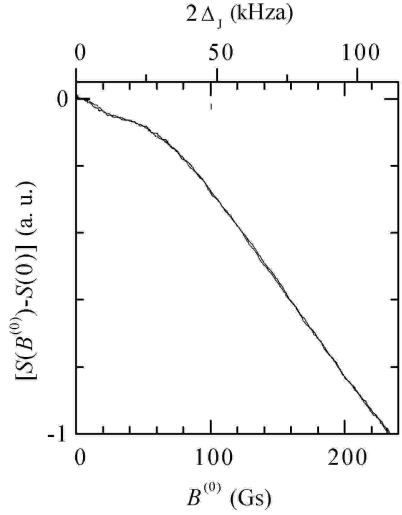}
 \caption{\label{12ch4_ex_l} Experimental
NOR/M$_\parallel$ (even function) in transmission of linearly
polarized radiation for component ($\omega_3$ $P(7)$
$F^{(2)}_{2(-)}$) of ${}^{12}\mathrm{CH}_4$ isotope; $P \sim
3\,\mathrm{mtorr}$, $S(0) \sim 1\,\mathrm{mW}\,\mathrm{cm}^{-2}$.}
\end{figure}

Our transition to research of NOR/M in methane molecules had
correlated with a paper \cite{TRSSh80}, where some collision
broadening of NOR/M in neon atoms was researched. From this paper
there was a tendency to research some collision properties of
these resonances. It allowed us to use the experimental parameters
tuned evidently far from that were used in the detection of HFS of
methane NOR/Fr \cite{HB73}. In particular, the working pressure in
our methane cell was on an order above and our signal of NOR/M
disappeared in noise with pressure decreasing to their magnitudes.
Thus, the subject of our work was born unexpectedly and
undeliberately, in the experiment originally oriented on research
of collision properties of methane NOR/M. At that time its
``anomalous'' sub-collision structure was found out.
NOR/M was observed in the intensity of linearly polarized infrared
(IR) radiation of He-Ne/${}^{12}\mathrm{CH}_4$ laser without
frequency tuning. This radiation has wave length $\lambda
=3.39\,\mu\mathrm{m}$ and well hits in the absorption bands of
vibration-rotation spectrum of methane isotopes,
namely,\footnote{Here vibration band and rotation line are
designated.} [$\omega_3(0\rightarrow 1)$ $P(7)$] of
${}^{12}\mathrm{CH}_4$ and [$\omega_3(0\rightarrow 1)$ $P(6)$] of
${}^{13}\mathrm{CH}_4$ \cite{DNPR79}. Our in-cavity cell was
completely elongated in the solenoidal magnetic scanner and
portionly filled with methane at pressure $P =1\div
10\,\mathrm{mtorr}$. The laser radiation was propagated through
the cell along scanner magnetic field $\bm{B}^{(0)}$. Its
amplitude $B^{(0)}$ slowly varied from some chosen level
$B^{(0)}_0$. The rectangularly impulse modulation of the field in
the limits between $B^{(0)}_0$ and $B^{(0)}$ with frequency $f$
($\simeq60\,\mathrm{Hza}$) brought to appearance on this
frequency\footnote{The product of modulation amplitude and
frequency is limited to adiabatic condition, see the footnote on
p.~\pageref{adiabat}.} of a rather small difference signal
$S_f(B^{(0)}) =\frac{2}{\pi}[S(B^{(0)}) -S(B^{(0)}_0)]$; here the
maximum factor $2/\pi$ corresponds to rectangular impulse
porousness $=2$. The typical record (scan) of this signal (being
symmetric with respect to $B^{(0)}=0$) is shown in
\figurename~\ref{12ch4_ex_l}. $S(0)$ is input radiation intensity.
Here the anomalously narrow structure clearly emerges from the
background represented by wide peak having frequency analog
(\ref{fm_analog}). The uncommonness of the situation is that we
are with pressure $P$, when collision half-width $\nu\sim
50\,\mathrm{kHza}$ \cite{BBCh72,BDDCh80}. It is much greater
$C^{\mathrm{H}}_{\mathrm{a}}$, the constant of hyperfine splitting
of the line, and any structure does not emerge in NOR/Fr; at us a
role of the last was played by the inverted Lamb dip, see \cite[\S
\,4.3]{Let76} and
\cite
{RSh91}. HFS of NOR/Fr \cite{HB73} begins to emerge with pressure
on an order below ours, when $\nu\lesssim
|C^\varsigma_{\mathrm{a}}|$. Thus we see that the collision
restriction is broken in NOR/M.

Then 
with the purpose to level the amplitudes of both structures we
went to scanning of field derivative $dS/dB^{(0)}$. By adding
small sine wave modulation to slowly varied magnetic field,
$B^{(0)} +\delta B^{(0)} \sin(ft)$, we extracted the first
harmonics in radiation intensity, $S_f(B^{(0)}) \simeq\delta
B^{(0)} dS/dB^{(0)}$. Due to this method we have found out one
more narrow structure in intensity of linearly polarized
radiation. It had an inverse sign, \ie\ was a dip. The next
important act is transition to external cell and use of circularly
polarized
radiation \cite{
BKKR:83,BKRSS:84}.

\begin{figure}\centering
\includegraphics[width=0.48\textwidth]{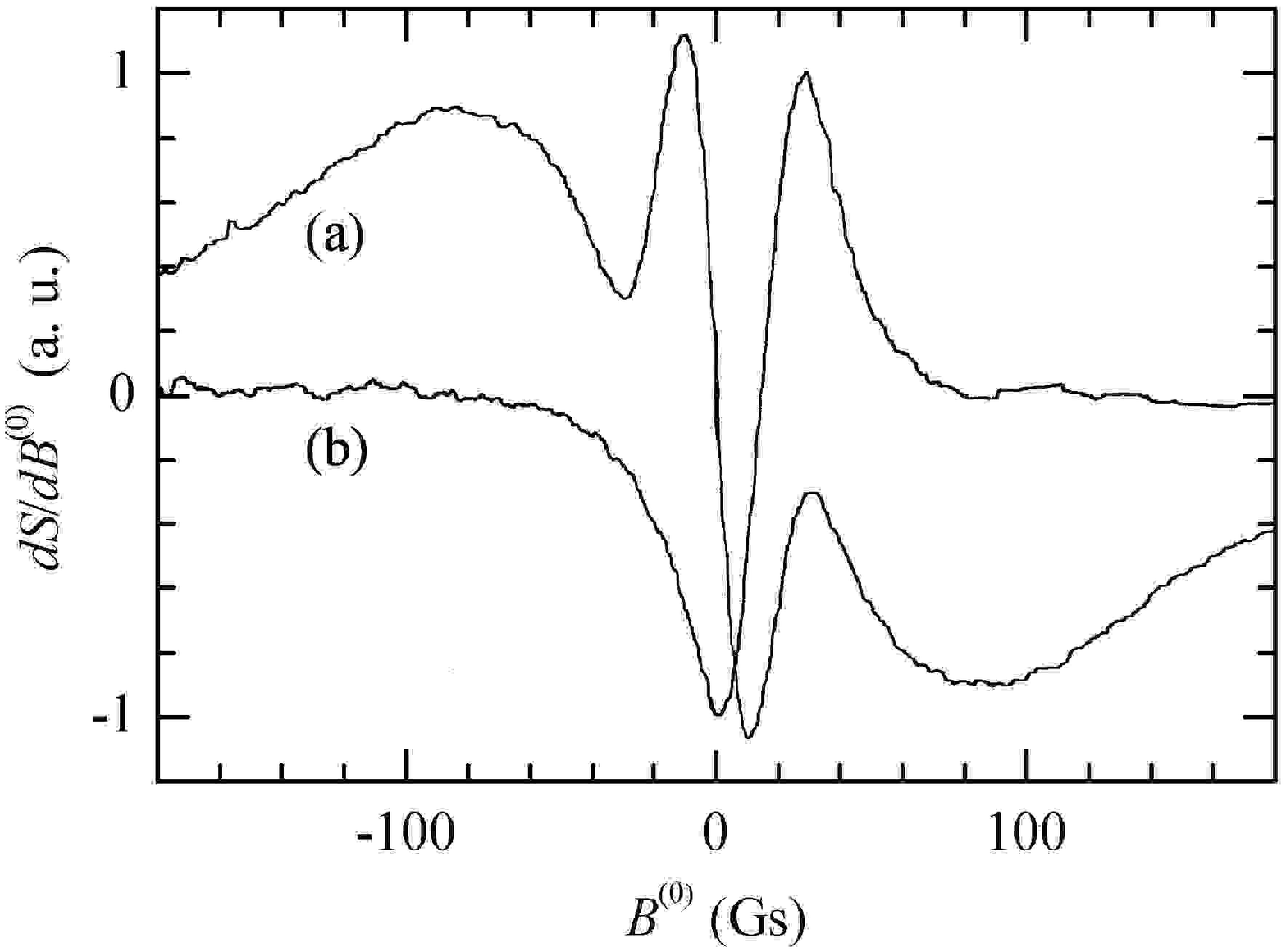} \hfill
\includegraphics[width=0.48\textwidth]{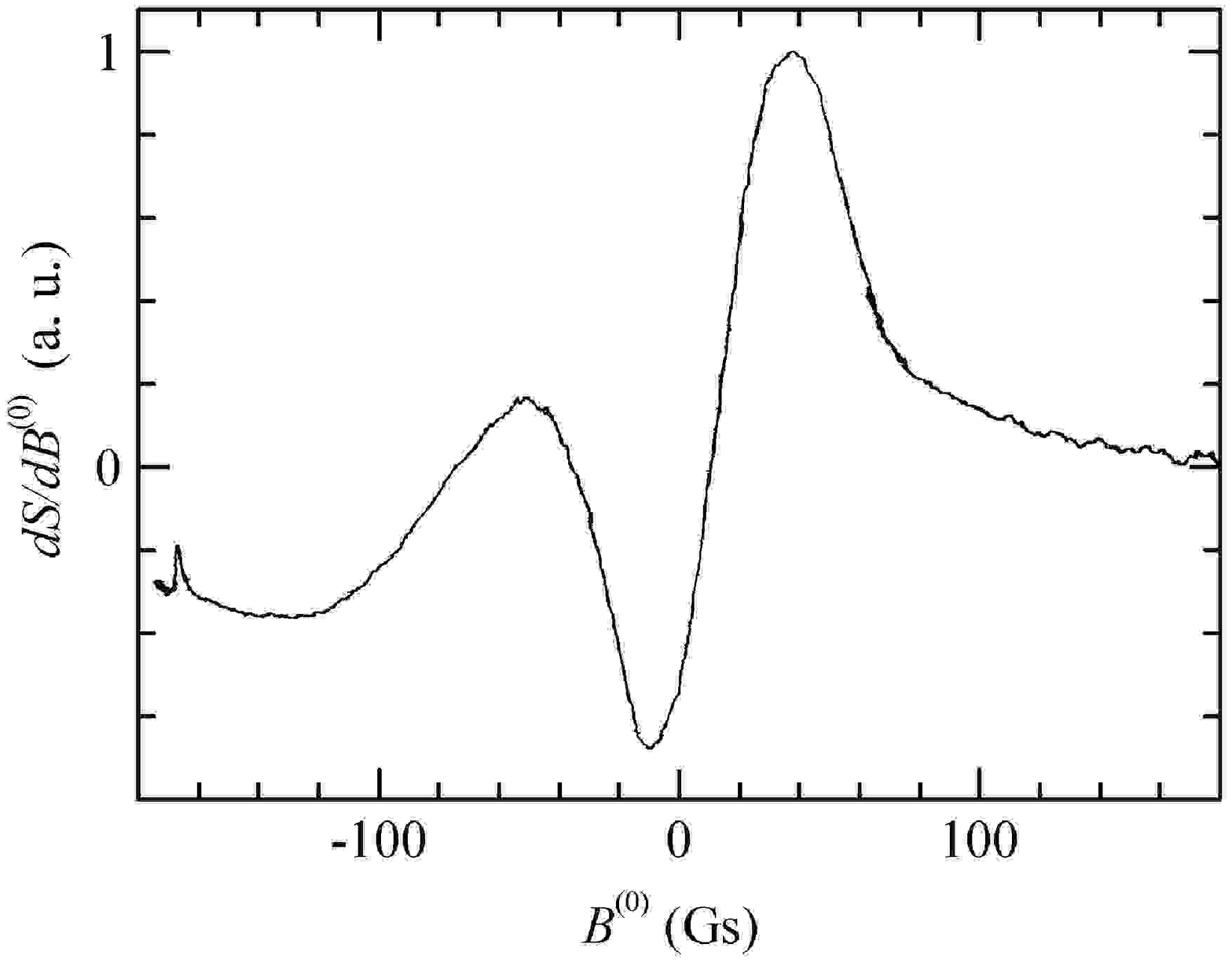}
\\
\caption{\label{12ch4_ex_dlc-13ch4_ex_dc} Derivatives of
experimental NOR/M$_\parallel$ in the transmission of linearly (a)
and right circularly (b) polarized radiations. \emph{At the left}
for component ($\omega_3$ $P(7)$ $F^{(2)}_{2(-)}$) of
${}^{12}\mathrm{CH}_4$ isotope; $\delta B^{(0)} \lesssim
10\,\mathrm{Gs}$. \emph{At the right} a curve is analogous to (b),
but for component ($\omega_3$ $P(6)$ $F^{(1)}_{2(-)}$) of
${}^{13}\mathrm{CH}_4$ isotope. For all curves $P\sim
2\,\mathrm{mtorr}$, $S(0) \sim 1\,\mathrm{mW}\,\mathrm{cm}^{-2}$.}
\end{figure}
\figurename~\ref{12ch4_ex_dlc-13ch4_ex_dc} at the left shows the
typical signal derivatives for both types of radiation
polarization [there are the similar experimental curves (without
combining) in
\cite{BKKR:83,BKRSS:84}]. In the latter cited paper there was
working pressure $P =3\div 6\,\mathrm{mtorr}$, when, according to
\cite{BDDCh80}, $\nu =55\div 100\,\mathrm{kHza}$; the recording of
curves was conducted with modulation amplitude $\delta B^{(0)}
=16\,\mathrm{Gs}$ and frequency $f =60\, \mathrm{Hza}$. Restoring
the initial output signal $S$, we shall see, that the dips are
present on both resonance curves, \ie\ there are symmetrically
displaced two dips [with weight $=0.25$, as it will be seen from
(\ref{s3par})] for linearly polarized radiation and there is
respectively displaced one of them (with weight $=1$) for every
circularly polarized one.

We investigated only those components of fine structure of methane
vibration-rotation spectrum, which hit in the limited frequency
area of magnetic detuning of He-Ne laser with respect to basic
component\footnote{The components are labeled by means of bottom
level symmetry with a distinguishing superscript and parity
subscript, both enclosed in parenthesises.} ($\omega_3$ $P(7)$
$F^{(2)}_{2(-)}$) of ${}^{12}\mathrm{CH}_4$ isotope:
$|\varOmega|<3.5\,\mathrm{GHza}$. The Doppler broadening of these
components will further show in (\ref{DopplerHW}). The next
component ($\omega_3$ $P(6)$ $F^{(1)}_{2(-)}$) belongs to
${}^{13}\mathrm{CH}_4$ isotope and is detuned on $-0.9
\,\mathrm{GHza}$ from basic one. One more component ($\omega_3$
$P(7)$ $E_{(\pm)}$) of ${}^{12}\mathrm{CH}_4$ isotope is detuned
on $-3.0 \,\mathrm{GHza}$ from basic one and has zero total spins
of the subsystems with nuclei of the same type. Any sub-collision
structure of its NOR/M were not experimentally revealed. At the
same time, previous two {mentioned} components belong to different
methane isotopes and, in accessible pressure range, their field
spectra are quite (even qualitatively) distinctive for circularly
polarized radiation and weakly (again qualitatively) distinctive
for linearly polarized one. It is evidence from the typical
experimental derivatives of NOR/M shown in the
\figurename~\ref{12ch4_ex_dlc-13ch4_ex_dc} (at the right the curve
for linearly polarized radiation is omitted).

It is necessary to note, that, except the methane transmission,
the sub-collision structure of NOR/M was recorded in its circular
birefringence \cite{BKMS:84}, see below an expression
(\ref{theta3}).

First published and unsuccessful\footnote{As we now understand it,
see section~\ref{COL_NOR_M}.} attempts \cite{BKRSS:84} to
interpret these structures were based on a model which is taking
into account the collision in-summands in kinetic equations
(\ref{kineq}) for both resonance levels. It was considered, that
the model of relaxational constants, which are taking into account
only the collision out-summands in the kinetic equations (without
in-summands), is insufficient for the interpretation. The authors
of cited paper tried to explain the narrow structures of NOR/M for
linearly polarized radiation by means of collision changing a
velocity of molecules and forming the collision interference
resonance
\cite{RSh91}. One more collision model was chosen in
\cite{SSSh85}, where the possibility of complete intermixing for
population of hyperfine sublevels by means of deorientation
collisions without velocity modification was taken into account.
However, it was gradually found out (see
section~\ref{COL_NOR_M}), that the collision model (as in itself
and with the account of HFS) is not capable to describe adequately
an observable resonance structure particularly for circularly
polarized radiation.

\section{\label{ExactHFS}Exact calculation of HFS of NOR/M$_\parallel$
in ${}^{12}\mathrm{CH}_4$}

Let us consider the resonancely absorptive molecular gas with an
operator density matrix
\begin{equation}
 \hat\rho(v_k,r_k,t)
=\left(
 \begin{array}{cc}\hat\rho_m(v_k,r_k,t) & \hat\rho_{mn}(v_k,r_k,t)\\
 \hat\rho^\dagger_{mn}(v_k,r_k,t) & \hat\rho_n(v_k,r_k,t) \end{array}
 \right). \label{rho}
\end{equation}
It depends on molecular velocity projection $v_k=\bm{v} \cdot
\bm{k}/k$ and, similarly, coordinate one $r_k$ for some time $t$.
The wave vector $\bm{k}$ sets the propagation direction of
absorbed light. The evolution of $\hat\rho(v_k,r_k,t)$ is
represented by the quantum kinetic equation with classical
description of translational motion of molecules
\cite{RSh91}:
\begin{equation}
(\partial_t +v_k\partial_{r_k})\hat\rho(v_k,r_k,t)
=\hat{R}(v_k,r_k,t)
 +\hat{S}(v_k,r_k,t)
 -\mathrm{i}\left[\hat{H}(r_k,t),\hat\rho(v_k,r_k,t)\right].
\label{kineq}
\end{equation}
The statistical and dynamic summands enter in the right side of
this equation. Statistical ones are represented by spontaneous one
$\hat{R}(v_k,r_k,t)$ and collision one $\hat{S}(v_k,r_k,t)$.
Dynamic ones are represented by a commutator with Hamiltonian
$\hat{H}(r_k,t)$. Here we are going to consider NOR/M for
vibration-rotation transitions of molecules. These transitions
usually hit in IR area of frequency spectrum, where it is possible
to do not take $\hat{R}(v_k, r_k,t)$ into account completely. Let
us remark only, that the spontaneous structure of NOR/M begins to
be revealed in visible frequency area, when the electronic
transitions are considered, and its analysis will be submitted in
paper \cite{GR90b}.\label{spontan}

The summand $\hat{S}(v_k,r_k,t)$ includes components of collision
excitation\footnote{Without deorientational in-summand, about it
see below (\ref{Col_int_with_tilde_nu}).} and relaxation for the
subsystem ${[}^m_n$, extracted by absorbed radiation.
\begin{equation}
 \hat{S}_{ij}(v_k,r_k,t) =\delta_{i,j} \nu_j
 \left[\tilde{N}_j W(v_k)\hat1_j - \hat\rho_j(v_k,r_k,t)\right]
  -(1 -\delta_{i,j})\nu\hat\rho_{ij}(v_k,r_k,t)
\label{Col_int}
\end{equation}
with $i$ and $j \in (m,n)$. In the presence of laser radiation
tuned to ${[}^m_n$ the sufficient condition for particle number
conservation is
\[\sum_{j=m,n}
\langle\langle\hat{S}_j(v_k,r_k,t) \rangle\rangle_{v_k} =0,
\]
and in its absence the mentioned condition is
$\langle\langle\hat{S}_j(v_k,r_k,t) \rangle\rangle_{v_k} =0$.
Angular brackets $\langle\ldots\rangle$ and
$\langle\ldots\rangle_x$ (or more detailed
$\langle\ldots\rangle^{(b)}_{x(a)}$)\label{bracketdef}
respectively mark trace and integration on $x$ (from $a$ up to
$b$). For our purposes it is enough to suppose that the collisions
are isotropic and excite only diagonal elements of
$\hat{\rho}_j(v_k,r_k,t)$, \ie\ sublevel populations of extracted
$j$-levels. $\hat{1}_j$ --- diagonal identity matrix with diagonal
size $[j]=\langle \hat{1}_j \rangle$. The molecular excitation is
characterized by Maxwell's distribution \vs\ velocity (or, with
reference to our field orientation, \vs\ its $k$-projection),
\begin{equation}
W(v_k)=\Exp^{-(v_k/v_{\mathrm{M}})^2}/\sqrt{\pi}v_{\mathrm{M}}
\label{maxWell}
\end{equation}
with integral normalization $\langle W(v_k) \rangle_{v_k} =1$.
Here most probable thermal velocity $v_{\mathrm{M}}
=\sqrt{2k_{\mathrm{B}}T/m}$, where $m$ --- molecular mass. For
methane, when $T$ is room temperature, the velocity $\left.
v_{\mathrm{M}}\right|_{20\,{}^o\mathrm{C}} \simeq 552 \,\mathrm{m}
\,\mathrm{s}^{-1}$. Components $\hat\rho_j(v_k,r_k,t)$ and
$\hat\rho_{mn}(v_k,r_k,t)$ relax with frequency velocities $\nu_j
\equiv\nu_{jj}$ and $\nu \equiv\nu_{mn}$, respectively.
\begin{equation}
\tilde{N}_j =N_j/[j]
 \label{tildeN}
\end{equation}
--- volumetric density of population for
each sublevel of $j$-level. It makes a part from $N_j$ ---
volumetric density of total population of $j$-level (term) with
sublevel number $[j]$. If to speak about methane \cite{BDDCh80}
with pressure $P \sim2\,\mathrm{mtorr}$, all its relaxational
constants $\nu_j =\nu\sim 40\,\mathrm{kHza} \ll
\omega_{\mathrm{D}}$, where Doppler half-width (at level
$\Exp^{-1}$ from maximum)
\begin{equation}
\omega_{\mathrm{D}} =kv_{\mathrm{M}} \simeq 0.16 \,\mathrm{GHza}
\label{DopplerHW}
\end{equation}
for light wave length $\lambda =3.39 \,\mu\mathrm{m}$. In these
conditions the spectra inhomogeneously broaden and the observation
of NOR is possible.

The last (dynamic) summand in the right side of the equation
(\ref{kineq}) is a commutator with Hamiltonian
\begin{equation}
 \hat{H}(r_k,t)
 =\hat{H}^{(0)} + \hat{H}^{(1)} + \hat{V}(r_k,t)
 =\left(\begin{array}{cc}
 \omega_m + \hat{H}^{(1)}_m  & \hat{V}_{mn}(r_k,t)\\
 \hat{V}^\dagger_{mn}(r_k,t)   & \omega_n + \hat{H}^{(1)}_n
 \end{array}\right).
 \label{hhv}
\end{equation}
As well as in (\ref{rho}), it is convenient for all operators to
keep the matrix representation with operator elements, where
$\hat{H}^{(0)}$ is diagonal. $\hat{H}^{(0)}$ describes a
two-component subsystem ${[}^m_n$, \ie\ $\hat{H}^{(0)} |j\rangle
=\omega_j |j\rangle$. The frequency splitting of these components
is optical and designated with $\omega_{mn} \equiv\omega_m
-\omega_n$. Each component has HFS and $\hat{H}^{(1)}$ describes
it and its nonlinear splitting in a magnetic field.
$\hat{V}(r_k,t)$ describes resonance electro-dipole interaction of
these multiplet components with light, which is generated as a
plane monochromatic travelling wave. In the given section we
intend to describe the effect of $\hat{H}^{(1)}$ on tensor
components of nonlinear-optical susceptibilities
$\bm{\chi}^{(3)}$, \ie\ on NOR/Fi at the end.

To exclude from consideration too large interaction of nuclear
quadrupole with molecule rotation, we shall be limited to a case
of molecules consisted of the half-spin nuclei. Some of these
nuclei can be identical among themselves and then the molecules
will have various spins modifications (\ie\  isomers), as in a
subsequently considered example of molecular symmetry
$\mathrm{T}_{\underline{\mathrm{d}}}$. The respective irreducible
spin representations $\Gamma_I$ of group $\mathrm{S}_4 \cong
\mathrm{T}_{\underline{\mathrm{d}}}$ are connected one-to-one with
rotation-inversion ones $\Gamma_{J_n}$ of group
$\mathrm{T}_{\underline{\mathrm{d}}}$ for components of a fine
structure of vibration-rotation spectrum in the correspondence
with Pauli exclusion principle, \ie\  $\Gamma_{J_n} \otimes
\Gamma_I \supseteq A_2$. For triply degenerated (\ie\ $F_2$-type)
$\omega_3$-vibration at its first exited level the Coriolis
interaction substantially gives splittings ($\sim10
\,\mathrm{GHza}$) of components ($m$ is one of them). At its
non-excited level the centrifugal perturbation remains only and
gives splittings of components ($n$ is one of them) on an order
less \cite{Shim76}. For either of the two optically connected
vibration-rotation components ($m$ and $n$) without parity
doubling, \ie\ excluding $E_{(\pm)}$-components,\footnote{Owing to
parity doubling, electric scanning of their NOR can be used in
place of magnetic one.} it is possible to represent the effective
Hamiltonian, combining hyperfine and Zeeman interactions, as
\begin{equation}
 \hat H^{(1)}_j = -\sum_\varsigma
 \left( C^{\varsigma}_{\mathrm{a}} \hat{\bm{J}}_j
 + \bm{\varDelta}^{\varsigma}_I \right)
 \cdot
 \hat{\bm{I}}{}^{\varsigma}
 - \bm{\varDelta}_J\cdot\hat{\bm{J}}_j.
\label{hfz:a}
\end{equation}
We use frequency\footnote {\Ie\ their components are measured by
means of frequency units.} Zeeman vectors: nuclear-spin ones
$\bm{\varDelta}^{\varsigma}_I =\gamma^{\varsigma}_{I}\bm{B}^{(0)}$
and rotational one $\bm{\varDelta}_J =\gamma_J \bm{B}^{(0)}$. Here
various gyromagnetic ratios are designated as $\gamma^\varsigma_I
=g^\varsigma_I \gamma_{\mathrm{N}}$ and $\gamma_J
=g_J\gamma_{\mathrm{N}}$. They are products of respective
$g$-factors on a standard nuclear gyromagnetic ratio
$\gamma_{\mathrm{N}} =\mu_{\mathrm{N}} /\hbar \simeq 0.762
\,\mathrm{kHza}\,\mathrm{Gs}^{-1}$. $\mu_{\mathrm{N}}$ --- nuclear
magneton. The values of superscript $\varsigma$ are ordered and
distinguish, \eg\ for methane, both ordinary spin subsystem (with
$I^{{}^{12}\mathrm{C}} =0$ or $I^{{}^{13}\mathrm{C}} =1/2$) and
combined ${}^{1}\mathrm{H}_4$-subsystem. Spin modifications of the
latter are respectively characterized by total spins
$\left.I^{\varsigma} \right|_{\varsigma =\mathrm{H}A_1,
\mathrm{H}F_2, \mathrm{H}E}=2,1,0$. $\hat{\bm{J}}_j$ with
$j\in(m,n)$ and $\hat{\bm{I}}{}^{\varsigma}$ with $\varsigma
\in(\varsigma_1, \ldots,\varsigma_k)$ (here $k$
--- the number of various nuclear subsystems) respectively designate
vectorial operators of rotation angular momentum of molecule and
spins of its nuclear subsystems (all are measured in units of
Planck constant $\hbar$) for two optically connected
$(J_jI^{\bm\cdot})$-terms, where the set $I^{\bm{\cdot}}
\equiv(I^{\varsigma_1}, \ldots,
I^{\varsigma_k})$.\label{bmcdot_set} These methane terms have
close magnetic properties, \ie\  average hyperfine constants
$C^{{}^1\mathrm{H}}_{\mathrm{a}} \simeq 12\,\mathrm{kHza}$
\cite{YOR71,HB73}, $C^{{}^{13}\mathrm{C}}_{\mathrm{a}} \simeq
-12\,\mathrm{kHza}$ [see below (\ref{hfconsts}) and (\ref{Cav})],
rotation $g$-factor $g_J \simeq 0.313$ \cite{UHB71} and certainly
both spin ones $g^{{}^1\mathrm{H}}_I \simeq 5.5854$ and
$g^{{}^{13}\mathrm{C}}_I \simeq 1.4042$ \cite{Fly78}.

To have an obvious model for Hamiltonian (\ref{hfz:a}), we shall
now write out the appropriate set of motion equations for dynamic
variables in Heisenberg ``representation'' \cite{Mes62}:
\begin{equation}
 \begin{split}
 \partial_t \hat{\bm{J}}_j (t)
&= \hat{\bm{J}}_j (t) \times \left(\sum_\varsigma
C^{\varsigma}_{\mathrm{a}}
 \hat{\bm{I}}{}^{\varsigma} (t) +\bm{\varDelta}^{\varsigma}_J \right),
\\
 \partial_t \hat{\bm{I}}{}^{\varsigma} (t)
&= \hat{\bm{I}}{}^{\varsigma} (t) \times \left(
C^{\varsigma}_{\mathrm{a}} \hat{\bm{J}}_j (t)
+\bm{\varDelta}^{\varsigma}_I \right),
\end{split} \label{hfz:c}
\end{equation}
and $\varsigma \in (\varsigma_1, \ldots, \varsigma_k)$. Here
$\hat{\bm{J}}_j(t) =\Exp^{\mathrm{i}\hat{H}^{(1)}_j t}
\hat{\bm{J}}_j \Exp^{\bar{\mathrm{i}}\hat{H}^{(1)}_j t}$ and
$\hat{\bm {I}}{}^{\varsigma}(t)$ is similar. The precession
($\bm{\varDelta}_J \times\hat{\bm{J}}_j$) is a deviation from the
interaction ($\bm{\varDelta}_J \cdot \hat{\bm{J}}_j$) and, as far
as
\[
(\bm{\varDelta}_J \times\hat{\bm{J}}_j)^2 +(\bm{\varDelta}_J\cdot
\hat{\bm{J}}_j)^2 =\bm{\varDelta}^2_J \hat{\bm{J}}^2_j,
\]
they are two complementary characteristics of the motion. The set
of equations (\ref {hfz:c}) reflects our notions about a
precession of nuclear-spin and rotation subsystems in a magnetic
field, their hyperfine connections and balancing, and also their
ruptures with growth of the magnetic field and unbalancing
mentioned subsystems.

Let us designate the vector operators of total angular momentum
and total nuclear spin of molecule as
\begin{equation}
\hat{\bm{F}}_j =\hat{\bm{J}}_j +\hat{\bm{I}}{}
\quad\text{and}\quad \hat{\bm{I}}{} =\sum_\varsigma
\hat{\bm{I}}{}^{\varsigma}. \label{FJI}
\end{equation}
For commutator $[\hat{H}^{(1)}_j, \bm{\varDelta}_J \cdot
\hat{\bm{F}}_j] =0$,  it is convenient to represent (\ref {hfz:a})
as
\begin{gather}
 \hat H^{(1)}_j= \hat h_j-\bm{\varDelta}_J\cdot\hat{\bm{F}}_j,
\label{hfz:b} \\
\intertext{where}
 \hat  h_j  =\sum_\varsigma \hat h^{\varsigma}_j =
 - \sum_\varsigma \left(C^{\varsigma}_{\mathrm{a}}\hat{\bm{J}}_j
 + \bm{\varDelta}^{\varsigma}_{IJ}\right)
 \cdot \hat{\bm{I}}{}^{\varsigma}
\nonumber
\end{gather}
and $\bm{\varDelta}^{\varsigma}_{IJ}\equiv
\bm{\varDelta}^{\varsigma}_I - \bm{\varDelta}_J$. As well as
earlier it is convenient to extract the appropriate gyromagnetic
ratios, then $\bm{\varDelta}^{\varsigma}_{IJ}
=\gamma^{\varsigma}_{IJ} \bm{B}^{(0)}$. Already here it is
possible to notice, that HFS of NOR/M can be observed only if some
$\bm{\varDelta}^{\varsigma}_{I}$ differs from $\bm{\varDelta}_J$.
Just in this case it is possible to speak about a rupture of
hyperfine connection of the appropriate nuclear spin with
molecular rotation momentum. In a weak magnetic field the
precession of total angular momentum takes place around of the
field direction. With increasing value of the field the precession
nuclear spins and molecular rotation momentum becomes more and
more independent, and they cease to form the conserved total
angular momentum.

We direct Cartesian unit vector $\bm{u}_z$ along $\bm{B}^{(0)}$,
therefore $\bm{\varDelta}^{\varsigma}_I =\varDelta^{\varsigma}_I
\bm{u}_z$, $\bm{\varDelta}_J =\varDelta_J \bm{u}_z$, and
$\bm\varDelta^{\varsigma}_{IJ}
=\varDelta^{\varsigma}_{IJ}\bm{u}_z$. The standard spherical basis
is defined by covariant unit vectors \cite{BL81}, \ie\
\begin{equation}
 \bm{u}_{\dot0{}} = \bm{u}_z
\quad\text{and}\quad \bm{u}_{\pm\dot1{}} =\mp(\bm{u}_x\pm
i\bm{u}_y)/\sqrt2. \label{stdspherbasis}
\end{equation}
Contravariant unit vectors $\bm{u}_{\underdot{q}{}}
\equiv\bm{u}^*_{\dot q{}} =(-1)^q \bm{u}_{\dot{\bar{q}}{}}$, so
$\bm{u}_{\dot{q}'{}} \cdot\bm{u}_{\underdot{q}{}} =\delta_{q',q}$.
Contra- and co-variant magnitudes\footnote{They are analogues of
bra- and ket-vectors.} have under- and over-dotted indices,
respectively. The complex conjugation and its generalization, the
Hermitian one, are respectively designated with superscript
asterisk ${}^*$ and dagger ${}^\dagger$.

Unlike \cite{SSSh85}, using $F_jM$-basis, we shall choose
completely split basis of wave functions,\footnote{It is clear
that the final result does not depend from this choice.} \ie\
\begin{subequations}
\begin{equation}
 |j\mu^{\bm\cdot}M\rangle
=|(J_j I^{\bm\cdot}) \mu^{\bm\cdot}M\rangle
\equiv
 |J_jM-\mu\rangle\prod_\varsigma|I^{\varsigma}\mu^{\varsigma}\rangle,
\label{split}
\end{equation}
where $M\equiv\langle j\mu^{\bm\cdot}M| \hat F_{jz}
|j\mu^{\bm\cdot}M\rangle$ and the set of nuclear spin
projections\footnote{We have adopted the set notation from
\cite{SW71}, also see
\cite[Chap.\,I \S\,1]{Gant60}.}
\begin{equation}
\mu^{\bm\cdot} \equiv(\mu^{\varsigma_1}, \ldots,
\mu^{\varsigma_k}),\quad
 \mu=\sum_\varsigma\mu^{\varsigma},\quad |\mu^\varsigma|\leq
 I^\varsigma; \label{mu_cdot}
\end{equation}
\end{subequations}
In this basis the Hamilton operator $\hat{H}^{(1)}_j$ has the
following matrix representation\footnote{The matrix representation
depends on basis and is marked with underline \underline{\ }.}
\begin{gather}
 \underline{\hat{H}}{}^{(1)}_{jM} = \sum_\varsigma
 \underline{\hat{h}}{}^{\varsigma}_{jM} - \varDelta_J M,
\nonumber\\ \intertext{where}
 \left(\underline{\hat{h}}{}^{\varsigma}_{jM} \right)_{\mu^{\bm\cdot
 \prime}\mu^{\bm\cdot}} \equiv \langle j\mu^{\bm\cdot \prime}M
 |\hat h{}^{\varsigma}_j|j\mu^{\bm\cdot}M\rangle
\nonumber\\ \intertext{with}
 \underline{\hat{h}}{}^{\varsigma}_{jM}
 =  -\left(C^{\varsigma}_{\mathrm{a}}
 \bm{J}^{\dagger}_{j,M-\underline{\hat{I}}{}_z}
 +\bm\varDelta^{\varsigma}_{IJ}\right) \cdot
 \underline{\hat{\bm{I}}}{}^{\varsigma}.
\label{hjM}
\end{gather}
Matrix elements of covariant rotation operator components,
$\hat{J}_{j\dot q{}} =\hat{\bm{J}}_j \cdot\bm{u}_{j\dot q{}}$, are
\begin{align}
 J_{j\dot q{}M}
 &\equiv  \langle J_jM'|\hat J_{j\dot q{}}|J_jM\rangle
  =\sqrt{\hat{\bm J}_j^2}\langle J_jM'|J_jM\,1q\rangle
\nonumber\\
 &=\delta_{M',M+q}\left(\delta_{q,0}M
   -q\delta_{|q|,1}\sqrt{(\hat{\bm{J}}_j^2-M'M)/2}\right).
\end{align}
Here $\hat{\bm{J}}{}^2_j \equiv J_j(J_j+1)$ and it is similar for
spin operators $\hat{\bm{I}}{}^{\varsigma}$.

\Eg, when $I=1$ (and $J\geq I$), the matrix (\ref{hjM}) is usually
obtained with sizes $3\times3$.  Its eigenvalues are numbered at
us by an index $\tilde\mu=1,0,\bar1$. For convenience we shall
cite explicit expressions\footnote{Cp.\ with \cite{BS97,KK68}.}
for three real roots of reduced (\ie\ without a quadratic summand)
cubic equation
\[
y^3 +py +q =0
\]
in ``irreducible'' case, \ie, when a discriminant $D =(p/3)^3
+(q/2)^2 <0$ and therefore $p<0$. The trigonometrical expression
for the roots, ordered in the way appropriated for us, is
\[
y_{\tilde\mu} =2R\cos\left( \frac{\alpha -2\pi\tilde\mu}{3} \right),
\]
where $R =\sqrt{\bar{p}/3}$ and $\cos\alpha =\bar{q}/2R^3$ with
$0\leq\alpha\leq\pi$.

When $I\geq2$ the exact eigenvalues for (\ref{hjM}) are determined
with a solution of the appropriate algebraic equation (of degree
$[I]$, if $J\geq I$), \eg, with the help of theta-functions (with
the displaced arguments) $\theta[{}^{\bm{a}}_{\bm{b}}](\bm{z},
\hat{\bm\varOmega})$, see \cite{Ume84}.

We shall designate the found eigenvalues as $z_{j_l}$, where a set
of quantum numbers $j_l \equiv j \tilde{\mu}^{\bm\cdot}_l M_l$. In
our case, for a determination of wave eigenfunctions appropriated
to them, one can use projective operators,
\begin{equation}
 \hat{P}^{(1)}_{j_l}
=\prod\limits_{j_k\neq j_l}
\frac{ z_{j_k} \hat{1}_j -\hat{H}^{(1)}_j
    }{ z_{j_k}           -z_{j_l} },
\end{equation}
constructed on the basis of minimum equation,\footnote{At us it
coincides with characteristic one.} see
\cite[appl.\ 5]{Fed79} and
\cite[Chap.\,IV]{Gant60}.
Another (equivalent) way of their explicit construction is based
on use for the operator (\ref{hfz:a}) a resolvent
\cite[\S\,5.8]{Lan69}:
\begin{equation}
\hat{R}^{(1)}_j(z) =(z\hat{1}_j -\hat{H}^{(1)}_j)^{-1} =\sum_{j_l}
 \frac{\hat{P}^{(1)}_{j_l} }{z -z_{j_l} }.
\label{Resolv_j}
\end{equation}
It is visible, that the projectors $\hat{P}^{(1)}_{j_l}$ are its
residues, \ie\
\begin{equation}
\hat{P}^{(1)}_{j_l} =\mathop{\mathrm{Res}}\limits_{j_l}\,
\hat{R}^{(1)}_j(z) =\oint\limits_{ {\scriptscriptstyle j_l}
\hspace{-0.65em}
\raisebox{-0.15em}{${\displaystyle\circlearrowleft}$} }
\frac{dz}{2\pi \mathrm{i}} \hat{R}^{(1)}_j(z). \label{Res_j}
\end{equation}
Here spectral parameter $z\in\mathbb{C}$, and integration path
$\circlearrowleft$ enclose only one point $z_{j_l}$ of spectrum
for operator $\hat{H}^{(1)}_j$.

As a result we come to a basis set of wave functions,
simultaneously diagonalizing the Hamiltonians of hyperfine and
Zeeman interactions, \ie\
\begin{subequations}
\begin{equation}
 |j\tilde{\mu}^{\bm\cdot}M\rangle =
  \sum_{\mu^{\bm\cdot}}|j\mu^{\bm\cdot}M\rangle
 (\hat{\mathcal{U}}_{jM})_{\mu^{\bm\cdot}\tilde{\mu}^{\bm\cdot}},
\label{diag}
\end{equation}
where the set of nuclear spin quasi-projections
\begin{equation}
 \tilde{\mu}^{\bm\cdot}
\equiv (\tilde{\mu}^{\varsigma_1},\ldots,
 \tilde{\mu}^{\varsigma_k}),\quad
 |\tilde{\mu}^{\varsigma}|\leqslant I^{\varsigma}.
\label{tildemu_cdot}
\end{equation}
\end{subequations}
It is convenient to choose them from the same set as
(\ref{mu_cdot}). As an example let us consider the methane isotope
${}^{12}\mathrm{CH}_4$, where there is only one spin subsystem (of
hydrogen nuclei). We shall be limited to a case, when its total
spin $I =I^{\mathrm{H}} =1$, therefore $\tilde{\mu}
=\tilde{\mu}^{\mathrm{H}}$ and $\mu =\mu^{\mathrm{H}}$. Adding
spin-spin interaction to the spin-rotation one, it is possible to
represent more precisely the reduced form of total Hamiltonian
\cite{YOR71,LL:QM77,Mich84} as
\begin{equation}
 \hat H^{(1)}_j
 =\hat{h}_{jM} -\bm{\varDelta}_J \cdot\hat{\bm{F}}_j
 =-C^{\mathrm{H}}_{1j}  \hat{\bm{J}}_j \cdot \hat{\bm{I}}{}
\\
 -C^{\mathrm{H}}_{2j} (\hat{\bm{J}}_j \cdot \hat{\bm{I}})^2
 -\gamma_{\mathrm{N}} (g_J \hat{\bm{J}}_j + g^{\mathrm{H}}_I
\hat{\bm{I}}) \cdot \bm{B}^{(0)}. \label{JI:II}
\end{equation}
While the magnetic dependence of its spectrum
\begin{equation}
 H^{(1)}_{j\tilde{\mu}M}
=h_{j\tilde{\mu}M} - \varDelta_J M \label{JJ:II:spectrum}
\end{equation}
is nearly linear from $B^{(0)}=0$ and $(g_J \hat{\bm{J}}_j +
g^{\mathrm{H}}_I \hat{\bm{I}})\simeq g_{F} \hat{\bm{F}}_j$, it is
possible to use $g$-factors\footnote{The subscript ``j'' is
omitted.}
\begin{equation}
 g_{F} =\frac{g^{\mathrm{H}}_I +g_J}{2}
 +g^{\mathrm{H}}_{IJ}\frac{I(I+1) -J(J+1)}{2F(F+1)},
 \label{g_F_exact}
\end{equation}
where $g^{\mathrm{H}}_{IJ} \equiv g^{\mathrm{H}}_I -g_J$, and
similarly for gyromagnetic ratios $\gamma_{F}=g_{F}
\gamma_{\mathrm{N}}$. From here, \eg, for simultaneously integer
(or half-integer) $J$ and $I$, $g$-factor difference
\begin{equation}
 g_{F=J} -g_J
 =\frac{g^{\mathrm{H}}_{IJ}I(I+1)}{2J(J+1)} \geq0.
\end{equation}

\begin{figure}\centering
\includegraphics[width=0.614\textwidth]{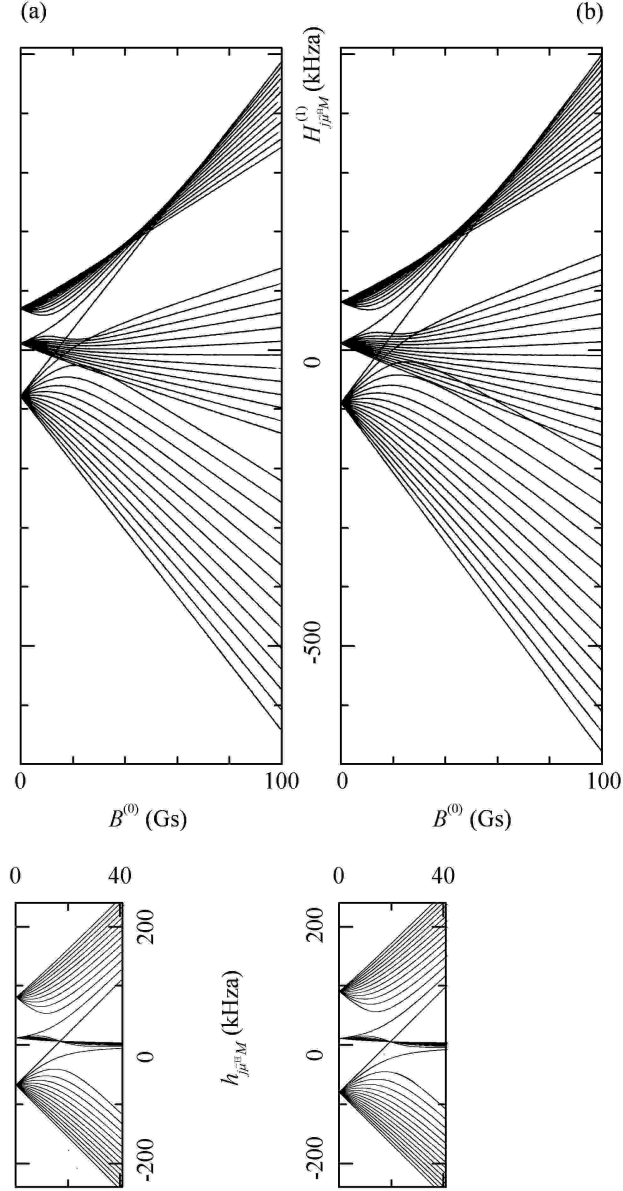}
\caption{\label{12ch4_th_hfs} Exact calculated magnetic splittings
(\ref{JJ:II:spectrum}) for both hyperfine multiplets
($J_jI^{\mathrm{H}}$) of component ($\omega_3$ $P(7)$
$F^{(2)}_{2(-)}$) of ${}^{12}{\mathrm{CH}}_4$ isotope:  (a) for
upper one with $J_{\mathrm{m}}=6$ and $I^{\mathrm{H}}=1$, (b) for
lower one with $J_{\mathrm{n}}=7$ and $I^{\mathrm{H}}=1$.}
\end{figure}

Taking hyperfine intervals (\ie\ experimentally measured
splittings in the upper and lower hyperfine multiplets with
$B^{(0)} =0$) from \cite{HB73}, we shall get the hyperfine
constants
\begin{equation}
\begin{split}
 C^{\mathrm{H}}_{1m} &=11.66\,\mathrm{kHza}, \quad
 C^{\mathrm{H}}_{2m} = 0.22\,\mathrm{kHza},
\\
 C^{\mathrm{H}}_{1n} &=11.58\,\mathrm{kHza}, \quad
 C^{\mathrm{H}}_{2n} = 0.17\,\mathrm{kHza}.
\end{split}
\label{hfconsts}
\end{equation}
Now let us see in \figurename~\ref{12ch4_th_hfs},
showing magnetic dependence of all $M$-sublevels in diagonalizing
basis. As we shall see further, the features of
$H^{(1)}_{j\tilde{\mu}M}$, shown on it above, and
$h_{j\tilde{\mu}M}$, shown on it below, correspond to the features
of spectrum of NOR/M$_\parallel$ in cases of linearly and
circularly polarized radiations respectively.
The graphs of sublevels only for positive fields are shown. It is
possible to restore a dependence of sublevels on negative fields
if we take into account that $H^{(1)}_{j\tilde{\mu}M}(-B^{(0)})
=H^{(1)}_{j\tilde{\mu}\bar{M}}(B^{(0)})$, \ie\  the whole
splitting picture comes from a mirror reflection with respect to
ordinate axis. It is visible that the one weakly differs at upper
(a) and lower (b) hyperfine multiplets. With large $B^{(0)}$
(after all crossings) the sets of sublevels $(M)_{\tilde{\mu}}$
are ordered from top to bottom, as
[$(-J_j-1,-J_j,\ldots,J_j-1)_{\bar1}$;
 $(-J_j, -J_j+1, \ldots,J_j  )_{0}$;
 $(-J_j+1,-J_j+2,\ldots, J_j+1)_{1}$] on graphs shown above,
and as [$( J_j-1,J_j-2, \ldots,-J_j-1)_{\bar1}$;
 $( J_j  , J_j-1,\ldots,-J_j  )_{    0}$,
 $(-J_j+1,-J_j+2,\ldots, J_j+1)_{    1}$] on graphs shown below.
In these sets there are repulsing (anti-crossing) of  sublevels
with equal $M$ and unequal $\tilde{\mu}$. For anyone $J_j$ all
magnetic sublevels with unequal $M$ and equal $\tilde{\mu}$ are
crossed with $B^{(0)} =0$, and separately for each $\tilde{\mu}$.
When the field $B^{(0)}$ is close to zero, we have an
approximation, $(F_j -J_j) \simeq \tilde{\mu}_j $, where
$\hat{\bm{F}}{}_j$ is defined in (\ref{FJI}). With increasing the
field all sublevels, shown in \figurename~\ref{12ch4_th_hfs}
above, diverge and we have another approximation, $\mu_j \simeq
\tilde{\mu}_j$.

In both upper graphs of \figurename~\ref{12ch4_th_hfs} the
greatest number of crossings in nonzero fields is observed in
single twist area\footnote{Under ``single twist area'' the compact
area of crossings for a fan of (almost direct) lines on a plane is
understood, which ordering inverts after passage of this area.}
for the fan of magnetic sublevels of (upper) component with
$\tilde{\mu} =-1$, exactly with
\begin{equation}
        {}^{\mathrm{H}}B^{(0)}_{\mathrm{cr}}
\simeq \pm C^{\mathrm{H}}_{\mathrm{a}}/\gamma_J, \label{cross}
\end{equation}
when $J \lesssim g^{\mathrm{H}}_{IJ}/g_J$ with
$g^{\mathrm{H}}_{IJ}\geq 0$. With greater $J$, when the factor
$g_{F=J-1}$ in (\ref{g_F_exact}) changes a sign, the twist area
disappears.

In both lower graphs of \figurename~\ref{12ch4_th_hfs} our
attention is attracted by the feature located at nonzero fields in
a point of twist for fan of magnetic sublevels of (middle)
component with $\tilde{\mu} =0$:
\begin{equation}
 {}^{\mathrm{H}}B^{(0)}_{\mathrm{Ncr}}
\simeq \pm C^{\mathrm{H}}_{\mathrm{a}} J/\gamma^{\mathrm{H}}_{IJ}
\label{Ncross}
\end{equation}
Non-repulsed magnetic sublevel of (lower) component with
$\tilde{\mu} =1$ also passes through it.

In the basis chosen by us for diagonalizing wave functions with
anyone $B^{(0)}$ all states would be stationary at absence of
collision pumping and relaxation. At us, at their presence, they
will be equilibrium, \ie\ balanced on these processes. All
possible transitions between these states will be purely optical,
due to resonance electro-dipole interaction with a light field
$\bm{E}(r_k,t)$, exactly
\begin{align} \hat V_{mn}(r_k,t)
&=-\hat{\bm{d}}{}^{mn}\cdot\bm{E}(r_k,t)/\hbar
 \nonumber\\
&=-\sum_q  \hat d^{mn}_{\dot q{}}
 E_{\underdot{q}{}}(r_k,t)/\hbar. \label{Vmn}
\end{align}

According to Wigner-Eckert factorizational theorem (see
\cite{BL81} or expression (62) in \cite{Gus95}), vector operator
of electro-dipole moment of a transition can be connected with
covariant components of standard\footnote{\Ie\ normalized and
determined only with rotation symmetry.} operator, affecting only
rotation variable, namely,
\begin{gather}
 \hat{\bm{d}}{}^{mn}
=\tilde d_{mn}
 \hat{\bm{T}}{}^{mn}=
 \tilde d_{mn}\sum_q
 \hat T^{mn}_{\dot{q}{}}\bm{u}_{\underdot{q}{}},
 \label{WignerEckert:a}
\\
\intertext{and trace normalization is}
 \langle\hat T^{mn\dagger}_{\dot q{}'}\hat T^{mn}_{\dot q{}}
 \rangle
\equiv \tr
 \left(\hat T^{mn\dagger}_{\dot q{}'}\hat T^{mn}_{\dot q{}}\right)
 =\delta_{q',q}. \label{WignerEckert:b}
\end{gather}
The physical characteristics of operator $\hat{\bm{d}}$ is
determined with its modified\footnote{Take note that at us any
modification is usually marked by tilde.} $\tilde{d}_{mn}$. The
matrix elements of covariant spherical components of standard
rotation operator in $JM_J$-basis are\footnote{Here $M=M_J$.}
\begin{align}
 T^{mn}_{\varkappa\dot{q}M}
 \equiv {}_{\mathrm{s}}T^{J_mJ_n}_{\varkappa\dot{q}M}
 &=\langle J_m M'|
   \hat{T}^{mn}_{\varkappa\dot{q}{}}|J_n M\rangle
\nonumber\\ &\equiv
 \frac{[J_m\|\hat{\bm{T}}^{(J)mn}_{\varkappa} \|J_n]}{\sqrt{[J_m]}}
 \langle J_m M'|\varkappa q~J_n M\rangle
\label{WignerEckert:cc}\\
 &=\sqrt{\frac{[\varkappa]}{[J_m]}}
 \langle J_m M'|J_n M~\varkappa q\rangle.
 \label{WignerEckert:c}
\end{align}
At us $[J_j]\equiv 2J_j+1$. Here we have defined another reduced
matrix element $[J_m\|\hat{\bm{T}}^{(J)mn}_{\varkappa}\|J_n]$. In
this particular case we have
\begin{equation}
[J_m\|\hat{\bm{T}}^{(J)mn}_{\varkappa}\|J_n]
 =(-1)^{-J_m +\varkappa +J_n}\sqrt{[\varkappa]}.
 \label{WignerEckert:ccc}
\end{equation}
It is convenient also to use extra operators (to standard one)
\begin{equation}
\begin{split}
 {}_{\mathrm{e}}\hat{\bm{T}}{}^{J_mJ_n}_{\varkappa}
&=(-1)^{-J_m +\varkappa +J_n}
 {}_{\mathrm{s}}\hat{\bm{T}}{}^{J_mJ_n}_{\varkappa},
\\
 {}_{\mathrm{e}}T^{J_mJ_n}_{\varkappa\dot{q}M}
&=\sqrt{\frac{[\varkappa]}{[J_m]}}
 \langle J_m M'|\varkappa q~J_n M\rangle,
\end{split}
\end{equation}
coordinated with the definition (\ref{WignerEckert:cc}) on phase.
Owing to that, we now have
\begin{equation}
[J_m\|{}_{\mathrm{e}}\hat{\bm{T}}^{(J)mn}_{\varkappa}\|J_n]=\sqrt{[\varkappa]}.
 \label{WignerEckert:cccc}
\end{equation}
The subscript ``s'' is sometimes omitted, as in
(\ref{WignerEckert:cc}) and (\ref{WignerEckert:ccc}). Also
sometimes the subscript ``$\varkappa$'', if it is equal $1$, is
generally omitted, as in (\ref{WignerEckert:a}) and
(\ref{WignerEckert:b}). Owing to the orthogonality of Wigner
coefficients a permutation is possible, namely, $\langle J_m
M'|J_n M~\varkappa q\rangle =\langle J_n M~\varkappa q|J_m
M'\rangle$. With large $J_j$ an approximation by means of
$d$-functions \cite{BL81} is possible:
\begin{equation}
 \langle J_m M'|J_n M~\varkappa q\rangle
 \simeq d^\varkappa_{q,J_{mn}}(\theta_{JM}) \delta_{M',M+q}
 \varepsilon^{(\varDelta)}_{J_m,J_n,\varkappa}.
\label{approxWig}
\end{equation}
Here $\cos\theta_{JM}=M/\sqrt{\hat{\bm{J}}{}^2_n}$. $J_{mn} \equiv
J_m-J_n = -1,0,1$ (or P,Q,R). If triangle condition for
($J_m,J_n,\varkappa$) is true, then function
$\varepsilon^{(\varDelta)}_{J_m,J_n,\varkappa}$ is equal $1$, else
$0$
\cite[Chap.\,5 \S\,8]{BL81}. The matrix elements of covariant
operator components of electro-dipole moment are now written as
\begin{equation}
 d^{mn}_{\dot{q}{}M}
=\tilde d_{mn}T^{mn}_{\dot{q}{}M} \simeq d_{mn}
d^1_{q,J_{mn}}(\theta).
\end{equation}
We have
\[
\tilde{d}_{nm} =(-1)^{J_{mn}}\tilde{d}^*_{mn}
\]
and, for $d_{mn} =\sqrt{3/[J_m]} \tilde{d}_{mn}$,
\[
d_{nm}=(-1)^{J_{mn}} d^*_{mn} \sqrt{[J_m]/[J_n]}.
\]
Take note, that for contravariant components we have
\[
 \hat d^{nm}_{\underdot{q}{}} \equiv \hat
d^{mn\dagger}_{\dot q{}}= (-1)^q \hat d^{nm}_{\dot{\bar{q}}{}},
 \]
but
\[
 \hat T^{nm}_{\underdot{q}{}}  = (-1)^{J_{mn}}  \hat
  T^{mn\dagger}_{\dot   q{}}=    (-1)^q    \hat T^{nm}_{\dot{\bar
 q}{}}.
\]

Before to consider selection rules for ours electro-dipole
transition in an arbitrary magnetic field, we shall write out the
reduced matrix element\footnote{It was defined in
(\ref{WignerEckert:c}).} of standard rotation operator for
$FM$-basis:
\begin{equation}
 [(J'I)F'\|T^{(J)mn}_{\varkappa}\|(JI)F]
 =(-1)^{J'+I+F'} \Pi_{F'\varkappa F}
 \left\{\begin{array}{ccc} F' & \varkappa & F \\
                           J  & I         & J'\end{array}\right\}
\label{TinFM}
\end{equation}
Here $\Pi_{xy\ldots} \equiv\sqrt{[xy \ldots]}$ with $[xy \ldots]
\equiv [x][y]\ldots$. The case $I =\varkappa =1$ is interesting to
us, when exact expression for
\begin{equation}
 \left([(J'1)F'\|T^{(J)mn}_{1}\|(J1)F]_{\tilde\mu',\tilde\mu}\right)
 =\sqrt3\left(\begin{array}{ccc}
  \sqrt{\frac{2J+3}{2J+1}}   & -\frac{1}{J}
                             & \frac{1}{J\sqrt{4J^2-1}} \\
  0 & \frac{\sqrt{J^2-1}}{J} & -\frac{1}{J} \\
  0 & 0                      & \sqrt{\frac{2J-3}{2J-1}}
 \end{array}\right).
\label{exactT1I1}
\end{equation}
Here both matrix subscripts, $\tilde{\mu}' =F' -J'$ and
$\tilde{\mu} =F -J$, accept values $1,0, -1$ beginning at left
upper angle of the matrix.

Now we shall define in (\ref{Vmn}) an electrical field of
resonance absorbed IR radiation by means of slowly varying
contravariant circular components
$E^{(\omega)}_{\underdot{q}{}}(r_k)$ of plane monochromatic
travelling wave [cp.\ with (\ref {bichrom})]
\begin{equation}
\begin{split} \bm{E}(r_k,t)
 &= \Real
 \left(\bm{E}^{(\omega)}(r_k)\Exp^{\mathrm{i}(k r_k-\omega t)}\right)
\\ \text{and}\quad \bm{E}^{(\omega)}(r_k)
 &= \sum_q E^{(\omega)}_{\underdot{q}{}}(r_k)
 \bm{u}_{\dot q{}}.
\end{split}
\label{monochrom}
\end{equation}
Here $E^{(\omega)}_{\underdot{q}{}}(r_k)=
E^{(\omega)}_{\underdot{q}{}}(0) \Exp^{\mathrm{i}2\pi \chi_q
kr_k}$ with wave number $k_q = n_q k \simeq (1 + 2\pi\chi_q )k$,
$k = 2\pi/\lambda$. Effective susceptibility $\chi_q$ and
refraction factor $n_q$ have imaginary components due to a small
absorption of gas medium. The factor (or inverse length) of this
absorption is\footnote{At us $z^{\prime} \equiv\Real z$ and $
z^{\prime\prime} \equiv\Imag z$.\label{primes}}
\begin{equation}
 \alpha_q  = 1/L_{\mathrm{a}} = 2kn''_q \simeq 4\pi k\chi''_q.
\label{absorpfactr}
\end{equation}
Normalized (on pressure) factor of linear absorption of methane
\cite{BH69} for $\lambda =3.39\,\mu\mathrm{m}$ is
\begin{equation}
 \alpha^{(1)P}
=\alpha^{(1)}/P \simeq 0.18\,\mathrm{cm}^{-1}\,\mathrm{torr}^{-1}.
\label{norm_absorp_l}
\end{equation}
Hence, when pressure $P \sim 1\,\mathrm{mtorr}$, the absorption
length $L_{\mathrm{a}} \sim 56\,\mathrm{m}$ and
$kL_{\mathrm{a}}\gg1$. The absolute value of light detuning
$|\varOmega| \ll \omega_{mn}$ and it is possible to use resonance
approximation.

With parallel orientation of fields \(\bm{e} \perp\bm{k} \parallel
\bm{B}^{(0)}\), where $\bm{e}$ is defined in (\ref{e_polar}). We
suppose, that the radiation polarized on right circle, has $q =1$,
\ie\ positive spirality. For a radiation polarized linearly, it is
convenient to define a rotation angle $\theta$ of polarization
plane, and a ratio $\psi$ of small semi-axis of polarization
ellipse to large one:
\begin{equation}
\theta +\mathrm{i} \psi = \pi k r_k (\chi_1 -\chi_{\bar1})
               =     k r_k (n_1 -n_{\bar1})/2.
\label{theta_psi}
\end{equation}
The equation can be expanded with the account of nonlinear-optical
corrections.

The interaction (\ref{Vmn}) brings to a small nonlinear absorption
of polarized laser radiation, having intensity (\ie\ surface
density of radiation power)
\[
 S_s(r_k)=|\bm{E}^{(\omega)}(r_k)|^2   c/8\pi.
\]
A variation of the intensity after passage of absorptive gas cell
is\footnote{Angular brackets have been defined on
p.~\pageref{bracketdef}.}
\begin{gather*} \delta S_s(L) = S_s(L)-S_s(0)
 =\langle \delta S_v(r_k) \rangle_{r_k}^{(L)}
\\
 =-\langle \langle \bm{E}(r_k,t) \cdot
   \frac{\partial_t}{T} \bm{P}(r_k,t)\rangle_t^{(T)} \rangle_{r_k}^{(L)}
 \simeq -\frac{\omega}{2} \Imag
   \langle \bm{E}^{(\omega)*}(r_k) \cdot \bm{P}^{(\omega)}(r_k)
   \rangle^{(L)}_{r_k} \leq0.
\end{gather*}
Here volume density of radiation power
\[
\delta S_v(r_k) =2\hbar\omega \Imag\langle
\hat{V}^{(\omega)\dagger}_{mn}(r_k) \langle
\hat{\rho}^{(\omega)}_{mn}(v_k,r_k) \rangle_{v_k} \rangle,
\]
as we shall see from (\ref {polar}) and (\ref {slowVmn}). The
integral with respect to $r_k$ and $t$-averaging are taken from
volume density of absorption power, $\bm{E}(r_k,t) \cdot
\partial_t \bm{P}(r_k,t)$, spent by light field on polarization
variation of gas medium. $L$ practically is a length of absorptive
cell and the average is taken on time period of light, \ie\ $T
=2\pi/\omega$.
\[
 \delta S_s(L) \simeq\delta S_s^{(1)}(L)+\delta S_s^{(3)}(L).
\]
Here the contributions of linear and cubic (on amplitude of
electric field of light) components of polarization of gas medium
are only retained. If relative linear absorption
$\tilde{\alpha}^{(1)}_{nm}r_k \ll 1$ and saturation parameter
$\varkappa \ll 1$, a relative variation of transmission is also
small and its expansion on $\varkappa$ looks as
\begin{subequations}
\begin{gather}
 \delta S_s(L)/S_s(0)\simeq -\alpha L <0
\label{DSonS}
\end{gather}
with effective absorption factor (\ie\ line density of absorption)
\begin{gather}
 \alpha = \sum_q \alpha_q |e_{\underdot{q}{}}|^2
 \simeq \alpha^{(1)}+\alpha^{(3)}
\nonumber\\
 =\sum_q
 (\alpha^{(1)}_q+\alpha^{(3)}_q) |e_{\underdot{q}{}}|^2
 \simeq -\tilde{\alpha}^{(1)}_{nm} (s^{(1)} +\varkappa s^{(3)}/2).
 \label{alpha}
\end{gather}
\end{subequations}
Factor $1/2$ at $\varkappa$ is extracted by analogy with
nondegenerate two-level case, when saturation expansion
$(1+\varkappa)^{-1/2}\simeq 1-\varkappa/2$; see
\cite{RSh91}. As well in our case, it is expected that the
amplitudes of amplification functions $s^{(i)}$ (modulo) are close
to $1$, \ie\  $s^{(i)}$ are normalized. Linear-optical resonance
with magnetic scanning (LOR/M) is\footnote{The dots between
tensors designate their contractions.}
\begin{gather}
s^{(1)} \equiv s^{(1)}_{(\bm{e})} = -\Real
[\tilde{\bm{X}}{}^{(1)mn} \,:\, \bm{e}^{*} \otimes\bm{e}]
\nonumber\\
= -\Real \sum_{(q_i)} \tilde{X}^{(1)mn}_{\underdot{q}{}_0
\dot{q}{}_1} e^*_{\underdot{q}{}_0} e_{\underdot{q}{}_1} = -\sum_q
\tilde{X}^{(1)mn\prime}_{\underdot{q} {}\dot q{}}
                |e_{\underdot{q}{}}|^2.
\end{gather}
NOR/M is (in vector designations and component by component)
\begin{align}
 s^{(3)}\equiv s^{(3)}_{(\bm{e})}
&= \tilde{\bm{X}}{}^{(3)mn}
 \,\underset{\displaystyle\cdot}{\vdots}\, \bm{e}^{*}
 \otimes\bm{e} \otimes\bm{e}^{*} \otimes\bm{e}
\nonumber\\
&=\sum_{(q_i)}
  \tilde{X}^{(3)mn}_{\underdot{q}{}_0 \dot{q}{}_1
                           \underdot{q}{}_2 \dot{q}{}_3}
e^*_{\underdot{q}{}_0} e_{\underdot{q}{}_1} e^*_{\underdot{q}{}_2}
e_{\underdot{q}{}_3} \quad (\in \mathbb{R}). \label{s3}
\end{align}
NOR/M$_\parallel$ is
\begin{gather}
 s^{(3)}_{\parallel} =\sum_{q=\pm1}
  \left[ \tilde{\bm{\between}}{}^{m}_{nq}
 |e_{\underdot{q}{}}|^2 +
  \left(\tilde{\bm{\vee}}{}^{m}_{nq}
+ \tilde{\bm{\wedge}}{}^{m}_{nq}\right)
 |e_{\underdot{\bar{q}}{}}|^2 \right]
 |e_{\underdot{q}{}}|^2
\nonumber\\
\equiv\sum_{q=\pm1}
  \left[ \tilde{X}^{(3)mn}_{\underdot{q}{}\dot q{}
                          \underdot{q}{}\dot q{}}
 |e_{\underdot{q}{}}|^2
 +\left(\tilde{X}^{(3)mn}_{\underdot{q}{}\dot q{}
       \underdot{\bar{q}}{}\dot{\bar{q}}{}}
+ \tilde{X}^{(3)mn}_{\underdot{q}{}\dot{\bar{q}}{}
       \underdot{\bar{q}}{}\dot q{}}\right)
 |e_{\underdot{\bar{q}}{}}|^2 \right]
 |e_{\underdot{q}{}}|^2
\nonumber\\
 =\tilde{\bm{\between}}{}^{m}_{n1}|e_{\underdot{1}{}}|^4
 +\tilde{\bm{\between}}{}^{m}_{n\bar1}|e_{\underdot{\bar{1}}{}}|^4
 +2|e_{\underdot{1}{}} e_{\underdot{\bar{1}}{}}|^2
 \left(\tilde{\bm{\vee  }}{}^{m\prime}_{n1}
           +\tilde{\bm{\wedge}}{}^{m\prime}_{n1}\right);
\label{s3par}
\end{gather}
the convenient graphic designations are here introduced.
NOR/M$_\perp$ is
\begin{equation}
 s^{(3)}_{\perp}=\tilde{\bm{\between}}{}^{m}_{n0}
\equiv\tilde{X}^{(3)mn}_{\underdot{0}{}\dot{0}{}
                         \underdot{0}{}\dot{0}{}}.
\label{s3perp}
\end{equation}
The meaning of subscripts $\parallel$ and $\perp$ is above defined
on p.~\pageref{par_per}. In both cases $\bm{u}_z =\bm{B}^{(0)}
/B^{(0)}$. Here we use the components of two modified $X$-tensors,
\begin{equation}
\begin{split}
 \tilde{\bm{X}}{}^{(1)mn}
&=\left. \bm{X}^{(1)mn} \right/ [p]\prod_\varsigma [I^\varsigma]
\\
\text{and}\quad
 \tilde{\bm{X}}{}^{(3)mn}
&=\left. \bm{X}^{(3)mn}\, [J_m] \right/ 3[p]\prod_\varsigma
[I^\varsigma].
\end{split}
\label{modXdef}
\end{equation}
It is visible that their normalizing factors are different.
Original $X$-tensors are defined slightly further in (\ref{Xdef}).
Their explicit construction turns out from the short Maxwell
equations in approximation of slowly varying amplitudes
\cite{Shen84}, \ie\
\[
 \partial_{r_k} \bm{E}^{(\omega)}(r_k)=
         \mathrm{i} 2\pi k \bm{P}^{(\omega)}(r_k).
\]
For that it is required to calculate the volume density of
light-induced (on light frequency) electro-dipole momentum (or
electrical polarization) of our gas medium, namely,
\begin{align}
 \bm{P}(r_k,t)
 &=\left\langle
 \hat{\bm{d}} \langle\hat\rho(v_k,r_k,t)\rangle_{v_k}
  \right\rangle
 =2\Real \left\langle
 \hat{\bm{d}}{}^{mn\dagger}
 \langle\hat\rho_{mn}(v_k,r_k,t)\rangle_{v_k} \right\rangle
\nonumber\\
 &=\Real \left(\bm{P}^{(\omega)}(r_k)
 \Exp^{\mathrm{i}(kr_k-\omega t)}\right)
\nonumber\\ \text{and}\quad
 \bm{P}^{(\omega)}(r_k)
 &=\sum_q P^{(\omega)}_{\underdot{q}{}}(r_k) \bm{u}_{\dot q{}}
 =2\left\langle \hat{\bm{d}}{}^{mn\dagger}
 \langle\hat\rho^{(\omega)}_{mn}(v_k,r_k)\rangle_{v_k}
 \right\rangle.
\label{polar}
\end{align}
Here $\hat{\rho}(v_k,r_k,t)$ is the operator density matrix of gas
medium and its light-induced non-diagonal component
$\hat{\rho}_{mn}(v_k,r_k,t) \simeq
\hat{\rho}^{(\omega)}_{mn}(v_k,r_k) \Exp^{\mathrm{i}(kr_k-\omega
t)} $. Passing in the equation (\ref{kineq}) to the representation
of interaction on
\[
 \hat{V}'(r_k,t)
 =\Exp^{\mathrm{i}(\hat{H}^{(0)} +\hat{H}^{(1)})t} \hat{V}(r_k,t)
  \Exp^{\bar{\mathrm{i}}(\hat{H}^{(0)} +\hat{H}^{(1)})t}
\]
(see
\cite[Chap.\,VIII \S\,14 and Chap.\,XVII \S\,1]{Mes62}), using
shortened form of the equation, when $v_k\partial_{r_k}\sim
v_{\mathrm{M}}/L_{\mathrm{a}}\ll\nu$, and being then limited its
slow component in resonance condition ($|\varOmega|
\ll\omega_{mn}$), when
\[
\hat{V}'_{mn}(r_k,t)\simeq
\hat{V}^{(\varOmega)}_{mn}(r_k,t)\Exp^{\mathrm{i}(kr_k-\varOmega
t)},
\]
but $\hat{V}'_{jj}(r_k,t)=0$, and
\[
\hat{V}^{(\varOmega)}(r_k,t) \equiv
\Exp^{\mathrm{i}\hat{H}^{(1)}t} \hat{V}^{(\omega)}(r_k)
\Exp^{\bar{\mathrm{i}}\hat{H}^{(1)}t}
\]
with
\begin{equation} \hat{V}^{(\omega)}_{mn}(r_k)
 =-\hat{\bm{d}}_{mn} \cdot \bm{E}^{(\omega)}(r_k)/2\hbar
 =-\tilde{\bm{G}}{}^{(\omega)}_{mn}(r_k) \cdot \hat{\bm{T}}{}^{mn},
 \label{slowVmn}
\end{equation}
and analogously for $\hat{\rho}'_{mn}(v_k,r_k,t)$, but
$\hat{\rho}'_{jj}(v_k,r_k,t) \simeq
\hat{\rho}^{(\varOmega)}_{jj}(v_k,r_k,t)$, it is not difficult to
receive an integral equation for slow component
$\hat{\rho}^{(\omega)}_{mn}(v_k,r_k)$ alone:
\begin{gather}
 \hat{\rho}^{(\omega)}_{mn}(v_k,r_k) =\int\limits^{\infty}_{0}
 \mathrm{d}\tau \Exp^{[\bar{\nu}+\mathrm{i}(\varOmega'
 -\hat{H}^{(1)}_{m})]\tau}
 \left\{\mathrm{i}\tilde{N}_{mn}(v_k)\hat{V}^{(\omega)}_{mn}(r_k)
 \vphantom{\int^{\infty}_{0}} \right.
\nonumber\\
 +\int\limits^{\infty}_{0} \mathrm{d}\tau_1 \left[
 \Exp^{(\bar{\nu}_m-\mathrm{i}\hat{H}^{(1)}_{m})\tau_1}\left(
 \hat{V}^{(\omega)}_{mn}(r_k) \hat{\rho}^{(\omega)\dagger}_{mn}(v_k,r_k)
 -\underline {\hat{\rho}^{(\omega)}_{mn}(v_k,r_k)
 \hat{V}^{(\omega)\dagger}_{mn}(r_k)} \right) \right.
\nonumber\\ \times
 \Exp^{\mathrm{i}\hat{H}^{(1)}_{m}\tau_1}\hat{V}^{(\omega)}_{mn}(r_k)
 +\hat{V}^{(\omega)}_{mn}(r_k)
 \Exp^{(\bar{\nu}_{n} -\mathrm{i}\hat{H}^{(1)}_{n})\tau_1}
\nonumber\\ \left. \vphantom{\int^{\infty}_{0}} \left.
 \times \left(
 \hat{\rho}^{(\omega)\dagger}_{mn}(v_k,r_k)
 \hat{V}^{(\omega)}_{mn}(r_k)
-\underline {\hat{V}^{(\omega)\dagger}_{mn}(r_k)
 \hat{\rho}^{(\omega)}_{mn}(v_k,r_k)} \right)
 \Exp^{\mathrm{i}\hat{H}^{(1)}_{n}\tau_1} \right] \right\}
 \Exp^{\mathrm{i}\hat{H}^{(1)}_{n}\tau}, \label {slow_rho_mn}
\end{gather}
where $\tilde{N}_{mn}(v_k)=\tilde{N}_{mn}W(v_k)$ with
$\tilde{N}_{mn}=\tilde{N}_m -\tilde{N}_n$; the rest of
designations has defined in (\ref {maxWell}) and (\ref {tildeN}).
The equation is solved by iterations on light field. Thus,
sequentially selecting  and $v_k$-integrating optical
nonlinearities, it is possible to determine linear and first
nonlinear susceptibilities of electrical polarization
\begin{equation}
 \bm{P}^{(\omega)}(r_k)
 \simeq \bm{\chi}^{(1)} \cdot \bm{E}^{(\omega)}(r_k)
 +\bm{\chi}^{(3)} \vdots \bm{E}^{(\omega) }(r_k)
 \bm{E}^{(\omega)*}(r_k) \bm{E}^{(\omega) }(r_k).
\end{equation}
In the second case alone the integration on $v_k$ makes the
underlined summands smaller then the previous ones by factor of
$\omega_{\mathrm{D}}/\nu$. It is convenient to extract
dimensionless tensor $X$-functions, defining
\begin{equation}
\begin{split}
 \bm{\chi}^{(1)} &= \mathrm{i}\tilde\chi_{nm} \bm{X}^{(1)mn}
 \\
 \text{and}\quad
 \bm{\chi}^{(3)} &=-\mathrm{i}\tilde\chi_{nm}
 \frac{|\tilde{d}_{mn}/\hbar\nu|^2}{2} \bm{X}^{(3)mn}.
 \end{split}
 \label{Xdef}
\end{equation}
Saturation parameter in (\ref{alpha}) is
\begin{equation}
\varkappa
= |2 G^{(\omega)}_{mn}/\nu|^2
= |d_{mn}E^{(\omega)}/\hbar\nu|^2. \label{saturpar}
\end{equation}
Here Rabi half-frequency $G^{(\omega)}_{mn}
=d_{mn}E^{(\omega)}/2\hbar$. It is possible differently to define
a saturation parameter, but concordantly with $s^{(3)}$ so that
their product did not vary. All our expansions on saturation
parameter concern just to (\ref{saturpar}), and it is supposed
what exactly it is small. \Eg, for light intensity $S(0)
=1\,\mathrm{mW}\,\mathrm{cm}^{-2}$ the Rabi half-frequency
$G^{(\omega)}_{mn} =12.7\,\mathrm{kHza}$. From here, with pressure
$P=6\,\mathrm{mtorr}$, when, according to \cite{BDDCh80}, $\nu
=100\,\mathrm{kHza}$, the saturation parameter $\varkappa
\simeq6.5\,\%$. For circularly polarized light in the distance
from our resonance the absorption factor with correction for
saturation is $\alpha_q \simeq\tilde{\alpha}^{(1)}_{mn}
(1-0.3\varkappa)$. We consider the fields of such intensity, that
it is possible to be limited to the first correction for
saturation.

In susceptibilities we extract the factor
\begin{equation}
 \tilde\chi_{nm}
=\tilde{N}_{nm}|\tilde d_{nm}|^2 \sqrt\pi/\hbar
 \omega_{\mathrm{D}}. \label{tildechi}
\end{equation}
Statistical weight of $j$-term, \ie\ total multiplicity of
rotation, spin, and parity degenerations, is
\begin{equation}
[j]\equiv \langle\hat1_j\rangle
   = \sum_{M_{J_j}  \mu^{\bm\cdot} p} 1
   = [J_j][p]\prod_\varsigma [I^{\varsigma}]. \label{statw}
\end{equation}
For methane
\begin{equation}
 p = -\delta_{\Gamma_J,A_1} +\delta_{\Gamma_J,A_2}   \pm
\delta_{\Gamma_J,E} +\delta_{\Gamma_J,F_1} -\delta_{\Gamma_J,F_2}
\label{ch4:p}
\end{equation}
and $[p]=\sum_p 1=1  +\delta_{\Gamma_J,E}$.

Using (\ref{absorpfactr}), we shall designate the maximum of
linear absorption factor as
\begin{align}  \tilde{\alpha}^{(1)}_{nm}
 \equiv \alpha^{(1)}_{q(\varOmega_q=0)}
&= 4\pi k\tilde\chi_{nm}
 X^{(1)mn\prime}_{\underdot{q}{}\dot q{}
 (\tilde\varOmega_q=0)}
\nonumber\\
&=\left(\pi/k\right)^2
 [m]A_{mn}\tilde{N}_{nm}/\sqrt\pi  \omega_{\mathrm{D}},
\label{maxabsorpindx}
\end{align}
where $A_{mn} =4k^3 |d_{mn}|^2/3\hbar$ is the rate of spontaneous
relaxation for ${[}^m_n$-transition ($[m]A_{mn}=[n]A_{nm}$).
Sometimes it is enough to know, that $\tilde{\alpha}^{(1)}_{nm}
\sim [n]$. The electro-dipole moments of vibration transition
$\omega_3$ (for methane isotopes ${}^{12,13}\mathrm{CH}_4$) are
${}^{12}d_{mn} =0.0534(3)\,\mathrm{deb}$ and ${}^{13}d_{mn}
=0.0530(3)\,\mathrm{deb}$ \cite{DNPR79}. From here $A_{mn} \simeq
4\,\mathrm{Hza}$ and it is only small part of all spontaneous
relaxation of $m$-term (without collisions) as the last is $\left.
\nu_m \right|_{P=0} \simeq 100\,\mathrm{Hza}$ \cite{BH69}.

If we use components
\begin{equation}
 X^{(1)mn}_{\underdot{q}{}_0\dot q{}_1}(\tau)
=\left\langle
  \hat T^{mn\dagger}_{\dot q{}_0}\Exp^{\bar{\mathrm{i}}
  \hat H^{(1)}_m\tau}
  \hat T^{mn       }_{\dot q{}_1}\Exp^{\mathrm{i}\hat H^{(1)}_n\tau}
 \right\rangle,
\label{X1:a:}
\end{equation}
the appropriate components of tensor $X$-functions
\begin{subequations}
  \begin{gather}
 X^{(1)mn}_{\underdot{q}{}_0\dot q{}_1}(B^{(0)})
 =\frac{\omega_{\mathrm{D}}}{\sqrt\pi}\int\limits_0^\infty
 \mathrm{d}\tau\,
 \Exp^{-(\omega_{\mathrm{D}}\tau/2)^2+(\mathrm{i}\varOmega -\nu)\tau}
 X^{(1)mn}_{\underdot{q}{}_0\dot q{}_1}(\tau)
 \label{X1:a}
\\
=\delta_{q_0,q_1}\sum_{\substack{m_2\\ n_1}}
 \mathrm{w}\left(\frac{\varOmega_{q_0}-\lambda_{m_2n_1}
  +\mathrm{i}\nu}{\omega_{\mathrm{D}}}\right) |T^{m_2n_1}_{\dot q{}_0}|^2
 \label{X1:b}
\\
\simeq\delta_{q_0,q_1}\mathrm{w}
 \left(\varOmega_{q_0}/\omega_{\mathrm{D}}\right)
 [p]\prod_\varsigma[I^{\varsigma}].
 \label{X1:c}
\end{gather}
\label{X1}
\end{subequations}
Here $ \varOmega_q = \varOmega + q\varDelta_J $ and
$\omega_{\mathrm{D}}$ has defined in (\ref{DopplerHW}).
$T^{m_2n_1}_{\dot q{}}  = \langle m_2|\hat T^{mn}_{\dot
q{}}|n_1\rangle$ and, according to equation (\ref{hfz:b}),
\[
H^{(1)}_j  \equiv  \langle  j|  \hat  H^{(1)}_j |j\rangle
=h_j-\varDelta_J M.
\]
For short we shall designate
\begin{equation}
 \Lambda_{ij} \equiv H^{(1)}_i-H^{(1)}_j
 \quad\text{and}\quad
 \lambda_{ij} \equiv h_i-h_j, \label{Lambda_lambda}
\end{equation}
where the set of subscripts (or quantum numbers) $j_l \equiv j
\tilde{\mu}^{\bm\cdot}_l M_l$, \eg,
$\Lambda_{m_2n_1}=-q\varDelta_J +\lambda_{m_2n_1}$.
\begin{equation}
 \mathrm{w}(\zeta)
 =\Exp^{-\zeta^2} \bigl( 1 + \frac{2\mathrm{i}}{\sqrt\pi}
 \int\limits_0^\zeta \mathrm{d}t \, \Exp^{t^2} \bigr), \label{wz}
\end{equation}
it is the error function (or probability integral) of a complex
variable \cite{AbSt68,Sheff75}. The transition to final expression
(\ref{X1:c}) is obtained with $\omega_{\mathrm{D}} \gg
|-\lambda_{m_2 n_1} +\mathrm{i}\nu |$. In conditions of isotropic
excitation of magnetic sublevels, HFS of Doppler contour for
linear susceptibility is discovered neither with frequency nor
with magnetic scanning.

\begin{figure}\centering
\includegraphics[scale=0.6]{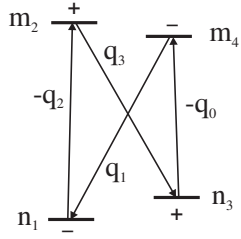}
\caption{\label{4leveldiagram} Four-level diagram corresponding to
resonance scattering of light.}
\end{figure}

HFS is discovered only with the account of correction from
saturation. Integrating it with respect to velocity of molecule
movement and at once with restriction $\nu =(\nu_m +\nu_n)/2$, we
obtain the appropriate spherical components of tensor
$X$-functions, containing all the necessary field structures:
\begin{subequations}
 \begin{gather}
    X^{(3)mn}_{\underdot{q}{}_0\dot q{}_1
    \underdot{q}{}_2\dot q{}_3}(B^{(0)})
 = \Exp^{-(\varOmega_{q_2}/\omega_{\mathrm{D}})^2}
    \iint\limits_0^{\phantom{\infty\infty}\infty}
    \nu^2 \mathrm{d}\tau'\, \mathrm{d}\tau \,
    \Exp^{\bar{\nu}_m\tau' +\bar{\nu}_n\tau}
\nonumber\\
 \quad \times \left\langle
    \hat T^{mn\dagger}_{\dot q{}_0}
    \Exp^{\bar{\mathrm{i}} \hat H^{(1)}_m\tau'}
    \hat T^{mn}_{       \dot q{}_1}
    \Exp^{\bar{\mathrm{i}} \hat H^{(1)}_n\tau }
    \hat T^{mn\dagger}_{\dot q{}_2}
    \Exp^{     \mathrm{i}  \hat H^{(1)}_m\tau'}
    \hat T^{mn}_{       \dot q{}_3}
    \Exp^{     \mathrm{i}  \hat H^{(1)}_n\tau } \right\rangle
\label{X3:a}
\\
 = \Exp^{-(\varOmega_{q_2}/\omega_{\mathrm{D}})^2}
    \sum_{\substack{m_2m_4\\ \,\,n_1\,\,n_3}}
    \frac{T^{m_4n_3*}_{\dot q{}_0}\,T^{m_4n_1}_{\dot q{}_1}\,
          T^{m_2n_1*}_{\dot q{}_2}\,T^{m_2n_3}_{\dot q{}_3}\,
    \nu^2}{\left[\nu_m
  -\mathrm{i}\left(q_{12}\varDelta_J +\lambda_{m_2m_4}
    \right)\right]\left[\nu_n
  -\mathrm{i}\left(q_{32}\varDelta_J +\lambda_{n_3n_1}
    \right)\right] }. \label{X3:b}
 \end{gather}
\label{X3}
\end{subequations}
Here $q_{ij} \equiv q_i-q_j$ and $q_{10} = q_{23}$. In general
case
the separate summands of last expression (\ref{X3:b}) are
pictorially represented with closed four-level diagrams of
$\Join$-type,\label{4leveldescript} see
\figurename~\ref{4leveldiagram}
[it is necessary to associate a oriented line segment
$\overset{q_k}{\longrightarrow}$ from $m_i$-(sub)level to
$n_j$-(sub)level with $ T^{m_i n_j}_{\dot{q}{}_k}$]. In cases
$J_{mn} =\pm1$, in accordance with selection rules [see
(\ref{exactT1I1}) and further between (\ref{select_tildemu}) and
(\ref{regular_select})], the spectral manifestations of four-level
diagrams with unequal (both lower and upper) indices are
essentially weakened. The situation is better, when there is index
coincidence only in one pair of lower or upper levels, that gives
the Raman three-level diagrams, respectively, $\bm{\vee}$- or
$\bm{\wedge}$-types. Their spectral manifestations, as ``Raman
HFS'' of the same types, are conditioned by appropriate complex
resonance co-factor from the pair disposed in denominator of
expression (\ref{X3:b}) (the complexity can be manifested only in
the diagrams with differently polarized light components). At last
the simultaneous index coincidences in both upper and lower level
pairs give the two-level diagrams of $\bm{\between}$-types. The
collision constants then enter only in a simple nonresonance
factor $\nu^2/\nu_m\nu_n$, and all resonance feature are
determined with field dependence of real numerator of expression
(\ref{X3:b}). We have introduced the name ``ballast HFS'' for
resonance structures of the last type. In the pure state they are
present in the transmission of circularly polarized radiation.
Their distinctive feature is always negative sign with respect to
Raman HFS. It is possible to make certain of it, \eg, using just
described expansion of expression (\ref{X3:b}) on components:
$\bm{\between} +\bm{\vee} +\bm{\wedge} +{\Join}$. It is easy to
see, that all the components, except last one, are positive. For
transitions with changing $J_j$, the last one can be rejected, as
it has the next smallness order $J^{-4}$ with respect to previous
components; when $B^{(0)}=0$, it is visible from
(\ref{exactT1I1}). As a result, using an inequality
\[
\left| \Real \prod_j (1 -\mathrm{i} x_j)^{-1} \right| \leq1
\]
with $x_j \in \mathbb{R}$, we have
\begin{equation}
\bm{\between} +\bm{\vee}  +\bm{\wedge} +{\Join} \leq \text{wing},
\label{compofdip}
\end{equation}
\ie, if the radiation is circularly polarized, the total
expression (\ref{X3:b}) with any $B^{(0)}$ near resonance, is
always less than its wing value, when $B^{(0)}$ is large. The
first summand in (\ref{compofdip}) is always resonance. The
residuals are completely or partially suppressed (in according to
whether $\nu =0$ or not) because of their nonresonance character
connected to anti-crossings of the diagram levels. With large
$\nu$ all four summands in sum (\ref{compofdip}) reproduce the
wing.

For multi-spin molecule the resonance for circularly polarized
radiation consists of several dips. These dips are manifested,
when light interacts with rotation subsystem and feels that the
hyperfine coupling of any spin subsystem (as a ballast) take
place. Thus, unlike Raman scattering, the gas medium property to
absorb light is increased. The light energy indirectly comes in
nuclear spin subsystems and is spent on flipping nuclear spins of
molecule.

An elementary mechanical model, imitating the ballast structure of
the spectrum, consists of two interacting tops. \Eg, let they will
be a pencil vertically clamped in hands and a gramophone plate
freely pined upon it. The pencil here corresponds to rotation
subsystem and the plate --- to ballast spin one. The small
friction between them corresponds to hyperfine coupling (without
precession). Twisting the pencil between hands we shall imitate an
influence of light to our molecule being capable of its
absorption. The frequency of twisting is analog of a collision
frequency $\nu$, restricting the influence of light. Comparing
these absorption capabilities determined with the low and high
frequencies, it is not difficult to see that it is higher in the
first case than in the second one. In the first case the pencil
and the plate rotate together as a whole. In the second case the
pencil rotates, practically not having time to pull about the
pined plate.\footnote {See also the derivation of formula
(\ref{SimpleModel}) for the elementary model of levels.}

In other (equivalent) interpretation the ballast structure --- the
manifestation of anti-crossings of hyperfine magnetic sublevels
\cite{Eck67}. For the first time this structure was observed in
resonance fluorescence of lithium atoms \cite{EFW63}.
As it is known
\cite
{RSh91}, the structure of field spectra in nonlinear transmission
of pumping (\ref{s3}) is analogous\footnote{If spontaneous
relaxation ($A_{\mathrm{mn}}$) is much less than collision one
($\nu_j$),
otherwise see p.~\pageref{spontan}.} to the structure in resonance
fluorescence \cite{AChKh93}
\begin{equation}
s^{(2)}_{(\bm{e},\bm{f})} =
\tilde{\bm{X}}{}^{(3)mn}_{(\hat{H}^{(1)}_{n} =0)}
\,\underset{\displaystyle\cdot}{\vdots}\, \bm{f}^{*} \otimes\bm{e}
\otimes\bm{e}^{*} \otimes\bm{f} \quad (\in \mathbb{R}).
\label{resfluor}
\end{equation}
Here nonlinear cubic susceptibility (\ref{Xdef}) or differently
fourth rank tensor (\ref{modXdef}) of gas medium is contracted
twice with $\bm{e}$ and twice with $\bm{f}$, respectively, the
unit polarization vectors of optical pumping (excitation) and
resonance fluorescence; also it is noted, that the multiplet
structure of lower level ($n$) in resonance fluorescence of upper
one ($m$) is not manifested (and that ensures the real value of
contracted expression). However, the analogy and comparison are
possible only theoretically, as both ordinary and resonance
fluorescences in our IR range are practically unobservable, unlike
nonlinear transmission of pumping.

\begin{figure}\centering
\includegraphics[width=0.614\textwidth]{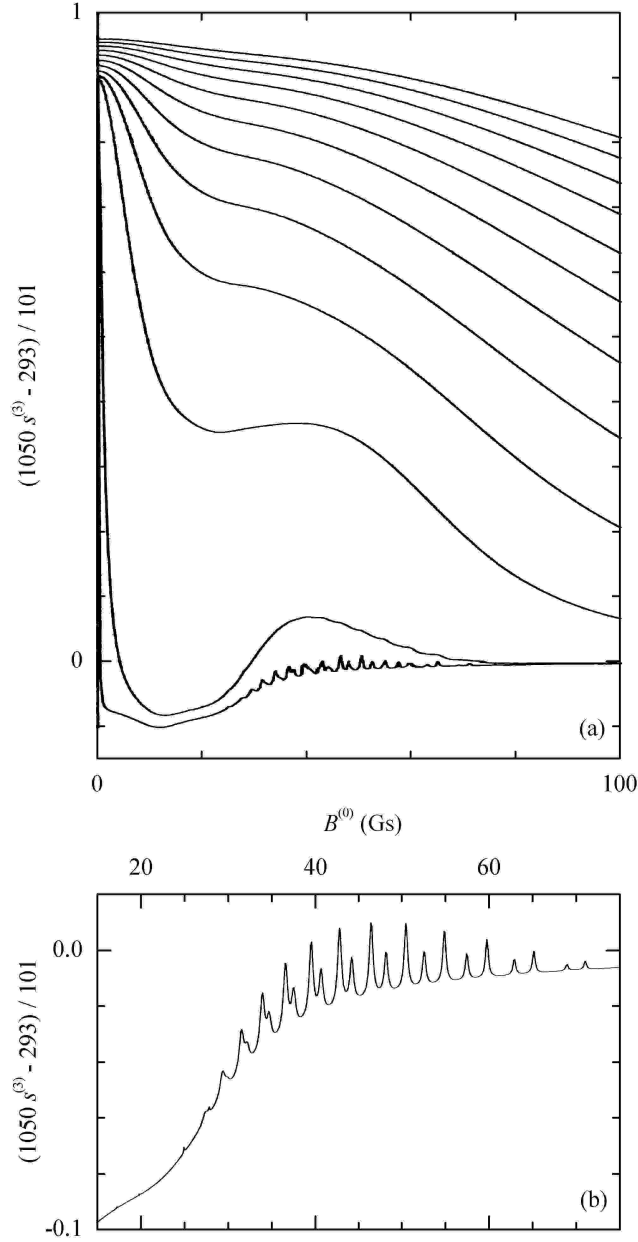}
\caption{\label{12ch4_th_l} Exact calculated NOR/M$_\parallel$
(even functions) in transmission of linearly polarized radiation
for component ($\omega_3$ $P(7)$ $F^{(2)}_{2(-)}$) of isotope
${}^{12}\mathrm{CH}_4$: (a) $\nu=0.1$, $1$, $10$, $20$, $\ldots$,
$100\,\mathrm{kHza}$ (the curves with greater $\nu$ pass higher);
(b) $\nu=0.1\,\mathrm{kHza}$.}
\end{figure}

\begin{figure}\centering
\includegraphics[width=0.614\textwidth]{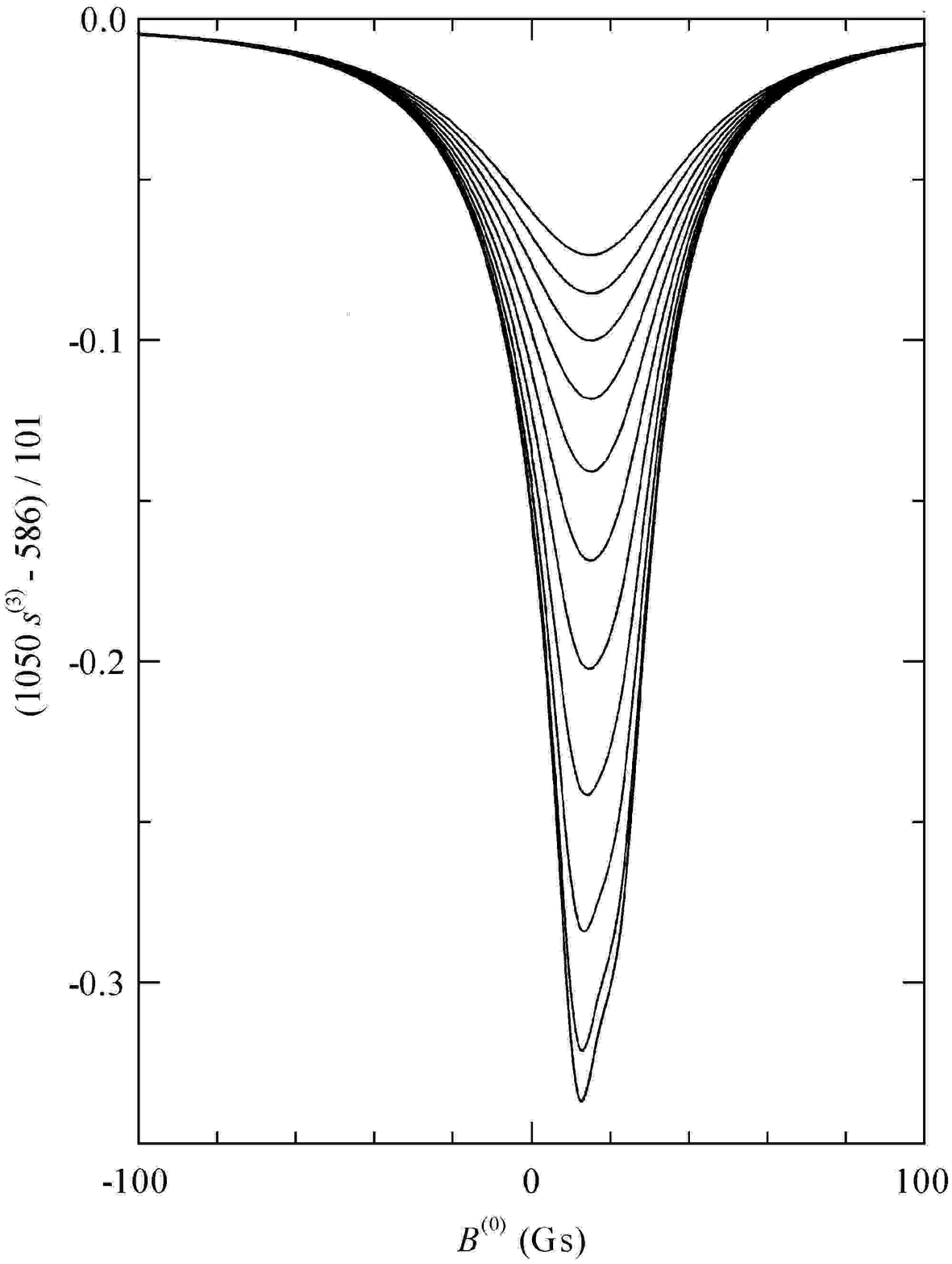}
\caption{\label{12ch4_th_c} Exactly calculated NOR/M$_\parallel$
in transmission of right circularly polarized radiation for
component ($\omega_3$ $P(7)$ $F^{(2)}_{2(-)}$) of isotope
${}^{12}\mathrm{CH}_4$: $\nu=0.1$, $1$, $10$, $20$, $\ldots$,
$100\,\mathrm{kHza}$ (the dip with maximal depth practically
corresponds to the first two cases, and the ones with greater
$\nu$ are shallower).}
\end{figure}

As an example to the described classification we shall represent
the results of exact calculations of formula (\ref{X3:b}) with use
of a computer, for NOR/M$_\parallel$ in
 radiation transmission of methane isotope ${}^{12}\mathrm{CH}_4$.
In \figurename~\ref{12ch4_th_l}, concerning to a linearly
polarized radiation, the Raman HFS is shown obviously with
impurity of ballast.
The purely Raman HFS consists of peaks only and never passes below
than wing level at large $B^{(0)}$. The purely ballast HFS is
shown in next \figurename~\ref{12ch4_th_c}, concerning to
circularly polarized radiation.
It is practically symmetric dip,\footnote{Its asymmetry is barely
visible.} displaced to the right\footnote{Where $\bm{B}^{(0)}
\cdot\bm {k}
>0$.} for rightly polarized radiation.\footnote{Our definition of
right circularly polarized radiation is from
\cite{FLS63:I}.}
The dip for contrary polarized radiation is obtained by mirror
reflection with respect to ordinate axis, \ie\ its shift has
opposite sign. The shift does not practically depend on pressure.
With increasing $\nu$ on an order, from $10$ to
$100\,\mathrm{kHza}$, the width on half-depth and the depth of the
dip are approximately doubled and quartered, respectively.

\begin{figure}\centering
\includegraphics[width=0.614\textwidth]{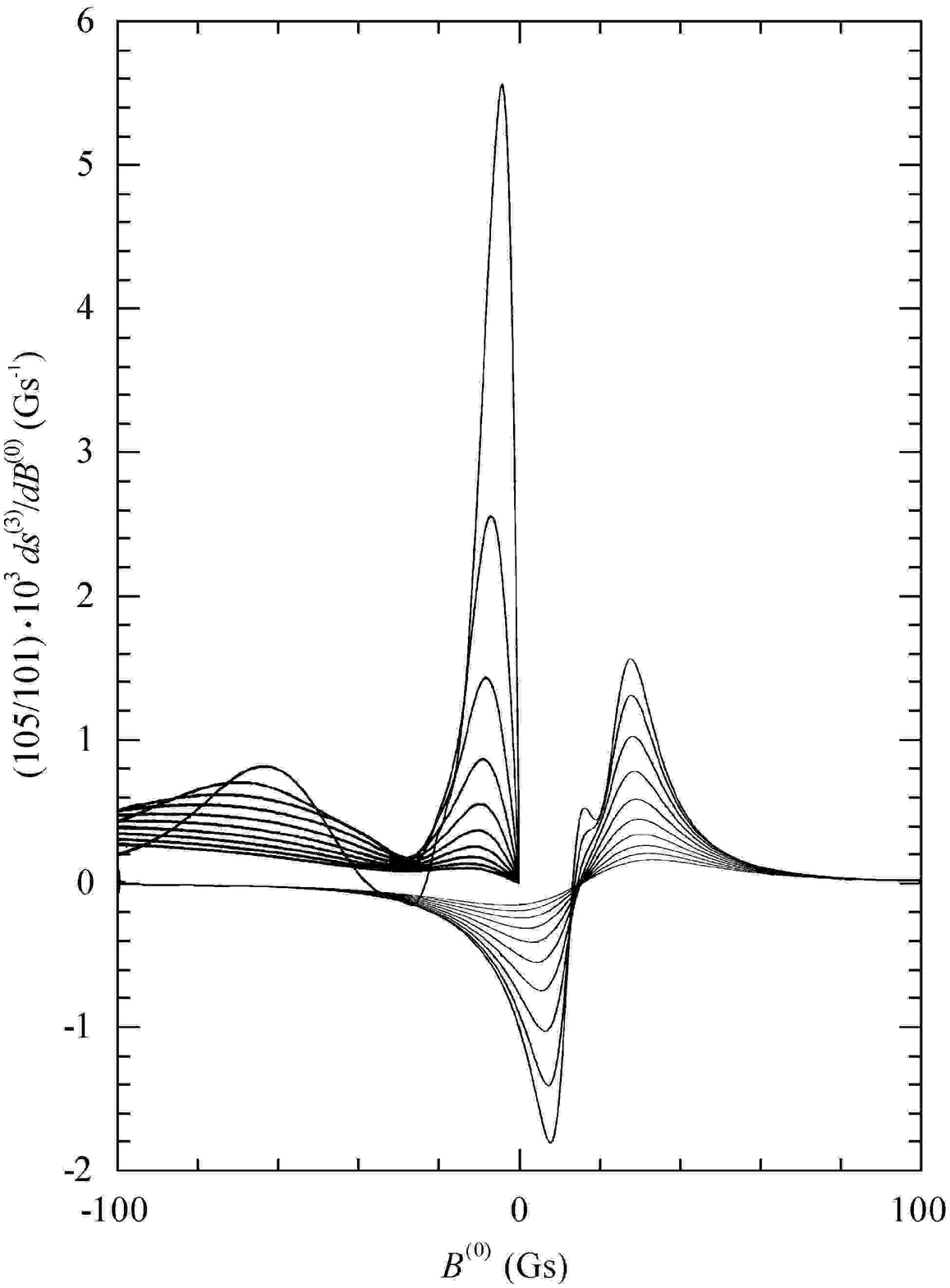}
\caption{\label{12ch4_th_dlc} Derivatives of exact calculated
NOR/M$_\parallel$ in transmission of radiations polarized linearly
(odd functions; their negative branches with $B^{(0)}>0$ are
omitted) and right-circularly for component ($\omega_3$ $P(7)$
$F^{(2)}_{2(-)}$) of isotope ${}^{12}\mathrm{CH}_4$:
$\nu=10,20,\ldots,100\, \mathrm{kHza}$ (here the curves with
greater $\nu$ have smaller amplitude).}
\end{figure}

To facilitate a comparison with the above-represented experimental
data, derivatives are shown in \figurename~\ref{12ch4_th_dlc}, and
they appropriate to the previous two FIGs.
Judging by amplitude ratio of narrow and wide structures for
linearly polarized radiation, the experimental curves in
\figurename~\ref{12ch4_ex_dlc-13ch4_ex_dc} at the left correspond
to case $\nu\simeq 40\,\mathrm{kHza}$. As we have seen, the dip in
\figurename~\ref{12ch4_th_c} for circularly polarized radiation is
practically symmetric and displaced. Respectively, the dip
derivative is antisymmetric and also displaced. For the more
detailed analysis it is convenient to describe an existing small
quantity of dip asymmetry for circularly polarized radiation in
the following way. On a curve, representing derivative, let us
mark (always existing) two ordered points, where the module of
amplitude of this derivative is maximum, by means of coordinates
$(B^{(0)},\partial s^{(3)}/\partial B^{(0)})$, namely, $(x_0,y_0)$
and $(x_1,y_1)$. An ordering condition for the points is $|x_0|<
|x_1|$. It is convenient for that allows to do not pay attention
to what circular polarization is considered. The quantity of dip
asymmetry is now determined as $a_y =1-|y_0/y_1|$. Judging by
\figurename~\ref{12ch4_th_dlc}, this quantity is small and also
changes its sign at $\nu\simeq 30\,\mathrm{kHza}$. There is a
feature observed in \figurename~\ref{12ch4_th_dlc} only with small
$\nu$, when the second field derivative in the area of resonance
(without wings) begins to have four zero points instead of two
ones. It is possible to estimate the position of the feature,
changing the asymmetry sign, as (\ref{Ncross}).\footnote{The
feature position sign is the same as the dip shift one.} It is
approximately greater by factor of $\sqrt2$ than the dip shift or
otherwise the position of its lower point, see below
(\ref{anticross}).

The described structures can be represented in zero-pressure
limit, when $\nu$ is determined only by spontaneous decay. In
\figurename~\ref{12ch4_th_l}(b) at nonzero fields we clearly
distinguish the two types of peaks grouped in pairs,
$\bm{\vee}^{m\prime}_{nq}$ and
$\bm{\wedge}^{m\prime}_{nq}$.\footnote{The superscript
${}^{\prime}$ has been defined on p.~\pageref{primes}.} Without
HFS, according to (\ref{without_hfs}), the amplitude ratio of
peaks at zero field is $\bm{\vee}^{m\prime}_{nq}
/\bm{\wedge}^{m\prime}_{nq}\simeq a^2_{m2} /a^2_{n2}$ and
$\lesseqqgtr1$, if $J_{mn} \lesseqqgtr0$, respectively. With HFS,
there is the similar amplitude ratio of peaks at nonzero fields.
We mention in passing that the approximation, represented in the
next section~\ref{Approx_J}, is not sufficient for real deriving
of this ratio.

In appendix
the classification, represented here for HFS of NOR/M, understands
on a simple model case, when both level angular momenta and spin
are equal $1/2$. Let us note, that the case does not concern
directly to the molecules considered by us, and what is more the
case does not concern also to atoms. In their visible frequency
area, instead of HFS an other structure goes out on the first
plan, namely, spontaneous one, connected with the induced
two-photon absorption through an intermediate spontaneous decay
\cite{GR90b}.

Coming back to the formulae (\ref{X3}), we shall note the
following. Here we use only
the expression (\ref{X3:b}), but the initial expression
(\ref{X3:a}) can be invariantly calculated, \ie\ without the help
of specific basis of wave functions. Properly using only
commutation relations for tensor components of operators included
in it,
at first the trace $\langle \ldots \rangle$ may be calculated, and
then time integrals. Another alternative calculation way is
possible also, permitting to interchange the sequence of these
operations. For that it is necessary to use a resolvent
representation of evolutionary operators
\cite[Chap.\,XXI \S\,13]{Mes62} (see also
\cite[Chap.\,V \S\,4]{Gant60}):
\begin{equation}
\Exp^{\bar{\mathrm{i}} \hat{H}^{(1)}_j t}
=\oint\limits_\circlearrowleft \frac{\mathrm{d}z}{2\pi \mathrm{i}}
\Exp^{\bar{\mathrm{i}} z t} \hat{R}^{(1)}_j(z). \label{resolv}
\end{equation}
The resolvent $\hat{R}^{(1)}_j(z)$ has defined in
(\ref{Resolv_j}). Its complex spectral parameter $z$ passes on
integration contour $\circlearrowleft$, enveloping all points of
spectrum for operator $\hat{H}^{(1)}_j$, \ie\  the poles
(singularities) of its resolvent. After calculation in
(\ref{X3:a}) the time integrals, we obtain
\begin{multline}
    X^{(3)mn}_{\underdot{q}{}_0\dot q{}_1
    \underdot{q}{}_2\dot q{}_3}(B^{(0)})
 = \Exp^{-(\varOmega_{q_2}/\omega_{\mathrm{D}})^2}
\left( \prod_{k=1,2,3,4}\oint\limits_\circlearrowleft
\frac{\mathrm{d}z_k}{2\pi \mathrm{i}} \right) \frac{\nu^2}{(\nu_m
-\mathrm{i} z_{24})(\nu_n -\mathrm{i} z_{31})}
\\ \times
\left\langle
    \hat T^{mn\dagger}_{\dot q{}_0} \hat{R}^{(1)}_m(z_4)
    \hat T^{mn}_{       \dot q{}_1} \hat{R}^{(1)}_n(z_1)
    \hat T^{mn\dagger}_{\dot q{}_2} \hat{R}^{(1)}_m(z_2)
    \hat T^{mn}_{       \dot q{}_3} \hat{R}^{(1)}_n(z_3)
\right\rangle \label{X3:aa}
\end{multline}
From here at once, using the basis (\ref{diag}), the equation
(\ref{X3:b}) is obtained. However it is possible to calculate the
expression (\ref{X3:aa}), as well as (\ref{X3:a}), without
concretizing the basis set of wave functions. In the given paper
we shall not go into more details of these two ways of
calculation.

The model, used by us, has simple and rather reliable basis in
itself. What concerns a comparison with methane experiment, here a
discussion can go only about details. \Eg, if someone will
interest collision components, their calculation in the model of
strong deorientational collisions is produced below in
section~\ref{COL_NOR_M}. One may assume that the transition to
other molecules is more actual. The main (hyperfine) structure of
the field spectrum of NOR will be thus involved. \Eg, in
section~\ref{DoubleHFS} the molecule of fluoromethane is
considered.

\section{\label{Approx_J}High-$J$ approximation of HFS of
NOR/M$_\parallel$ in ${}^{13}\mathrm{CH}_4$}

To simplify the analysis of the problem, when spin subsystems more
than one, we consider the states, in which the molecule rotates
rather fast. Owing to that, in (\ref{JI:II}) one can take into
consideration only isotropic (scalar) spin-rotation interaction,
as in (\ref{hfz:a}).
When $J\gg I^{\varsigma}$ for all $\varsigma$, it is possible to
put down in (\ref{hjM}) for zero approximation\footnote{Indexing
shows the order of Cartesian components of the vector.}
\begin{equation}
\bm{J}_{j,M-\underline{\hat{I}}{}_z} \simeq\bm{J}_{jM}\simeq
\left( \sqrt{\hat{\bm{J}}{}_j^2-M^2}, 0, M \right)_{x,y,z}.
\label{JjMxyz}
\end{equation}
In the last expression we used that $J_{y,jM}\simeq0$. Thus we
obtain a linearization of Hamiltonian (\ref{hjM}) on spins
$\underline{\hat{\bm{I}}}{}^{\varsigma}$, \ie\
\begin{subequations}
\begin{gather}
 \underline{\hat{h}}{}^{\varsigma}_{jM}   \simeq
 -\left(C^{\varsigma}_{\mathrm{a}} \bm{J}^*_{jM}
 +\bm\varDelta^{\varsigma}_{IJ}\right)
  \cdot \underline{\hat{\bm{I}}}{}^{\varsigma} \simeq
 -\bm\omega^{\varsigma}_{jM}
 \cdot \underline{\hat{\bm{I}}}{}^{\varsigma},
\label{lineHs}
\\ \intertext{where vector}
 \bm\omega^{\varsigma}_{jM} \simeq \left( C^{\varsigma}_{\mathrm{a}}
 \sqrt{\hat{\bm{J}}{}_j^2-M^2},  0,  C^{\varsigma}_{\mathrm{a}} M
 +\varDelta^{\varsigma}_{IJ} \right)_{x,y,z}.
\label{omega_jM}
\end{gather}
\end{subequations}
In this approximation the commutator
$[\underline{\hat{h}}{}^{\varsigma_1}_{jM_1},
\underline{\hat{h}}{}^{\varsigma_2}_{jM_2}] =0$, therefore it is
possible to consider independently all interactions of rotation
subsystem with spin ones. We shall use approximately factored
transforming operators in (\ref{diag}), namely
\begin{equation} \hat{\mathcal{U}}_{jM} \simeq
 \prod_\varsigma \hat{\mathcal{U}}{}^{\varsigma}_{jM} \simeq
 \prod_\varsigma \Exp^{\bar{\mathrm{i}}\beta^{\varsigma}_{jM}
 \hat{I}^{\varsigma}_y}
\end{equation}
with $\hat{I}{}^{\varsigma}_{y}= \bm{u}^{\varsigma}_{y} \cdot
\hat{\bm{I}}{}^{\varsigma}$. Now this transformation is the set of
independent rotations in spin function spaces around basis vectors
$\bm{u}^{\varsigma}_{y}$ on angles $\beta^{\varsigma}_{jM}$, which
are determined through unit vectors
\begin{equation}
 \left( \sin\beta^{\varsigma}_{jM}, 0, \cos\beta^{\varsigma}_{jM}
   \right)_{x,y,z} = \bm{n}^{\varsigma}_{jM}  \equiv
   \bm\omega^{\varsigma}_{jM} /\omega^{\varsigma}_{jM}.
\label{norm_omega}
\end{equation}
The rotation is arranged so, that all matrices
$\underline{\hat{h}}{}^{\varsigma}_{jM}$ in (\ref{hjM}),
originally defined by us in old basis (\ref{split}), will be
transformed in unitarily similar diagonal ones
$\underline{\hat{h}}{}^{\varsigma\prime}_{jM}$, defined in new
basis (\ref{diag}), \ie\
\begin{gather}
 \left( \underline{\hat{h}}{}^\prime_{jM}
 \right)_{\tilde{\mu}^{\bm\cdot}\tilde{\mu}^{\bm\cdot}} =
 \langle j\tilde{\mu}^{\bm\cdot}M| \hat h_j
 |j\tilde{\mu}^{\bm\cdot}M \rangle
\nonumber\\ \intertext{and}
 \underline{\hat{h}}{}^\prime_{jM}=\hat{\mathcal{U}}^\dagger_{jM}
 \underline{\hat{h}}{}_{jM}\hat{\mathcal{U}}_{jM}=
 -\sum_\varsigma \omega^{\varsigma}_{jM}
 \underline{\hat{I}}{}^{\varsigma}_z. \label{hdiag}
\end{gather}
Thus the (hyperbolic) approximation of spectrum of total
Hamiltonian is
\begin{equation}
 H^{(1)}_{j\tilde{\mu}^{\bm\cdot}_jM}
= -\sum_\varsigma  \tilde{\mu}^\varsigma_j \omega^\varsigma_{jM}
  -\varDelta_J M. \label{approx_hfs}
\end{equation}
The expansion near $B^{(0)}=0$ is
\begin{subequations}
\begin{equation}
 H^{(1)}_{j\tilde{\mu}^{\bm\cdot}_jM}(B^{(0)})
 \simeq -\sum_\varsigma \tilde\mu^\varsigma_j
 C^\varsigma_{\mathrm{a}} (J_j +1/2)
 -M\gamma^{(\mathrm{eff})}_{j\tilde\mu^{\bm\cdot}_j}B^{(0)}.
\label{approx_hfs_B0}
\end{equation}
Effective gyromagnetic ratio
\begin{equation}
 \gamma^{(\mathrm{eff})}_{j\tilde\mu^{\bm\cdot}_j}
 =\left. \gamma^{(\mathrm{eff})}_{j\tilde\mu^{\bm\cdot}_j M}
 \right|_{B^{(0)}=0}
 =\left. -M^{-1}\partial_{B^{(0)}} H^{(1)}_{j\tilde\mu^{\bm\cdot}_j M}
 \right|_{B^{(0)}=0}
 \simeq \gamma_J +\sum_\varsigma \tilde\mu^\varsigma_j
 \gamma^\varsigma_{IJ} / (J_j +1/2), \label {gyrom_eff_approx}
\end{equation}
\end{subequations}
where there is summation on spin subsystems $\varsigma$ with
$C^{\varsigma}_{\mathrm{a}} \neq0$ only. From here one can
determine the total ordered\footnote{Respectively, from top to
bottom hyperfine components of \figurename~\ref {12ch4_th_hfs}.}
set of $g$-factors,
$g^{(\mathrm{eff})}_{j\tilde{\mu}^{\mathrm{H}}}
=\gamma^{(\mathrm{eff})}_{j\tilde{\mu}^{\mathrm{H}}}
/\gamma_{\mathrm{N}}$, for $P(7)$-branch of methane with
$I^{\mathrm{H}} =1$:
\begin{equation}
\begin{split}
 g^{(\mathrm{eff})}_{m;\bar1,0,1} &\simeq -0.5,\, 0.3,\, 1.1
\\
\text{and}\quad
 g^{(\mathrm{eff})}_{n;\bar1,0,1} &\simeq -0.4,\, 0.3,\, 1.
 \end{split}
\end{equation}
The exact expressions for $g$-factors (\ref{g_F_exact}) can be
compared with their approximation (\ref{gyrom_eff_approx}) in the
form of gyromagnetic ratios.

In split basis of wave functions (\ref{split}), the matrix
elements of covariant components of standard vectorial operator of
${[}^m_n$-transition are
\begin{subequations}
\begin{gather}
  \left(\underline{\hat{T}}{}^{mn}_{\dot q{}M}
 \right)_{\mu^{\bm\cdot}\mu^{\bm\cdot}} \equiv
 \langle m\mu^{\bm\cdot}M+q|\hat T^{mn}_{\dot q{}}|
         n\mu^{\bm\cdot}M\rangle
\nonumber\\ \intertext{with}
 \underline{\hat{T}}{}^{mn}_{\dot q{}M}=
 T^{mn}_{\dot q{}(M-\underline{\hat{I}}{}_z)}.
\label{amplitudes:a}
\end{gather}
In new basis (\ref{diag})
\[
\left(\underline{\hat{T}}{}^{mn\prime}_{\dot q{}M}
 \right)_{\tilde{\mu}^{\bm\cdot}_m\tilde{\mu}^{\bm\cdot}_n} \equiv
 \langle m\tilde{\mu}^{\bm\cdot}_mM+q|\hat T^{mn}_{\dot q{}}|
         n\tilde{\mu}^{\bm\cdot}_nM\rangle
\]
with
\begin{equation}
 \underline{\hat{T}}{}^{mn\prime}_{\dot q{}M}=
 \hat{\mathcal{U}}^\dagger_{m,M+q}\underline{\hat{T}}{}^{mn}_{\dot q{}M}
 \hat{\mathcal{U}}_{nM}
 \simeq \Exp^{\mathrm{i}\sum_\varsigma
 \beta^{\varsigma mn}_{qM}
 \underline{\hat{I}}{}^{\varsigma}_y }
 T^{mn}_{\dot q{} \bigl( M-\sum_\varsigma
 (n^{\varsigma}_{z,nM} \underline{\hat{I}}{}^{\varsigma}_z
 -n^{\varsigma}_{x,nM} \underline{\hat{I}}{}^{\varsigma}_x) \bigr) }
 \label{amplitudes:b}
\end{equation}
Using here the Taylor expansion on
$\underline{\hat{\bm{I}}}{}^{\varsigma}$ and being limited
quadratic summands, we obtain
\begin{multline}
 \underline{\hat{T}}{}^{mn\prime}_{\dot q{}M}
 \Rightarrow T^{mn}_{\dot q{} M}
 \left\{\hat1 -\sum_\varsigma
 \left[ \frac{\partial_M T^{mn}_{\dot q{} M}}{T^{mn}_{\dot q{} M}}
        n^\varsigma_{z,nM} \underline{\hat{I}}{}^\varsigma_z
+\frac{1}{\sqrt2} \left( \beta^{\varsigma mn}_{qM}
 +\frac{\partial_M T^{mn}_{\dot q{} M}}{T^{mn}_{\dot q{} M}}
    n^\varsigma_{x,nM} \right)
    \underline{\hat{I}}{}^\varsigma_{\dot1}
 \right.\right.
\\ \left.\left.
+\frac{1}{\sqrt2} \left( \beta^{\varsigma mn}_{qM}
 -\frac{\partial_M T^{mn}_{\dot q{} M}}{T^{mn}_{\dot q{} M}}
    n^\varsigma_{x,nM} \right)
    \underline{\hat{I}}{}^\varsigma_{\dot{\bar1}}
+\frac12 \beta^{\varsigma mn 2}_{qM}
   \underline{\hat{I}}{}^{\varsigma 2}_y
\right] \right\}. \label{amplitudes:c}
\end{multline}
\end{subequations}
The arrow ($\Rightarrow$) here marks that from quadratic
corrections only one is kept, which gives the contribution to
resonance.

Here the covariant spherical components
$\underline{\hat{I}}{}^{\varsigma}_{\dot{q}}$ are standardly
defined and similar with (\ref{stdspherbasis}). The difference
$\beta^{\varsigma mn}_{qM} \equiv\beta^{\varsigma}_{m,M +q}
-\beta^{\varsigma}_{nM}$ and small, therefore below, in
(\ref{all_circ}), it is possible to substitute
\begin{equation}
     \beta^{\varsigma mn}_{qM} \simeq
 \sin\beta^{\varsigma mn}_{qM}
=-[\bm{n}^{\varsigma}_{m,M+q}\times
   \bm{n}^{\varsigma}_{n M}]_y.
\label{betamn}
\end{equation}

As it was above noted, when the field $B^{(0)}$ is close to zero,
$F^{\varsigma}_j -J_j \simeq\tilde{\mu}^{\varsigma}_j$, where
$\hat{\bm{F}}{}^{\varsigma}_j =\hat{\bm{J}}_j
+\hat{\bm{I}}{}^{\varsigma}$. Here there are selection rules for
the matrix elements of electro-dipole transition ${[}^m_n$
\cite{Sob72}, namely,\footnote{In case of only one spin
subsystem.}
\begin{equation}
|F^\varsigma_{mn}|=|J_{mn}+\tilde\mu^\varsigma_{mn}|\leq1
\label{select_tildemu}
\end{equation}
with $\tilde{\mu}^{\varsigma}_{mn} \equiv
\tilde{\mu}^{\varsigma}_m - \tilde{\mu}^{\varsigma}_n$, also
$F^\varsigma_m +F^\varsigma_n \geq1$ and for parities $p_mp_n
=-1$.

When the field $B^{(0)}$ is arbitrary, we can use the expansion
(\ref{amplitudes:c}). The corrections, giving the contributions
only in resonance wing, have been omitted though they are
important for selection rules of weaker (satellite) transitions.
The given expansion gives us selection rules in arbitrary magnetic
field, when there are the main transitions with an order $\epsilon
\equiv |\tilde{\mu}^{\varsigma}_{mn}| =0$ and satellite ones with
$\epsilon =1,2,\ldots$. The satellite transitions are weaker,
$\propto J^{-\epsilon}$, in respect to main ones. Let us remark,
that in the correspondence with (\ref{select_tildemu}) for $J_{mn}
=-1$, \ie\  in $P$-branch, there should be an asymmetry of
satellites [see also (\ref{exactT1I1})]. Projecting it on
(\ref{amplitudes:c}), we see that in the first order at satellites
it should be
\begin{equation}
\left( \beta^{\varsigma mn}_{qM}
 -\frac{\partial_M T^{mn}_{\dot q{} M}}{T^{mn}_{\dot q{} M}}
    n^\varsigma_{x,nM} \right)_{B^{(0)}=0} =0
\label{regular_select}
\end{equation}
and similarly in the second one (of course taking into account the
dropped wing corrections). At $B^{(0)} =0$, according to
(\ref{select_tildemu}), in $P$-branch the satellites with more
than second order should not be at all. Within the framework of
our approximation it is applied not for all $M$, though the
tendency to that is kept and conducts to partial suppression of
$\Join$-type summands\footnote{See below (\ref{Join}).} for
NOR/M$_\parallel$ in $P$-branch.

To describe structural features of NOR/M$_\parallel$ in linearly
polarized radiation, it is enough to take into account the
fixed\footnote{\Ie\ independent from the field $B^{(0)}$.}
contribution of main transitions in Taylor expansion for
(\ref{amplitudes:c}) with $|\tilde{\mu}^{\varsigma}_{mn}| =0$ and
even without the contributions $\sim J^{-1}$ with respect to fixed
one. As a result we obtain
\begin{gather}
 \bm{\vee}^m_{nq}
 +\bm{\wedge}^m_{nq}
\equiv
 X^{(3)mn}_{\underdot{q}{}\dot q{}\underdot{\bar{q}}{}\dot{\bar{q}}{}}
 +X^{(3)mn}_{\underdot{q}{}\dot{\bar{q}}{}\underdot{\bar{q}}{}\dot q{}}
\nonumber\\
 \simeq \Exp^{-(\varOmega_{\bar{q}}/\omega_{\mathrm{D}})^2}
 \frac{\nu^2}{\nu_m \nu_n}
 \sum_{\tilde{\mu}^{\bm\cdot}M} \left[
 \frac{|T^{mn}_{\dot q{}M}T^{mn}_{\dot{\bar{q}}{}M}|^2\,
 \nu_m}{\nu_m -\mathrm{i}\left(2q\varDelta_J +\sum_\varsigma
 \tilde{\mu}^{\varsigma} \omega^{\varsigma mm}_{M+q,M-q} \right)}
 +(m\leftrightarrow n) \right].
\label{line}
\end{gather}
Here $\omega^{\varsigma j'j}_{M'M} \equiv\omega^\varsigma_{j'M '}
-\omega^\varsigma_{jM}$. The summands in square brackets of the
formula are ordered and obtained by mutual permutation of indices,
\ie\ $\bm{\wedge}^m_{nq} =\bm{\vee}^n_{mq}$ (however $ \omega _
{mn} $ in $ \varOmega _ {\bar{q}} $ by this permutation is not
affected). The formula describes a set of peaks of Raman
scattering of $\bm{\vee}$- and $\bm{\wedge}$-types\footnote{In
these designations of types it is necessary to associate the
photon of certain polarization (spirality) with each of two
components.} at crossings of hyperfine magnetic sublevels,
therefore we have named this structure of NOR/M as Raman one.
Because of complexity it can be found out with both amplitude and
phase measurements. The complexity is connected with polarization
opposition of photons, forming a combining pair in scattering. It
is visible, that $\bm{\vee}^{m*}_{nq} =\bm{\vee}^m_{n\bar{q}}$ and
$\bm{\wedge}^{m*}_{nq} =\bm{\wedge}^m_{n\bar{q}}$. The condition
of manifestation of Raman HFS of NOR/M in nonzero fields by means
of a set of peaks is $\nu\leq |C^\varsigma_{\mathrm{a}}|$.
Especially it is necessary to draw attention\footnote{Here it is
possible to see in appropriate \figurename~\ref{12ch4_th_l} from
the previous section.} to zero peak,\footnote{\Ie\ in zero of
field.} when $\nu\gg |C^\varsigma_{\mathrm{a}}|$. It continues to
look as a cusp on the background of collision Raman non-hyperfine
NOR/M, if the degree of its quadratic sharpness with respect to
the background is ratio\footnote{In the beginning the upper
operation is fulfilled and then the lower one.}
\begin{equation}
 \frac{\partial^{2}_{B^{(0)}}
 \left[\bm{\vee}^{m\prime}_{n,1}+\bm{\wedge}^{m\prime}_{n,1}
 \right]^{(\text{all }C^{\varsigma}_{\mathrm{a}}\neq0)}_{B^{(0)}=0}}%
{\partial^{2}_{B^{(0)}}
 \left[\bm{\vee}^{m\prime}_{n,1}+\bm{\wedge}^{m\prime}_{n,1}
 \right]^{(\text{all }C^{\varsigma}_{\mathrm{a}}   =0)}_{B^{(0)}=0}}
 \simeq(\prod_\varsigma [I^\varsigma])^{-1}
 \sum_{j\tilde\mu^{\bm\cdot}}
  g^{(\mathrm{eff})2}_{j\tilde\mu^{\bm\cdot}} / 2g^2_J >1,
\label{rambound}
\end{equation}
where effective $g$-factors (or respective gyromagnetic ratios
$\gamma$) and the condition of summation on spin subsystems
$\varsigma$ are the same with (\ref{gyrom_eff_approx}). For
methane, when $P(7)$-branch is considered and $I=1$, the ratio
$\simeq 5$. Thus, if collision Raman non-hyperfine NOR/M is
observed, then its sub-collision HFS is observed all the more. The
structure in the pure kind is inconvenient for observation in
transmission, and convenient in birefringence, as it was made in
\cite{BKMS:84}. There was registration of field derivative of
rotation angle\footnote{See its definition (\ref{theta_psi}).} of
light polarization plane. In conditions, when saturation parameter
$\varkappa\gg \nu/\omega_{\mathrm{D}}$, the angle
\begin{equation}
 \theta \simeq \theta^{(3)}
\propto
\bm{\vee}^{m\prime\prime}_{n,1}+\bm{\wedge}^{m\prime\prime}_{n,1}.
\label{theta3}
\end{equation}
The sub-collision Raman HFS (without ballast one) was observed on
background of Raman non-hyperfine one as a cusp of
$\partial_{B^{(0)}}\theta$. The degree of its sharpness can be
found from ratio
\begin{equation}
 \frac{\partial^{3}_{B^{(0)}}
 \left[\bm{\vee}^{m\prime\prime}_{n,1}+\bm{\wedge}^{m\prime\prime}_{n,1}
 \right]^{(\text{all }C^{\varsigma}_{\mathrm{a}}\neq0)}_{B^{(0)}=0}}%
{\partial^{3}_{B^{(0)}}
 \left[\bm{\vee}^{m\prime\prime}_{n,1}+\bm{\wedge}^{m\prime\prime}_{n,1}
 \right]^{(\text{all }C^{\varsigma}_{\mathrm{a}}   =0)}_{B^{(0)}=0}}
 \simeq(\prod_\varsigma [I^\varsigma])^{-1}
 \sum_{j\tilde\mu^{\bm\cdot}}
  g^{(\mathrm{eff})3}_{j\tilde\mu^{\bm\cdot}} / 2g^3_J >1,
\label{anglebound}
\end{equation}
and for methane ($P(7)$-branch and $I=1$) the ratio $\simeq 14$.
If we judge by the signal approximation (\ref{line}), there is
just its cusp, (\ref{rambound}) or (\ref{anglebound}), in field
zero, but its amplitude here does not vary. The signal amplitude
in \figurename~\ref{12ch4_th_l} is mainly varied through both
$\bm{\between}$-type summands in (\ref{s3par}).

Thus, after passage by linearly polarized radiation of absorbing
cell, its field spectrum acquires components, which look as peaks
in amplitude measurements. They are conditioned by process of
resonance scattering with crossings of hyperfine sublevels in
magnetic field, when $[2q\varDelta_J +\sum_\varsigma
\tilde{\mu}^{\varsigma} \omega^{\varsigma jj}_{M+q,M-q}] =0$ with
$j\in(m,n)$. The greatest amplitude has peak\footnote{Here it is
again possible to see in appropriate \figurename~\ref{12ch4_th_l}
from the previous section.} at field zero, due to large number of
sublevel crossings. Peaks are grouped in area of nonzero fields,
where (\ref{cross}), from single crossings in sublevel pairs with
$|M^\prime_j-M_j|=2$ and $\tilde{\mu}^{\mathrm{H}\prime}_j
=\tilde{\mu}^{\mathrm{H}}_j =-1$. For ${}^{12}\mathrm{CH}_4$ with
rather low pressure there are only two such areas located mutually
symmetrically from field zero. For ${}^{13}\mathrm{CH}_4$ with
similar pressure these areas of crossings are split on pairs
($\tilde{\mu}^{\mathrm{H}}, \tilde{\mu}^{\mathrm{C}}$); in the
beginning ($-1,-\frac12$) and further ($-1,\frac12$), if we scan
from field zero (appropriate FIGs.\ are omitted). The half-width
of these peaks is determined by magnitude $\nu/2\gamma$ with
slightly distinguishing $\gamma$-factors varying near $\gamma_J$.
If we direct $\nu$ to zero, when magnetic field is non-peak, we
shall receive $(\bm{\vee}^m_{nq} +\bm{\wedge}^m_{nq}) \rightarrow
0$. This property is unconnected with the approximation, used in
(\ref{line}), and kept in exact calculation. In
\figurename~\ref{12ch4_th_l}, \ie\ in amplitude measurements
(\ref{s3par}), the Raman peaks is observed on the background of
both ballast dips. In any way the resonance curve below than wing
level could not be lowered without theirs. As we have already
noted, the last ones disappear in phase measurements. When $\nu\gg
|C^{\varsigma}_{\mathrm{a}}|$ for all $\varsigma$, the Raman
structure of NOR/M smooths out and we see single (practically
Lorentz) contour with half-width $\nu/2\gamma_J$ (see
\figurename~\ref{12ch4_ex_l}). Its top is nevertheless sharpened,
according to (\ref{rambound}). Because of its small amplitude it
is better to observe the cusp in the field derivatives of
transmission, as in \figurename~\ref{12ch4_th_dlc}, or\footnote{It
is even more better.} circular birefringence (\ref{theta3}), as in
\cite{BKMS:84}.

Without hyperfine interaction for NOR/M in linearly polarized
radiation there is a frequency analog [see (\ref{without_hfs})],
\ie\ $\varDelta_J \leftrightarrow \omega_\varDelta$, and also
analog of Kramers-Kronig relation for amplitude and phase
\cite{Alt72}. HFS breaks this relation and that also can be used
for extracting of its contribution in linearly polarized
radiation.

To describe field structures of NOR/M$_\parallel$ in circularly
polarized radiation it is necessary to expand (\ref{amplitudes:b})
in a series (\ref{amplitudes:c}), \ie, in addition to main
transitions with $|\tilde{\mu}^\varsigma_{mn}| =0$, to take into
account weaker satellite ones with $|\tilde{\mu}^\varsigma_{mn}|
=1,2$. The magnetic and hyperfine properties of both $j$-terms are
close among themselves, therefore $|\beta^{\varsigma mn}_{qM}|\ll
1$ and it is possible to use (\ref{betamn}). Taking into account
all that, we obtain
\begin{subequations}
\begin{gather}
 \bm{\between}^m_{nq}
\equiv
 X^{(3)mn}_{\underdot{q}{}\dot q{}\underdot{q}{}\dot q{}}
\simeq \Exp^{-(\varOmega_q/\omega_{\mathrm{D}})^2}
\frac{\nu^2}{\nu_m \nu_n} \prod_{\varsigma'} [I^{\varsigma'}]
 \sum_M |T^{mn}_{\dot q{}M}|^4
  \left\{ 1 - \frac23 \sum_\varsigma \hat{\bm{I}}{}^{\varsigma 2}
\vphantom{\left(
  \frac{\partial_M T^{mn}_{\dot q{}M}}{T^{mn}_{\dot q{}M}}
        n^{\varsigma}_{x,nM}\right)^2} \right.
\nonumber\\ \times \left[
 (\bm{n}^{\varsigma}_{m,M+q}\times\bm{n}^{\varsigma}_{nM})^2
+3\left(\frac{\partial_M T^{mn}_{\dot q{}M}}{T^{mn}_{\dot q{}M}}
        n^{\varsigma}_{x,nM}\right)^2 \right. \label{between}
\\
-\left((\bm{n}^\varsigma_{m,M+q}\times\bm{n}^\varsigma_{nM})^2
  +\left(\frac{\partial_M T^{mn}_{\dot q{}M}}{T^{mn}_{\dot q{}M}}
         n^\varsigma_{x,nM}\right)^2\right)
 \left(\frac{\nu_m^2}{\nu_m^2 +\omega^{\varsigma 2}_{m,M+q}}
      +\frac{\nu_n^2}{\nu_n^2 +\omega^{\varsigma 2}_{n,M  }}\right)
\label{vee_wedge}
\\ \left. \left.
+\left((\bm{n}^{\varsigma}_{m,M+q}\times\bm{n}^{\varsigma}_{nM})^2
  -\left(\frac{\partial_M T^{mn}_{\dot q{}M}}{T^{mn}_{\dot q{}M}}
         n^{\varsigma}_{x,nM}\right)^2\right)
 \frac{\nu_m^2\,\nu_m^2}{(\nu_m^2 +\omega^{\varsigma 2}_{m,M+q})
                         (\nu_n^2 +\omega^{\varsigma 2}_{n,M  })}
\right] \right\}  \label{Join}
\\
\simeq \Exp^{-(\varOmega_q/\omega_{\mathrm{D}})^2}
\prod_{\varsigma'} [I^{\varsigma'}]
 \sum_M |T^{mn}_{\dot q{}M}|^4
  \left\{ 1 - \frac23 \sum_\varsigma \hat{\bm{I}}{}^{\varsigma 2}
  \right.
\nonumber\\ \times \left.
  \left[\frac{ (\bm\omega^{\varsigma}_{m,M+q}\times
                \bm\omega^{\varsigma}_{nM})^2}{
 (\nu^2+\omega^{\varsigma2}_{m,M+q})(\nu^2+\omega^{\varsigma2}_{nM})}
+\left(\frac{\partial_M
  T^{mn}_{\dot q{}M}}{T^{mn}_{\dot q{}M}}\right)^2
  \left(3+\frac{\nu^2}{\nu^2+\omega^{\varsigma2}_{nM}}\right)
  \frac{\omega^{\varsigma2}_{x,nM}}{\nu^2+\omega^{\varsigma2}_{nM}}
  \right] \right\}. \label{circ}
\end{gather} \label{all_circ}
\end{subequations}
The unit vectors $\bm{n}^{\varsigma}_{jM}$ are defined in
(\ref{norm_omega}). At first, in (square) brackets, we have
disjointed the structures of $\bm{\between}$-type (\ref{between}),
$(\bm{\vee} +\bm{\wedge})$-types (\ref{vee_wedge}), and
$\Join$-type (\ref{Join}), and then joined them, taking into
account that collision constants $\nu_j\simeq\nu$. The resonance
structure from each spin subsystem of $\varsigma$-sort is always
the dip with respect to Raman peak. The sign change is exactly
determined by $\bm{\between}$-type structures. The dip is
observed,\footnote{The approximation behaves oneself almost as
well as the exact solution in \figurename~\ref{12ch4_th_c} from
the previous section.} and both structures of $\bm{\vee}$- and
$\bm{\wedge}$-types only decrease its depth. Also the structure of
$\Join$-type obviously decreases it, when $J_{mn} =0$. When
$J_{mn} \neq0$, the structure is suppressed with the factor
situated before it, and that conforms with the selection rule
(\ref{select_tildemu}). All the summands from these structures
(even of Raman type) are real, \ie\ $\bm{\between}^{m*}_{nq}
=\bm{\between}^m_{nq}$. They are detected only with amplitude
measurements when the spin subsystem, being directly incapable to
interact with light, interacts indirectly, connects\footnote{We
have seen, the term ``connection'' naturally arises from the
interpretation of the equations (\ref{hfz:c}).} to rotation
subsystem and increasing its ability to absorb light. We have
therefore name these inverse structures ``ballast'' ones. It is
enough to deal with only one circular polarization of light at
NOR/M$_\parallel$, for always $\bm{\between}^m_{n\bar{q}}(B^{(0)})
=\bm{\between}^m_{nq}(-B^{(0)})$.

For circularly polarized radiation in basis of wave functions
(\ref{diag}), diagonalizing (\ref{hfz:a}), we have a picture, in
which resonance decreasing of scattering\footnote{All our first
correction on saturation corresponds to purely induced process of
scattering.} arises only due to field dependence of main (\ie\
with $|\tilde{\mu}^\varsigma_{mn}| =0$) electro-dipole matrix
elements between $m$- and $n$-terms with their anticrossing
hyperfine $\tilde{\mu}^{\varsigma}_j$-components in the sets with
equal $M$. The field tuning on the maximum of the interaction of
rotation and $\varsigma$-spin subsystems is connected to
decreasing of absorption saturation and appearance of
$\varsigma$-dip in magneto-field dependence of output radiation
intensity. For detailed description\footnote{Practically it is
just the same one in exact calculation of the previous section.}
of the ballast structure in diagonalizing basis (\ref{diag}), we
should consider the diagrams with two, three or four optically
connected $\tilde{\mu}^{\varsigma}_j$-sublevels. Magneto-field
dependence of these diagrams determines all the main spectral
features, cp.\ with (\ref{between_half}). When $\nu\ll
|C^{\varsigma}_{\mathrm{a}}|$, $\varsigma$-dip is formed in main
transitions of $\bm{\between}$-type (two-level diagrams with
$|\tilde{\mu}^{\varsigma}_{mn}| =0$), \ie\
\begin{equation}
 \bm{\between}^m_{nq}(B^{(0)})-\bm{\between}^m_{nq}(\infty)
 \simeq \sum_{\tilde{\mu}^{\bm\cdot}M}
 \Bigl|\left(\underline{\hat{T}}{}^{mn\prime}_{\dot q{}M}
 \right)_{\tilde{\mu}^{\bm\cdot}\tilde{\mu}^{\bm\cdot}}\Bigr|^4
  -\sum_{\mu^{\bm\cdot}M}
 \Bigl|\left(\underline{\hat{T}}{}^{mn}_{\dot q{}M}
       \right)_{\mu^{\bm\cdot}\mu^{\bm\cdot}}\Bigr|^4 \leq 0.
 \label{dipsign}
\end{equation}
The dip depth $\simeq J^{-2}$ with respect to the wing
$\bm{\between}^m_{nq}(\infty)$. With increasing $\nu$ the summands
of equal order from transitions $\bm{\vee}$-, $\bm{\wedge}$-, and
$\Join$-types (with $|\tilde{\mu}^{\varsigma}_{mn}| \neq0$) also
increase and the dips become shallower and wider. With our
approximation the last summand\footnote{It would be good to
specify (\eg, numerically) its behavior for nonzero fields with
exact calculation.} from the specified types with $J_{mn} \neq0$,
already begins to be manifested. It keeps some tendency to
suppression because of presence of two components with different
sign. The $\varsigma$-dip half-width is
$(\omega^{\varsigma2}_{x,jM_{\mathrm{eff}}} +\nu^2)^{1/2}
/\gamma^{\varsigma}_{IJ}$. It is determined by the greatest
approach (connecting) of magnetic repulsed $M$-sublevels, \ie\
$\omega^{\varsigma}_{x,jM_{\mathrm{eff}}}\simeq
C^{\varsigma}_{\mathrm{a}} \sqrt{(J_j +1/2)^2
-M^2_{\mathrm{eff}}}J$, and their collision broadening $\nu$. The
$\varsigma$-dip takes place about
\begin{equation}
        {}^{\varsigma}B^{(0)}_{\mathrm{acr}}
\simeq -C^{\varsigma}_{\mathrm{a}}  M_{\mathrm{eff}} /
\gamma^{\varsigma}_{IJ}, \label{anticross}
\end{equation}
where there is the anticrossing of magnetic sublevels connected by
the most strong optical transitions with $M_{\mathrm{eff}}\simeq q
J_{mn}(J_n +1/2) /\sqrt2$; it corresponds to condition
(\ref{sin_cos}) in the next section. For the positive
spin-rotation constant $C^{\varsigma}_{\mathrm{a}}$ on $P$-line
the $\varsigma$-dip in intensity of right circularly polarized
radiation is displaced from zero of magnetic field to the right.
Its depth $\simeq C^{\varsigma2}_{\mathrm{a}} /[\nu^2
+(C^{\varsigma}_{\mathrm{a}}J)^2/2]$ with respect to peak height
on field zero in linearly polarized radiation.

The form of observed field spectrum essentially depends on the
choice of mutual orientation of varied magnetic and fixed laser
fields and on the polarization of the latter. The ballast
structure from spin subsystem $\varsigma$ is manifested as
sub-collision one (again with respect to collision Raman
non-hyperfine one for linearly polarized radiation), when
\begin{equation}
\nu \geq |C^\varsigma_{\mathrm{a}}|J \left/
 \sqrt{2[(g^\varsigma_{IJ}/2g_J)^2 -1]}. \right.
\label{balbound}
\end{equation}
There is such choice of mutual orientation of fields, with which
the field spectroscopy reflects just the rupture of spin-rotation
connections and the ballast structure is visible without Raman one
at all. For that the amplitude-scanned magnetic field should be
directed or along propagation of circularly polarized radiation
($\bm{B}^{(0)} \parallel \bm{k}$; it is longitudinal (Faraday)
field orientation), or across propagation of linearly polarized
radiation as its vector of electrical field
($\bm{B}^{(0)}\parallel \bm{E}^{(\omega)}$; it is transverse
(Voigt) field orientation). It is visible that there are two
summands in square brackets of (\ref{circ}), describing the dip
for longitudinal field orientation ($q =\pm1$). At $J_{mn} =0$
[$Q(J_n)$-lines] there is only second one for transverse field
orientation ($q =0$).

Let us note, that it is sufficient to complete the expansion on
$J^{-2}$ by the first (main) summand with field structure, and
this sufficient precision of expansion is various for linear and
circular polarization. Just the same we act further in
section~\ref{DoubleHFS}, in case of $\mathrm{CH}_3\mathrm{F}$, and
there the expansion precision is various for $K=1$ and $K\neq1$.
At last, as it was noted in previous section, the transition to
field spectral derivatives is convenient by that allows to level
amplitudes of structures from different orders of expansion
because of occasionally appropriate distinction of their widths.

If we leave aside the numerical comparison,\footnote{Seemingly it
is more reasonable to struggle with the selection rule violation
(\ref{regular_select}).} the represented approximation reproduces
practically all the qualitative features of magnetic spectrum of
NOR, obtained under the exact formula (\ref{X3:b}) [at least so
long as we are not interested in such details, as small
asymmetries (of dip in particular) noted by us under analysis of
FIGs.\ attending the formula]. For simple estimations it is
possible to use the undermentioned formula (\ref{circ:estim})
representing a combination of hyperfine and collision structures.

\begin{figure}\centering
\includegraphics[width=0.48\textwidth]{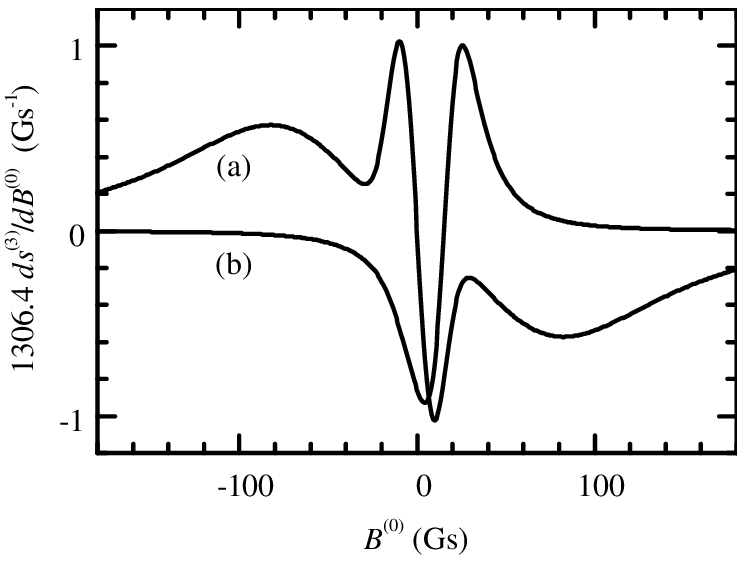} \hfill
\includegraphics[width=0.48\textwidth]{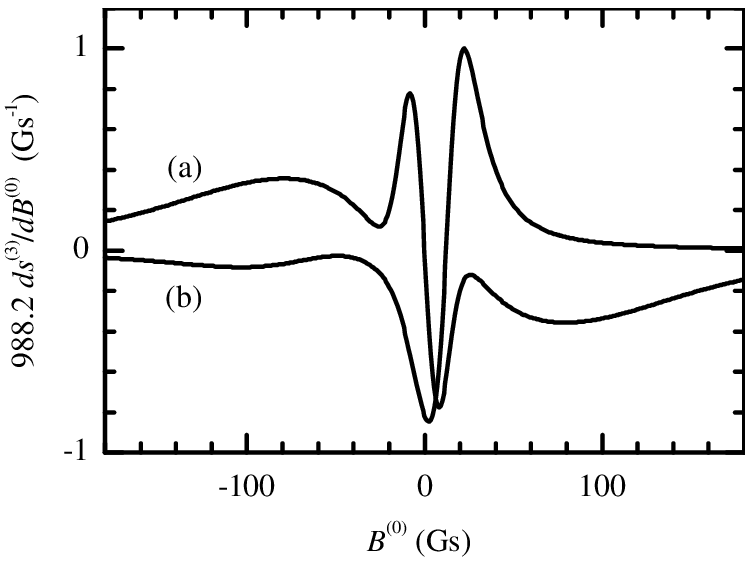}
\\
\caption{\label{12ch4_th_ap_dlc-13ch4_th_ap_dlc} Derivatives of
approximately calculated NOR/M$_\parallel$ in transmission of
linearly (a) and right circularly (b) polarized radiations.
\emph{At the left} for component ($\omega_3$ $P(7)$
$F^{(2)}_{2(-)}$) of ${}^{12}\mathrm{CH}_4$ isotope. \emph{At the
right} for component ($\omega_3$ $P(6)$ $F^{(1)}_{2(-)}$) of
${}^{13}\mathrm{CH}_4$ isotope. For all curves
$C^{\mathrm{H}}_{1j}= 12\,\mathrm{kHza}$,
$C^{\mathrm{C}}_{1j}=-12\,\mathrm{kHza}$, and $\nu=40\,
\mathrm{kHza}$.}
\end{figure}

Field derivatives of NOR/M (in just described approximation) for
two carbon-substituted methane isotopes are represented in
\figurename~\ref{12ch4_th_ap_dlc-13ch4_th_ap_dlc}.
The evaluation of $\nu$ 
was made by selection of a curve in \figurename~\ref{12ch4_th_dlc}
with amplitude ratio of narrow and wide resonance structures in
intensity of linearly polarized radiation, approaching to similar
ratio in \figurename~\ref{12ch4_ex_dlc-13ch4_ex_dc} at the left.
In the approximation of resonance scattering of contrarily
polarized photons we have taken into account only (most strong)
main transitions with $|\tilde{\mu}^\varsigma_{mn}| =0$. In order
that the amplitude ratio of appropriate curves for linearly and
circularly polarized radiations in
\figurename~\ref{12ch4_th_ap_dlc-13ch4_th_ap_dlc} at the left was
the same as in \figurename~\ref{12ch4_ex_dlc-13ch4_ex_dc} at the
left, it is apparently required side by side with (\ref{line}) to
take into account already next corrections $\sim J^{-2}$ . For
both isotopes we use the same evaluation of $\nu$, since their
working areas of pressures coincided.

Now let us adduce the data from which the hyperfine constants were
estimated. The ballast structures place in
\figurename~\ref{12ch4_ex_dlc-13ch4_ex_dc} at the left near to
field\footnote{The underline labels atom, forming the ballast
structure under anticrossing.}
$B^{(0)}_{\mathrm{acr}}({}^{12}\mathrm{C}
\underline{{}^1\mathrm{H}}{}_4) \simeq 15\,\mathrm{Gs}$ and in
\figurename~\ref{12ch4_ex_dlc-13ch4_ex_dc} at the right near to
fields $B^{(0)}_{\mathrm{acr}}({}^{13}\mathrm{C}
\underline{{}^1\mathrm{H}}{}_4) \simeq 13\,\mathrm{Gs}$ and
$B^{(0)}_{\mathrm{acr}}(\underline{{}^{13}\mathrm{C}}{}^1\mathrm{H}_4)
\simeq -61\,\mathrm{Gs}$. From here we obtain by formula
(\ref{anticross})\footnote{Here it is important that the one does
not depend on $\nu$.} that
\begin{subequations}
\begin{equation}
 C^{{}^{1}\mathrm{H}}_{\mathrm{a}} \simeq 12\,
\mathrm{kHza}, \quad
 C^{{}^{13}\mathrm{C}}_{\mathrm{a}} \simeq-12\,
\mathrm{kHza}. \label{Cav:a}
\end{equation}
These estimated data are obtained on separately taken curve. The
work on increasing precision in the determination of hyperfine
constants by our method was not carried out. The approximating
curves in \figurename~\ref{12ch4_th_ap_dlc-13ch4_th_ap_dlc}
correspond just to these values of constants. The opposite signs
of average spin-rotation constants indicate that the
intramolecular magnetic fields near to nuclei ${}^{13}$C and
${}^1$H have opposite directions. The sign change of the constant
for nucleus ${}^{13}$C indicates that its negative electronic
component prevails over positive nuclear one \cite{Fly78}.

It is possible to compare these our data with more indirect ones,
on chemical shifts in NMR spectra of methane nuclei ${}^1$H and
${}^{13}$C. As it is known \cite{FG68,GF72,AR66}, even for more
general, than (\ref{hfz:a}), symmetrized interaction [see
expression (5) in \cite{Gus99}] the spin-rotation tensor $\pmb{\sf
C}^{\varsigma} =g^\varsigma_I \tilde{\pmb{\sf B}}\cdot \pmb{\sf
R}^{\varsigma}$ with a dimensionless tensor $\pmb{\sf
R}^{\varsigma}\simeq m_{\mathrm{e}}
\bm\sigma^{\varsigma}_{\mathrm{m,a}} /m_{\mathrm{p}}$. Here tensor
$\bm\sigma^{\varsigma}_{\mathrm{m,a}}
=\bm\sigma^{\varsigma}_{\text{molecule}}
-\bm\sigma^{\varsigma}_{\text{atom}}$, \ie\ a difference of
magnetic shielding of $\varsigma$-nuclei compounding molecule from
free atoms, and $m_{\mathrm{e}} /m_{\mathrm{p}}$ is
electron-proton mass ratio. $\tilde{\pmb{\sf B}}/\hbar =\pmb{\sf
I}^{-1}$ is inverse tensor of molecular inertial moment.
$\tilde{B} \simeq 314.2(1)\,\mathrm{GHza}$ for spherical tops of
carbon-substituted methane isotopes. In general case, the tensor
$\pmb{\sf C}^{\varsigma}$ can be asymmetric \cite{Gus99}. In the
usual experiments we determine only average difference of
shielding, \ie\ chemical shift\footnote{The one is scalar. The
angular brackets have been defined on p.~\pageref{bracketdef}.}
$\sigma^{\varsigma}_{\mathrm{m,a}}\equiv \frac13
\langle\bm\sigma^{\varsigma}_{\mathrm{m,a}}\rangle$. According to
\cite{Fly78,AD74}, $\sigma^{{}^1\mathrm{H}}_{\mathrm{m,a}}/10^{-6}
\simeq 30.8-17.7 =13.1$ and
$\sigma^{{}^{13}\mathrm{C}}_{\mathrm{m,a}} /10^{-6} \simeq
196-260.7 =-64.7$. From here we obtain also average hyperfine
constants
\begin{equation}
 C^{{}^{ 1}\mathrm{H}}_{\mathrm{a}} \simeq  12.5\,
\mathrm{kHza},\quad
 C^{{}^{13}\mathrm{C}}_{\mathrm{a}} \simeq -15.5\,
\mathrm{kHza}, \label{Cav:b}
\end{equation}
\label{Cav}
\end{subequations}
that practically corresponds to our estimated data (\ref{Cav:a}).

\section{\label{COL_NOR_M}Collision structure of NOR/M$_\parallel$
by high-$J$ approximation}

Since the analysis of paper \cite{SSSh85}, both formulae and
conclusions, is complicated by mistakes existing there, we are
here forced to adduce more right (in our opinion) formulae, on
which our conclusions about unacceptability of the collision
interpretation ``anomalous'' structures are based.

Taking into account deorientational in-summand for level
populations \cite{RSh91,SSSh85} the collision summand
(\ref{Col_int}) turns in
\begin{align}
 \hat S_{ij}(v_k,r_k,t)
 =&\hphantom{+}\delta_{i,j}
 \left\{ \left[(\nu_j-\tilde\nu_j)\tilde{N}_jW(v_k)
 +\Tilde{\Tilde\nu}_j
 \langle\hat\rho_j(v_k,r_k,t)\rangle\right]\hat1_j - \nu_j
 \hat\rho_j(v_k,r_k,t)
 \vphantom{(\nu_j-\tilde\nu_j)\tilde{N}_jW(v_k)}\right\}
\nonumber\\
 &-(1-\delta_{i,j})\nu\hat\rho_{ij}(v_k,r_k,t).
\label{Col_int_with_tilde_nu}
\end{align}
Now here there are deorientational constant $\tilde{\nu}_j$ and
its modification $\Tilde{\Tilde{\nu}}_j =\tilde{\nu}_j /[j]$. As
far as the deorientational in-summand from other sublevels of
level $j$ has been extracted we have to subtract it in the
pumping, where $\nu_j \rightarrow (\nu_j-\tilde\nu_j)$, so that
$\langle\langle \hat{S}_j(v_k,r_k,t) \rangle\rangle_{v_k} =0$ in
absence of laser light. As a result of the extraction there is a
collision addend\footnote{It is labelled with breve $\breve{}$.}
to (\ref{X3}):
\begin{subequations}
\begin{gather}
 \breve{X}^{(3)mn}_{\underdot{q}{}_0\dot{q}{}_1
 \underdot{q}{}_2\dot{q}{}_3}(B^{(0)})
=\Exp^{-(\varOmega_{q_2}/\omega_{\mathrm{D}})^2}
 \int\limits_0^\infty \nu \mathrm{d}\tau\, \Exp^{2\bar{\nu} \tau}
 X^{(1)mn }_{\underdot{q}{}_0 \dot{q}{}_0}(\tau)
 X^{(1)mn*}_{\underdot{q}{}_2 \dot{q}{}_2}(\tau)
\nonumber\\ \times
 \nu \left( \frac{\delta_{q_0,q_3}\delta_{q_2,q_1}
  \Tilde{\Tilde\nu}_m}{\nu_m(\nu_m-\tilde\nu_m)}
+ \frac{\delta_{q_0,q_1}\delta_{q_2,q_3}
  \Tilde{\Tilde\nu}_n}{\nu_n(\nu_n-\tilde\nu_n)} \right)
\label{X3col:a}
\\
=\Exp^{-(\varOmega_{q_2}/\omega_{\mathrm{D}})^2}
 \sum_{\substack{m_2m_4\\ \,\,n_1\,\,n_3}}
\frac{|T^{m_4n_3}_{\dot q{}_0}\,T^{m_2n_1}_{\dot q{}_2}|^2\,
 \nu^2}{2\nu -\mathrm{i}( q_{02}\varDelta_J
   +\lambda_{m_2n_1} -\lambda_{m_4n_3} )}
\nonumber\\ \times
 \left( \frac{\delta_{q_0,q_3}\delta_{q_2,q_1}
 \Tilde{\Tilde\nu}_m}{\nu_m(\nu_m-\tilde\nu_m)}
+\frac{\delta_{q_0,q_1}\delta_{q_2,q_3}
 \Tilde{\Tilde\nu}_n}{\nu_n(\nu_n-\tilde\nu_n)} \right).
\label{X3col:b}
\end{gather}
\label{X3col}
\end{subequations}
Here again there was the function \(X^{(1)mn}_{\underdot{q}{}\dot
q{}}(\tau)\) defined in (\ref{X1:a:}). If in (\ref{X3col}) the
dependence of the first exponential factor from $q_2$ is
neglected, the collision addend to (\ref{s3}) is
\begin{gather}
 \breve{s}^{(3)} 
 =\Breve{\Tilde{\bm{X}}}^{(3)mn}
 \,\underset{\displaystyle\cdot}{\vdots}\, \bm{e}^{*} \otimes\bm{e}
 \otimes\bm{e}^{*} \otimes\bm{e}
\nonumber\\
 =\Exp^{-(\varOmega/\omega_{\mathrm{D}})^2}
 \int\limits_0^\infty \nu \mathrm{d}\tau\, \Exp^{2\bar{\nu}\tau}
 \left|\sum_{q} X^{(1)mn}_{\underdot{q}{}\dot q{}}(\tau)
 |e_{\underdot{q}{}}|^2 \right|^2
 \sum_j \frac{\nu\Tilde{\Tilde\nu}_j}{\nu_j(\nu_j-\tilde\nu_j)}
 \frac{[J_m]}{3[p]\prod\limits_\varsigma [I^\varsigma]}.
\label{breve_s3}
\end{gather}
The superscript $\tilde{}$ is defined in (\ref{modXdef}).
Practically all the features in NOR/M depend here on convergence
(for peak in NOR/M) and divergence (for dips in NOR/M) of
one-photon components in stepped two-photon processes of
$\bm{\sqcup}$- and $\bm{\sqcap}$-types through intermediate
collisional deorientation. In the conventional type designations
given here both side vertical segments and intermediate horizontal
segment, $\underline{\phantom{\sqcup}}$ or
$\overline{\phantom{\sqcap}}$, correspond to them, respectively.

As well as in (\ref{line}), with high $J_j$ of resonance levels,
the calculation of collision structure of NOR/M is possible,
taking into consideration only main optical transitions with
$|\tilde{\mu}^{\varsigma}_{mn}| =0$ (even without
$J^{-1}$-corrections dependent on magnetic field). The exact
expression (\ref{X3col:b}) becomes simpler and for linearly
polarized radiation we obtain the collision addend to
(\ref{line}):
\begin{gather}
 \breve{\bm{\vee}}{}^m_{nq}+\breve{\bm{\wedge}}{}^m_{nq}
\equiv \breve{X}{}^{(3)mn}_{\underdot{q}{}
 \dot{q}{}\underdot{\bar{q}}{}\dot{\bar{q}}{}}
+\breve{X}{}^{(3)mn}_{\underdot{q}{}
 \dot{\bar{q}}{}\underdot{\bar{q}}{}\dot{q}{}}
\nonumber\\
\simeq \Exp^{-(\varOmega_{\bar{q}}/\omega_{\mathrm{D}})^2}
 \sum_{\tilde\mu^{\bm\cdot}_1 M_1 \tilde\mu^{\bm\cdot}_3 M_3}
 \frac{|T^{mn}_{\bar{q}M_1}T^{mn}_{qM_3}|^2 \nu^2
 [\Tilde{\Tilde\nu}_n/\nu_n(\nu_n-\tilde\nu_n)
+(m\leftrightarrow n)]}{2\nu - \mathrm{i}[2q\varDelta_J
 +\sum_\varsigma (\tilde\mu^\varsigma_1
  \omega^{\varsigma mn}_{M_1-q,M_1} - \tilde\mu^\varsigma_3
  \omega^{\varsigma mn}_{M_3+q,M_3} ) ]}.
\label{line:col}
\end{gather}
As well as in (\ref{line}) the two summands are here ordered. The
appropriate \figurename\ is omitted.

\begin{figure}\centering
\includegraphics{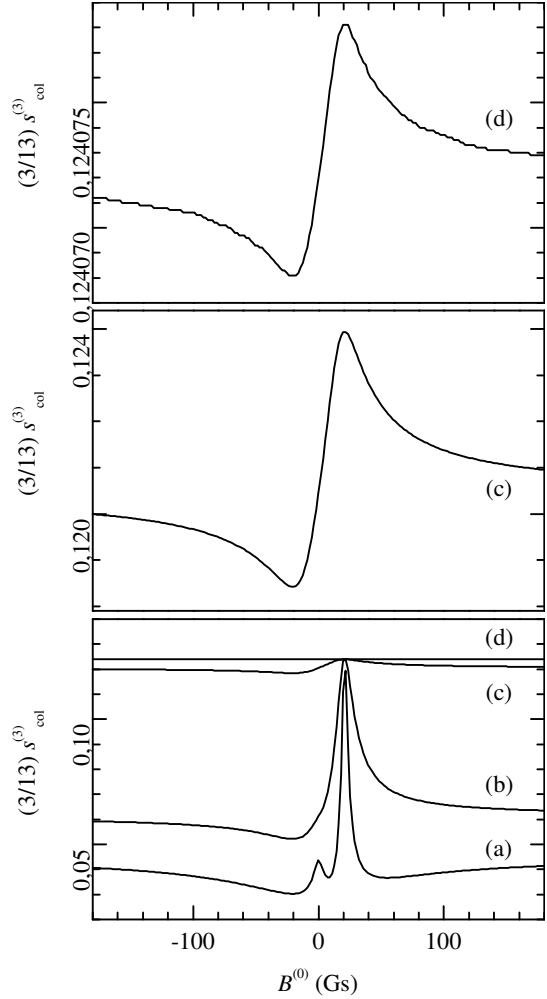}
\caption{\label{12ch4_th_ap_circ_col} Approximation of collision
addend to NOR/M$_\parallel$ in transmission of right circularly
polarized radiation for component ($\omega_3$ $P(7)$
$F^{(2)}_{2(-)}$) of isotope ${}^{12}\mathrm{CH}_4$: $[\nu_j,
\tilde\nu_j] /\mathrm{kHza} =[1, 0.67]$ (a), $[4, 2.67]$ (b),
$[40, 26.67]$ (c), $[1000, 667]$ (d). For all curves the ratio
$\tilde\nu_j/(\nu_j -\tilde\nu_j)=2$.}
\end{figure}

In the same approximation for circularly polarized radiation the
collision addend to (\ref{circ}) is\footnote{Here only real part
is retained, since imaginary one always is zero.}
\begin{equation}
\breve{\bm{\between}}{}^m_{nq} \equiv
\breve{X}{}^{(3)mn}_{\underdot{q}{}
 \dot{q}{}\underdot{q}{}\dot{q}{}}
\simeq \Exp^{-(\varOmega_q/\omega_{\mathrm{D}})^2}
 \sum_{\tilde\mu^{\bm\cdot}_1 M_1 \tilde\mu^{\bm\cdot}_3 M_3}
 \frac{|T^{mn}_{qM_1}T^{mn}_{qM_3}|^2 \, 2\nu^3
 \sum_j \Tilde{\Tilde\nu}_j/\nu_j(\nu_j-\tilde\nu_j)}{(2\nu)^2
+[\sum_\varsigma
 (\tilde\mu^\varsigma_1 \omega^{\varsigma mn}_{M_1+q,M_1}
-\tilde\mu^\varsigma_3 \omega^{\varsigma mn}_{M_3+q,M_3} )]^2}.
\label{circ:col}
\end{equation}
In \figurename~\ref{12ch4_th_ap_circ_col} we give only this
collision addend designated as $s^{(3)}_{\mathrm{col}}$, \ie\ as
addend to (\ref{s3par}) for right circularly polarized radiation.
There are four curves (a,b,c,d; from them the last two are
duplicated in expanded scale) with various $\nu_j =\nu$ but equal
ratio $\tilde{\nu}_j /(\nu_j -\tilde{\nu}_j)$. On the lower graph
of the \figurename\ there is a specific point\footnote{There is
another specific point by the same field (\ref{Ncross}).} by field
\begin{equation}
        {}^{\mathrm{H}}B^{(0)}_{\mathrm{Xcr}}
\simeq C^{\mathrm{H}}_{\mathrm{a}} J /\gamma^{\mathrm{H}}_{IJ},
\label{Xcross}
\end{equation}
where the amplitudes of all four curves (a,b,c,d) are equal
\begin{equation}
 \breve{\bm{\between}}{}^m_{n1}({}^{\mathrm{H}}
 B^{(0)}_{\mathrm{Xcr}})
=\left.\Exp^{-(\varOmega_q/\omega_{\mathrm{D}})^2}
  \right|_{B^{(0)}={}^{\mathrm{H}}B^{(0)}_{\mathrm{Xcr}}}
 \prod_{\varsigma'}[I^{\varsigma'}]
 \frac{\nu}{2}\sum_j \frac{\tilde\nu_j/[J_j]}{\nu_j(\nu_j
-\tilde\nu_j)}.
\end{equation}
It is connected that in this point the difference
$(\lambda_{m_2n_1} -\lambda_{m_4n_3}) $ in denominator of the
formula (\ref{X3col:b}) is equal to zero with any values of its
four subscripts $j_l$. Such point occurs only for circularly
polarized radiation.

In paper \cite{SSSh85} the formula (5.9) is similar to ours
(\ref{circ:col}) but has slightly different differences in
denominator, namely (5.10) \ibid. To represent their formula in
our designations, in the denominator of our formula
(\ref{circ:col}) it is necessary to make replacement
$\omega^{\varsigma mn}_{M_1+q,M_1} \rightarrow \omega^{\varsigma
mn}_{M_1,M_1} +(\omega^{\varsigma mm}_{M_1+q,M_1}
+\omega^{\varsigma nn}_{M_1+q,M_1})/2$
and similarly for $M_3$. With this replacement the qualitative
interpretation of the formula based on crossing of lines is lost.
At last, their approximation of HFS does not reflect the important
property, namely, in any model with $g^\varsigma_I =g_J$, magnetic
splitting on the graphs shown at the bottom of
\figurename~\ref{12ch4_th_hfs} and NOR/M itself for circularly
polarized radiation should absolutely vanish away.

Now let us adduce simplified estimated formula for
$\bm{\between}_q(B^{(0)}) =\bm{\between}_1(qB^{(0)})$, describing
both hyperfine (ballast) and collision\footnote{Or, more exactly,
collision-hyperfine one.} structures in transmission of circularly
polarized radiation ($q=\pm1$) on transition with $J_{mn}=-1$ and
$J_j\gg I=1$:
\begin{multline}
 \bm{\between}_1(B^{(0)})
 =\prod_{\varsigma'}[I^{\varsigma'}] J^{-1} \left\{1
 -\frac{C^{\mathrm{H}2}_{\mathrm{a}}}{\nu^2 +\omega^2_{\theta_0}}
 \right.
\\ \left.
 +\frac{\tilde\nu_j}{\nu_j -\tilde\nu_j} \left[1
 +\frac23\left( \frac{2(2\nu)^2}{(2\nu)^2 +\Pi_1^2}
 +\frac{\nu^2}{\nu^2 +\Pi_1^2} \right)\right] \right\}.
\label{circ:estim}
\end{multline}
Here $\omega_{\theta_0} =\left[(C^{\mathrm{H}}_{\mathrm{a}} J
\sin\theta_0)^2 +(\varDelta^{\mathrm{H}}_{IJ}
-C^{\mathrm{H}}_{\mathrm{a}} J \cos\theta_0)^2 \right]^{1/2}$ and
$\cos\theta_0 \equiv M_{\mathrm{eff}}/(J+1/2) =1/\sqrt2$, where
effective value $M_{\mathrm{eff}}$ is determined from condition:
\begin{equation}
\sin\theta_0 = \cos\theta_0. \label{sin_cos}
\end{equation}
We have also put $M_1 =M_3 =M$ in (\ref{circ:col}) and designated
the line difference
\begin{gather}
 \Pi_q =\omega^{\mathrm{H}mn}_{M+q,M}
  \simeq C^{\mathrm{H}}_{\mathrm{a}}(q\varDelta^{\mathrm{H}}_{IJ}
  +J_{mn}C^{\mathrm{H}}_{\mathrm{a}} J)/\omega_{\theta_0},
\label{line_diff}
\\ \intertext{when absolute value of number difference
$|\tilde\mu^{\mathrm{H}}_{13}|=1$. One can see that}
 |\Pi_1(B^{(0)}=0)| =|\Pi_1(B^{(0)}=\infty)|.
\label{ldiff_zero_infty}
\end{gather}

At last by separating wing level\footnote{It is meant for
simplicity, that it is reached with fields, where Doppler factor
varies still insignificantly.} we obtain
\begin{multline}
 \delta \bm{\between}_1^{(\mathrm{hfs})}
 +\delta \bm{\between}_1^{(\mathrm{col})}
 =\bm{\between}_1(B^{(0)}) -\bm{\between}_1(\infty)
\\
 =\prod_{\varsigma'}[I^{\varsigma'}] J^{-1}\left[
 -\frac{C^{\mathrm{H}2}_{\mathrm{a}}}{\nu^2 +\omega^2_{\theta_0}}
 +\frac{2\tilde\nu_j /3}{\nu_j -\tilde\nu_j}
 \left( \frac{2(2\nu)^2}{(2\nu)^2 +\Pi_1^2}
 \right.\right.
\\ \left.\vphantom{\frac{C^{\mathrm{H}2}_{\mathrm{a}}}{\nu^2
 +\omega^2_{\theta_0}}} \left.
 -\frac{2(2\nu)^2}{(2\nu)^2 +C^{\mathrm{H}2}_{\mathrm{a}}}
 +\frac{\nu^2}{\nu^2 +\Pi_1^2}
 -\frac{\nu^2}{\nu^2 +C^{\mathrm{H}2}_{\mathrm{a}}}
 \right) \right].
\end{multline}
From here the amplitude ratio of collision and hyperfine
structures is
\begin{equation}
 \frac{\left. \delta \bm{\between}_1^{(\mathrm{col})}
 \right|_{\varDelta^{\mathrm{H}}_{IJ}=C^{\mathrm{H}}_{\mathrm{a}}J}}{\left.
 \delta \bm{\between}_1^{(\mathrm{hfs})}
 \right|_{\varDelta^{\mathrm{H}}_{IJ}=C^{\mathrm{H}}_{\mathrm{a}}J/\sqrt2}}
 \simeq -\frac{\tilde\nu_j(2\nu^2
 +C^{\mathrm{H}2}_{\mathrm{a}}J^2)}{(\nu_j -\tilde\nu_j)3\nu^2}
 \left[ \frac{(2\nu)^2/2}{(2\nu)^2 +C^{\mathrm{H}2}_{\mathrm{a}}}
 +\frac{\nu^2}{\nu^2 +C^{\mathrm{H}2}_{\mathrm{a}}} \right].
\label{coltohfs}
\end{equation}
The simplified expression for the ratio is
\[
-\frac{\tilde\nu_j (C^{\mathrm{H}}_{\mathrm{a}}J/\nu)^2}{(\nu_j
-\tilde\nu_j)2}.
\]
It corresponds to evaluation (5.15) from \cite{SSSh85} and can be
used only if $C^{\mathrm{H}}_{\mathrm{a}}\ll \nu\ll
C^{\mathrm{H}}_{\mathrm{a}}J$.

Summing up, we can say the following. A certain dip [see the curve
(d) in \figurename~\ref{12ch4_th_ap_circ_col}; there, where
$B^{(0)}<0$], connected with divergence of ${[}^m_n$ lines in
diagrams of $^{m_2}_{n_1}\bm{\sqcup}^{m_4}_{n_3}$- and
$^{m_2}_{n_1}\bm{\sqcap}^{m_4}_{n_3}$-types, is also obtained in
this model, when the collision half-width of rotational $J$-levels
is much greater half-width of their hyperfine
splitting.\footnote{\Ie\ $\nu\gg C^{\mathrm{H}}_{\mathrm{a}}J_j$
in contrast to our experiments, where they were usually
comparable.} However its shift is opposite to the shift of
experimentally observed dip. Our analysis shows, that the
collision structure (\ref{circ:col}) is appreciably asymmetric.
Owing to (\ref{ldiff_zero_infty}), the level of the collision
structure component in (\ref{circ:estim}) for $B^{(0)}=0$
practically is at the level of its wings for large $B^{(0)}$ (by
more exact consideration, first one more and more approaches to
second one with increase of $J$), and this property in
\figurename~3 of \cite{SSSh85} is not looked through. The almost
symmetric shifted dip is experimentally observed in transmission
of circularly polarized radiation (one can see its derivative in
\figurename~\ref{12ch4_ex_dlc-13ch4_ex_dc} at the left). As the
formulae (\ref{circ:col}) and (\ref{circ:estim}) are shown, on the
place of this observed dip the collision model with anyone $\nu$
gives peak connected to the convergence of all main (with
difference $|\tilde{\mu}^\varsigma_{mn}| =0$) lines in pairs,
having every possible (\ie\ both equal and different ones) nuclear
spin quasi-projections $\tilde{\mu}^\varsigma_n$ [see
(\ref{tildemu_cdot})] and $M_n$. This peak should be manifested
more and more distinctly with pressure decreasing, as well as
usual peaks of resonance scattering of $\bm{\vee}$- and
$\bm{\wedge}$-types for linearly polarized radiation, connected to
crossings of hyperfine $M$-sublevels with $|\varDelta M|=2$.
Besides for circularly polarized radiation one more relatively
smaller peak should be manifested in zero of magnetic field,
connected with convergence of main ${[}^m_n$ lines in pairs having
equal $\tilde{\mu}^\varsigma_n$ and different $M_n$.

\section{\label{DoubleHFS} Parity doubling for HFS of NOR/M$_\perp$
and /E$_\perp$ in $\mathrm{CH}_3{\mathrm{F}}$ by high-$J$
approximation}

Here still unobserved field spectra of NOR of ballast type in
radiation transmission of fluoromethane molecule gas are briefly
considered (in other details, the molecules of fluoromethane
symmetry are considered in \cite{Gus95}). A feature of
fluoromethane, easily manifested in these spectra, is the parity
doubling for rotation $JK$-levels with $K\neq0$.

As we have seen, the field spectrum is nonlinear-optical resonance
is additively formed by high-$J$ approximation. The spin
subsystems of molecules are characterized by their total
eigen-spin $\bm{I}^\varsigma$ and interacts with rotation angular
momentum $\bm{J}$ independently from each other. Magnetic
spectrum, described by the formula (85) from \cite{Gus95},
corresponds to the case of linearly polarized radiation,
propagated across varied magnetic field (basis vector
$\bm{u}_z\parallel \bm{B}^{(0)}\parallel \bm{E}^{(\omega)}$) and
resonancely absorbed on molecular rotation-vibration transition of
${}^{Q}Q_K(J)$-type. Now we adapt it to fluoromethane.

Effective Hamiltonian of hyperfine and Zeeman interactions
\begin{align}
\hat{H}^{(1)}_{j,JK} &= -\sum_\varsigma \left[
 \left( C_{JK}^{\varsigma(\mathrm{a})}\hat{{\sf 1}}_2
+C^{\varsigma(z)}_{JK} \hat{\sigma}_{z}\right)
 \hat{\bm{J}}_j + \bm{\varDelta}_I^\varsigma \hat{\sf 1}_2
 \right] \cdot \hat{\bm{I}}{}^\varsigma
 -\bm{\varDelta }_{JK}^{r(\mathrm{a})}\cdot \hat{\bm{J}}_j
 \hat{\sf 1}_2
\nonumber\\ &= -\sum_\varsigma \left(
 \hat{\bm{\omega}}{}^{\varsigma(\mathrm{a})}_{j,JK}\hat{\sf1}_2
+\hat{\bm{\omega}}{}^{\varsigma(z)}_{j,JK}\hat\sigma_z\right)\cdot
 \hat{\bm{I}}{}^{\varsigma} - \bm{\varDelta}^{r(\mathrm{a})}_{JK}
 \cdot \hat{\bm{F}}_j \hat{\sf1}_2
=\hat{h}_{j,JK} - \bm{\varDelta}^{r(\mathrm{a})}_{JK}
 \cdot \hat{\bm{F}}_j \hat{\sf1}_2. \label{hfZ}
\end{align}
In the same way, as from (\ref{hjM}) to (\ref{lineHs}), we shall
use the linearization of the Hamiltonian on spins
$\hat{\bm{I}}{}^\varsigma$. We here introduce\footnote{Subscript
``j'' is omitted, since we assume the same properties of both
$j$-levels.} $\hat{h}^{(p)}_{JKM}\equiv \langle
j_pM|\hat{h}_{j,JK}|j_pM \rangle$ and
\begin{subequations}
\begin{equation}
 \hat{h}^{(p)}_{JKM} \simeq
-\sum_\varsigma \bm\omega^{\varsigma(p)}_{JKM}
 \cdot \hat{\bm{I}}{}^{\varsigma} \label{h_JKM}
\end{equation}
with
\begin{align}
 \bm\omega^{\varsigma(p)}_{JKM} &=
 \left( \omega^{\varsigma(p)}_{x,JKM},0,
 \omega^{\varsigma(p)}_{z,JKM}\right)_{x,y,z}
\nonumber\\
&=\bm\omega^{\varsigma(\mathrm{a})}_{JKM}+
 p\bm\omega^{\varsigma(z)}_{JKM}
\nonumber\\
&=C^{\varsigma(\mathrm{a})}_{JK}\bm{J}_M
 +\bm{\varDelta}^{\varsigma,r(\mathrm{a})}_{I,JK}
 +pC^{\varsigma(z)}_{JK}\bm{J}_M. \label{omega_JKM}
\end{align}
\end{subequations}
Here $\varsigma\in(\mathrm{F,C,H})$,
$\hat{\bm{I}}{}^{\mathrm{H}}=\sum_h \hat{\bm{I}}{}^h$,
$h\in(\mathrm{H}^1,\mathrm{H}^2,\mathrm{H}^3)$. If $K \equiv
0\,(\mathrm{mod}\,3)$ then $I^{\mathrm{H}}=3/2$ else
$I^{\mathrm{H}}=1/2$. $\hat\sigma_z$ is diagonal Pauli matrix
defined on states with definite parity, \ie\
$\hat\sigma_z|j_p\rangle = p|j_p\rangle$ where $j\in(m,n)$.  For
fluoromethane,
\begin{equation}
  p  =(-1)^{J+1} \delta_{K,0} \pm (1-\delta_{K,0})
\quad\text{and}\quad
 [p] =2-\delta_{K,0}. \label{ch3f:p}
\end{equation}
Zeeman frequencies $\bm{\varDelta}^{r(\mathrm{a})}_{JK} =
\gamma^{r(\mathrm{a})}_{JK} \bm{B}^{(0)}$ and
$\bm{\varDelta}^{\varsigma,r(\mathrm{a})}_{I,JK}
=\bm{\varDelta}^\varsigma_I -\bm{\varDelta}^{r(\mathrm{a})}_{JK} =
\gamma^{\varsigma,r(\mathrm{a})}_{I,JK} \bm{B}^{(0)}$.  The rest
of designations is the same as in formulae (\ref{hdiag}) to
(\ref{approx_hfs}). With $J\gg1$ average constants
\begin{equation}
\begin{split}
   C_{JK}^{\varsigma(\mathrm{a})}
&\simeq C_{\underline{\bot }}^{\varsigma A_1}
 +(C_{\underline{z}}^{\varsigma A_1}
 -C_{\underline{\bot}}^{\varsigma A_1})K^2/\hat{\bm{J}}{}^2,
\\ C^{\varsigma(z)}_{JK}
&\simeq \delta_{K,1}\delta_{\varsigma,\mathrm{H}}
 (-1)^{J+1}C_{\underline{\bot }}^{\mathrm{H}E}.
\end{split}
\end{equation}
$C^{\mathrm{H}(z)}_{J1}$ characterizes hyperfine $K$-doubling for
$JK$-levels with $K=1$. Average rotation gyromagnetic ratio
\begin{equation}
 \gamma_{JK}^{r(\mathrm{a})} = \gamma^r_{\underline{\bot}}
+(\gamma^r_{\underline{z}}
-\gamma^r_{\underline{\bot}})K^2/\hat{\bm{J}}{}^2.
\end{equation}
The shape of the spectrum is defined by ratio of spin-rotation
constants \cite{Gus99}, and also nuclear spin and rotation
$g$-factor (\cite{Fly74} and \cite{Fly78}, respectively). For our
estimations it is possible to suppose
\begin{gather*}
 C_{JK}^{\mathrm{F}(\mathrm{a})}\simeq
-4C_{JK}^{\mathrm{C}(\mathrm{a})}/3\simeq
 4C_{JK}^{\mathrm{H}(\mathrm{a})}\simeq 4\,
 \mathrm{kHza},
\\
 |C_{\underline{\bot }}^{\mathrm{H}E}| /
 |C_{\underline{\bot }}^{\mathrm{H}A_1}| \simeq 2,
\\ \intertext{and}
 g_I^{\mathrm{F}} \simeq 4 g_I^{\mathrm{C}} \simeq
 g_I^{\mathrm{H}}\simeq 5.6 \gg |g_{JK}^{r(\mathrm{a})}|.
\end{gather*}

To facilitate a comparison with Raman nonhyperfine structure of
fluoro\-me\-thane, manifested in NOR/M$_\parallel$ [see
(\ref{without_hfs})], we adduce from \cite{Fly78} the appropriate
gyromagnetic ratio for fluoromethane, when $J\gg1$:
$2\gamma^{r(\mathrm{a})}_{JK} \simeq 2\gamma^r_{\underline{\bot}}
\simeq -93\, \mathrm{Hza}\,\mathrm{Gs}^{-1}$.

\begin{figure}\centering
\includegraphics{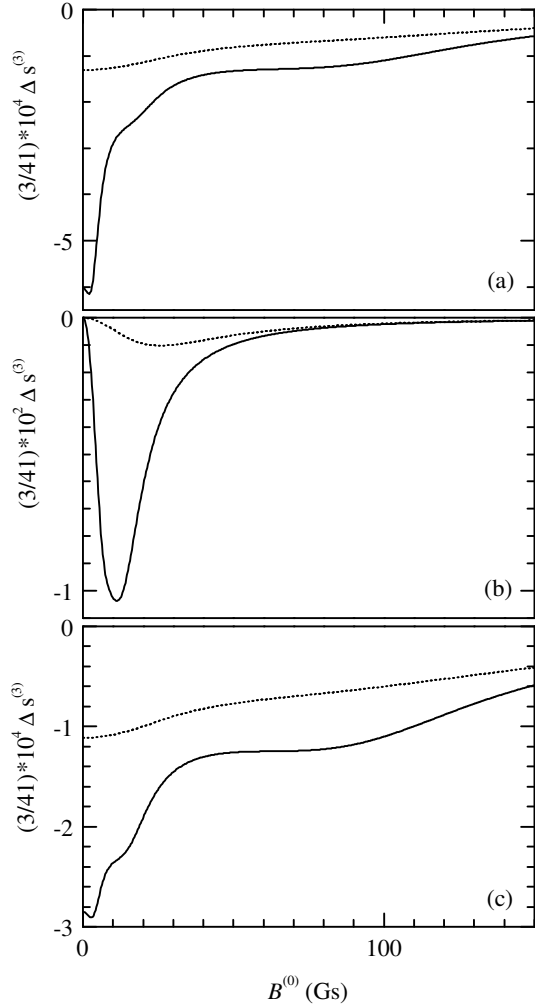}
\caption{\label{13ch3f_th_ap_lb} Approximation of NOR/M$_\perp$
[(\ref{doubling}), even function] in transmission of radiation
linearly polarized along $\bm{B}^{(0)}$ for transition
$^{Q}Q_{K}(20)$ of ${}^{13}\mathrm{CH}_3\mathrm{F}$ isotope: $K=0$
(a), $1$ (b), $2$ (c); $\nu_{20,K}/\mathrm{kHza}=10$ (solid
curve), $100$ (dotted one).}
\end{figure}

Using formula (\ref{X3:a}) with
\begin{equation}
 \hat{T}^{mn}_{0M} = T^{JJ}_{0M} \hat\sigma_x,
 \quad\text{where}\quad
 T^{JJ}_{0M}=\sqrt{3/\hat{\bm{J}}{}^2[J]}M
\label{TJJzeroM}
\end{equation}
[see (\ref{WignerEckert:c}) and (\ref{JjMxyz})], we write out the
required component of $X^{(3)}$-tensor, representing
NOR/M$_\perp$:
\begin{multline}
\left. X^{(3)mn}_{\underdot{0}{}\dot{0}{}\underdot{0}{}\dot{0}{}}
\right|_{\varOmega=0}
=\iint\limits_0^{\phantom{\infty\infty}\infty} \nu_{JK}^2\,
  \mathrm{d}\tau'\, \mathrm{d}\tau\, \Exp^{\bar{\nu}_{JK}(\tau'+\tau)}
\\ \times
\sum_{Mp}\left\langle
 T^{JJ}_{0,M-\hat I_z}\Exp^{\bar{\mathrm{i}}\hat h^{(\bar p)}_{JKM}\tau'}
 T^{JJ}_{0,M-\hat I_z}\Exp^{\bar{\mathrm{i}}\hat h^{(p)}_{JKM}\tau}
 \right.
\\ \left.\times
 T^{JJ}_{0,M-\hat I_z}\Exp^{\mathrm{i}\hat h^{(\bar p)}_{JKM}\tau'}
 T^{JJ}_{0,M-\hat I_z}\Exp^{\mathrm{i}\hat h^{(p)}_{JKM}\tau }
 \right\rangle.
\end{multline}
From here we obtain normalized amplification function
(\ref{s3perp}):
\begin{multline}
  \left.  s^{(3)}(B^{(0)})\right|_{\varOmega=0}
 =\left[ s^{(3)}(\infty) + \varDelta s^{(3)}(B^{(0)}) \right]_{\varOmega=0}
\\ \simeq \frac{[J]}{3}
 \sum_{M} |T^{JJ}_{0M}|^4 \left\{
 1-\frac23\sum_\varsigma \hat{\bm{I}}{}^{\varsigma2}
 \left[\frac{\delta_{K,1}\delta_{\varsigma,\mathrm{H}}
 \delta_{I^{\mathrm{H}},1/2}(2\varDelta^{\varsigma,r(\mathrm{a})}_{I,JK}
 \omega^{\varsigma(z)}_{x,JKM})^2
 }{(\nu^2_{JK} +\omega^{\varsigma(+)2}_{JKM})
   (\nu^2_{JK} +\omega^{\varsigma(-)2}_{JKM})}
 \right.\right.
\\ \left.\left. +\left(\frac{\partial_M
T^{JJ}_{0M}}{T^{JJ}_{0M}}\right)^2
 \frac{(1-\delta_{K,1})\omega^{\varsigma(\mathrm{a})2}_{x,JKM}
 }{\nu_{JK}^2+\omega^{\varsigma(\mathrm{a})2}_{JKM}}
 \left(3+\frac{\nu_{JK}^2}{\nu_{JK}^2
+\omega^{\varsigma(\mathrm{a})2}_{JKM}}\right)
 \right] \right\}.
\label{doubling}
\end{multline}
Here the relation between $C^{\varsigma(z)}_{JK}$ and
$C^{\varsigma(\mathrm{a})}_{JK}$ can be arbitrary, but
$I^{\mathrm{H}}=1/2$. In similar formula (85) from \cite{Gus95},
the latter restriction is removed, but only for $\left|
C^{\varsigma(z)}_{JK} / C^{\varsigma(\mathrm{a})}_{JK} \right| \ll
1$. In the spectrum of  fluoromethane with low enough pressure,
when $\nu_{JK}\ll C^{\varsigma(\mathrm{a})}_{JK}J$, in $Q$-branch
the triplex symmetric (with respect to field zero) structure is
observed. In \figurename~\ref{13ch3f_th_ap_lb}(a) and
\ref{13ch3f_th_ap_lb}(c) it is well visible, as dips with
different ratios. In \figurename~\ref{13ch3f_th_ap_lb}(b) the
structure being greater on two order (of amplitude) is added to
them in the form of split dip.
Starting with $B^{(0)}=0$ and up to $B^{(0)\rm H}\simeq
C^{\mathrm{H}(\mathrm{a})}_{JK}J/ \gamma^{\mathrm{H}}_I\simeq
0.33J\, \mathrm{Gs}$, the structure connected with spin-rotation
constants of nuclei H [namely, with
$C^{\mathrm{H}(\mathrm{a})}_{JK}$ and $C^{\mathrm{H}(z)}_{J1}$;
the latter is manifested only in
\figurename~\ref{13ch3f_th_ap_lb}(b)] is recorded, then up to
$B^{(0)\rm F}\simeq 3B^{(0)\rm H}$ the structure from nucleus F
($C^{\mathrm{F}(\mathrm{a})}_{JK}$), and at last up to $B^{(0)\rm
C}\simeq 9B^{(0)\rm H}$ the widest structure from nucleus C
($C^{\mathrm{C}(\mathrm{a})}_{JK}$). The amplitudes of the dips
connected with $C^{\varsigma(\mathrm{a})}_{JK}$ are approximately
equal and account $(C^{\mathrm{H}(\mathrm{a})}_{J1}/
C^{\mathrm{H}(z)}_{J1}J)^2$ a part from split (with respect to
field zero) dip connected with
$C^{\mathrm{H}E}_{\underline{\bot}}$. In cases, when
$C^{\mathrm{H}(z)}_{J1}$ and $C^{\mathrm{H}(\mathrm{a})}_{J1}$ are
much various, the bottoms of split dip are in symmetric points,
for which there is simple approximation, namely,
\[
 \left. B^{(0)\mathrm{H}} \right|_\pm \simeq
 \pm \frac{\sqrt{
 \nu^2_{J1} +\left( C^{\mathrm{H}(\mathrm{a})2}_{J1}
 +C^{\mathrm{H}(z)2}_{J1}  \right) J^2}}{
 \gamma^{\mathrm{H},r(\mathrm{a})}_{I,J1}}.
\]
Small splitting of $\varsigma$-dip with $K\neq1$ is also possible,
when $\nu_{JK} /C^{\varsigma(\mathrm{a})}_{JK}J \lesssim 0.96$
[see solid lines in FIGs.~\ref{13ch3f_th_ap_lb}(a) and
\ref{13ch3f_th_ap_lb}(c)].

Practically in the same conditions for $K=1$ it is possible to
observe NOR/E$_\perp$, \ie\ using Stark interaction by formula
(75) from \cite{Gus95}. Normalized amplification function (with
$J\gg1$)
\begin{gather}
  \left.  s^{(3)}(E^{(0)})\right|_{\varOmega=0}
 =\left[ s^{(3)}(\infty) + \varDelta s^{(3)}(E^{(0)}) \right]_{\varOmega=0}
\nonumber\\
\simeq \frac{[J]}{3} \sum_{M} |T^{JJ}_{0M}|^4
 \left\{\vphantom{\left[\frac{2\tilde{G}^{(0)}_{JK,z} T^{JJ}_{0M}
 C^{\mathrm{H}(z)}_{JK} \sqrt{\hat{\bm{J}}{}^2}
 }{ \nu^2_{JK} +C^{\mathrm{H}(z)2}_{JK} \hat{\bm{J}}{}^2
   +(2\tilde{G}^{(0)}_{JK,z} T^{JJ}_{0M})^2 }
 \right]^2}1 -2\delta_{K,1} \delta_{I^{\mathrm{H}},1/2}
 \left[\frac{2\tilde{G}^{(0)}_{JK,z} T^{JJ}_{0M}
 C^{\mathrm{H}(z)}_{JK} \sqrt{\hat{\bm{J}}{}^2}
 }{ \nu^2_{JK} +C^{\mathrm{H}(z)2}_{JK} \hat{\bm{J}}{}^2
   +(2\tilde{G}^{(0)}_{JK,z} T^{JJ}_{0M})^2 }
 \right]^2 \right\}.
 \label{doubling:e}
\end{gather}
Here (modified) Stark frequency
$\tilde{G}^{(0)}_{JK,z}=\tilde{d}_{JK} E^{(0)}_z /\hbar$ and basis
vector $\bm{u}_z \parallel \bm{E}^{(0)} \parallel
\bm{E}^{(\omega)} \perp \bm{k}$. $T^{JJ}_{0M}$ is determined in
(\ref{TJJzeroM}). It is clear that the split dip will be again
observed.\footnote{There is another (centrifugal) parity doubling
for NOR /E$_\perp$ with $K=3$, see formula (87) in \cite{Gus95}.}
Thus, for measurement of constant
$C^{\mathrm{H}E}_{\underline{\bot}}$ it is possible to use
NOR/E$_\perp$ in addition to NOR/M$_\perp$.

\section{Conclusion}

In the present work we have tried to find out the features
inherited to NOR/Fi for molecular levels with HFS. It is possible
to divide HFS of NOR/Fi into two essentially distinguishing types
--- Raman ($\bm{\vee}$ or $\bm{\wedge}$) and ballast ($
\bm{\between}$) ones. We think that it is important to note
sub-collision character of already observed HFS of field spectra,
namely, ballast and, in zero field, Raman ones. Still unobserved
Raman HFS in nonzero fields has not similar property. It can be
only found out with the same lower pressure as for HFS of NOR/Fr.
It is possible to observe Raman HFS separated from ballast one in
circular birefringence. Ballast HFS is always added to existing
Raman one in transmission, but can be observed in the absence of
the latter by choosing properly geometry and polarization of
fields. The existence of the field spectral structure then
directly depends on (average) hyperfine constants
$C^{\varsigma}_{\mathrm{a}}$, describing magneto-dipole connection
of spin $\varsigma$-subsystems with molecule rotation as whole.

The considered purely hyperfine model of NOR/Fi practically does
not contain free (\ie\ still indeterminate) parameters. If we set
aside the collision details (corrections) with their not quite
defined parameter $\tilde{\nu}_j/ (\nu_j -\tilde{\nu}_j)$ [because
of their functional behavior does not correlate with experimental
data at all], the model gives quite acceptable description of
(anomalous) sub-collision structure of NOR/M for methane isotopes.

Under exact account of hyperfine interaction, realized by us
before its approximation and only for ${}^{12}\mathrm{CH}_4$
isotope, we imply the account of both spin-rotation constants
$C^{\mathrm{H}}_{1j}$ ($\simeq C^{\mathrm{H}}_{\mathrm{a}}$) and
less significant\footnote{With the spectral contribution $\sim
10\,\%$.} spin-spin ones $C^{\mathrm{H}}_{2j}$ in appropriate
Hamiltonian reduced to the total spin $I^{\mathrm{H}}$. The
account gives small qualitative\footnote{And quantitative one, if
we deal with precision $\sim 10\,\%$.} changes of the graphs
represented in \figurename~\ref{12ch4_th_ap_dlc-13ch4_th_ap_dlc}
at the left.

It is well known, that the requirements to frequency control of
laser radiation are essentially weakened with field recording of
NOR spectra in comparison with frequency one. We have now added to
it another new (and important enough) its property, namely,
sub-collision character of observed HFS. A comparison with paper
\cite{HB73} (see also \cite{Che85}) here suffices, where the
frequency recording of NOR was used. For detection of HFS of NOR
they needed to decrease pressure $\lesssim 0.1\,\mathrm{mtorr}$
and increase the radius of laser beam $\gtrsim 1.8\,\mathrm{cm}$.

The set of molecules, which HFS can be researched with the help of
NOR/M, is numerous enough, see review \cite{MH76}. At that it is
necessary to distinguish HFS under electro-quadrupole and
magneto-dipole interactions. In the mentioned review the authors
have collected the data on super-collision\footnote{\Ie\ exceeding
the collision relaxation under working gas pressures of our
range.} HFS of NOR/Fr for halogen-substituted methanes (except
fluoromethane, which spin-rotation HFS is only magneto-dipole and
sub-collision and therefore practically not observed with NOR/Fr).
The manifestation of HFS under field recording of molecule
spectrum is based on rupture of spin-rotation connection, and
there should be half-spin nuclei (or their sets) in the molecule.
The nuclei with greater spin have the hardly ruptured connection,
conditioned their too strong electro-quadrupole interaction with
rotation of molecule as whole, and are not included in our
consideration.

We shall now describe our basic results some more:

\textbullet{} The analysis of both purely hyperfine and
collision-hyperfine models has shown that it is sufficient to have
only one collision constant $\nu$, describing relaxation of levels
and transition extracted by light, for calculation of
``anomalous'' structure of methane NOR/M. The account of the
features of collision populating of the levels (\eg, through the
deorientation of their hyperfine magnetic sublevels) is apparently
important only for details and corrections. More definitely the
statement can be refered to investigated area of methane pressure,
between $3$ and $6\,\mathrm{mtorr}$, and the used laser Gaussian
beam with radius $r_{1/\Exp}\simeq 0.5\,\mathrm{cm}$. The
anomalous structure of methane NOR/M \cite{BKKR:83,BKRSS:84} looks
as sub-collision one, \ie\ observed inside homogeneous width of
levels (taking into account the gyromagnetic scale $2\gamma_J$).
It is conditioned by difference (in methane case, on an order)
between nuclear spin $g$-factors and molecular rotation
$g$-factor. By nature this structure is basically hyperfine and
not collision or collision-hyperfine. The field spectra of NOR
break collision limit and facilitate the investigation of
sub-collision HFS of molecules.

\textbullet{} Since the theoretical model of observed structures
is based on the account of hyperfine structure of levels,
mathematical means, simplifying the analysis and making it
independent from basis set of wave functions, are operator ones.
Carrying out analytic calculations, it is necessary, as far as
possible, to keep away from any representation of operators. Some
ways have proposed for it in the end of section~\ref{ExactHFS}.

\textbullet{} The approximation of HFS of NOR/M for multi-spin and
fast rotated molecules is obtained. Using the approximation and
appropriate experimental data, it was possible to obtain (at
estimation level for the present time) the hyperfine constant
$C^{{}^{13}\mathrm{C}}_{\mathrm{a}}$ of ${}^{13}\mathrm{CH}_4$
isotope, which value is well coordinated with its NMR data. In the
same way both NOR/M$_\perp$ and NOR/E$_\perp$, conditioned with
hyperfine doubling of levels have calculated for fluoromethane,
that can facilitate their real observation.

\begin{ack} 

We would like to thank S.~G. Rautian for the useful discussions.

\end{ack} 

\appendix*
\section{\label{JJIhalf}HFS of NOR/M$_\parallel$ for the case
$J_m =J_n =I =1/2$}

Here we represent HFS of NOR/M$_\parallel$ for simple model case,
when $J_m=J_n=I=1/2$. The case was already considered in
\cite{SSSh85} but with some mistakes. In the mentioned paper the
formula (3.9) for circularly polarized radiation is uncorrect
(namely, with $\Gamma_j^2 +g_j^2 \varDelta^2 \leq A_j^2$, its
nonlinear addend, as a function of magnetic field, changes its own
sign), but further in the formula (3.10) for linearly polarized
radiation the similar addend from circularly polarized component
of the radiation is already written out correctly. Most principal,
their formula (3.10) for linear polarized radiation does not take
into account the diagram of $\Join$-type in the summand for
oppositely polarized photons. As a result, with $B^{(0)}=0$, when
we make $C_{\mathrm{a}}=0$, the summand does not disappear, that
is incorrect. The more right special expression for function
(\ref{X3}) are
\begin{subequations}
\begin{gather}
 \bm{\vee}^{m}_{nq} +\bm{\wedge}^{m}_{nq}
 = \Exp^{-(\varOmega_{\bar{q}}/\omega_{\mathrm{D}})^2}
 \frac{\nu^2}{\nu_m \nu_n} \,
 \frac{C_{\mathrm{a}}^2/2}{C_{\mathrm{a}}^2 +\varDelta_{IJ}^2}
\nonumber\\ \times
 \left[ \left(1 -\frac{\nu_n^2}{\nu_n^2 +C_{\mathrm{a}}^2
 +\varDelta_{IJ}^2} \right) \frac{\nu_m}{\nu_m -\mathrm{i} q (\varDelta_I
 +\varDelta_J)} +(m \leftrightarrow n) \right] \label{vee_wedge_half}
\\
 =\Exp^{-(\varOmega_{\bar{q}}/\omega_{\mathrm{D}})^2}
 \frac{\nu^2}{\nu_m \nu_n} \left[
 \frac{C_{\mathrm{a}}^2/2}{\nu_n^2 +C_{\mathrm{a}}^2 +\varDelta_{IJ}^2} \,
 \frac{\nu_m}{\nu_m -\mathrm{i} q (\varDelta_I +\varDelta_J)}
 +(m \leftrightarrow n) \right]
\end{gather}
\end{subequations}
and
\begin{subequations}
\begin{gather}
\bm{\between}^m_{n q} =
\Exp^{-(\varOmega_q/\omega_{\mathrm{D}})^2}
 \frac{2\nu^2}{\nu_m \nu_n} \left[1
-\frac{C_{\mathrm{a}}^2/2}{C_{\mathrm{a}}^2 +\varDelta_{IJ}^2}
 \left(1
-\sum_j \frac{\nu_j^2/2}{\nu_j^2 +C_{\mathrm{a}}^2
+\varDelta_{IJ}^2} \right)\right] \label{between_half}
\\
= \Exp^{-(\varOmega_q/\omega_{\mathrm{D}})^2}
 \frac{2\nu^2}{\nu_m \nu_n}
 \left[1 -\frac{[(\nu_m^2 +\nu_n^2)/2 +C_{\mathrm{a}}^2 +\varDelta_{IJ}^2]
  C_{\mathrm{a}}^2/2}{(\nu_m^2 +C_{\mathrm{a}}^2 +\varDelta_{IJ}^2)
                 (\nu_n^2 +C_{\mathrm{a}}^2 +\varDelta_{IJ}^2)}\right].
\end{gather}
\end{subequations}
It is visible, that in (\ref{vee_wedge_half}) there are summands
of $\bm{\vee}$-, $\bm{\wedge}$-, and $\Join$-types,\footnote{Just
the last summands of $\Join$-type are absent in the formula (3.10)
of \cite{SSSh85}.} and in (\ref{between_half}) there are summands
of $\bm{\between}$-, $\bm{\vee}$-, and $\bm{\wedge}$-types.
Substituting the adduced expressions in (\ref{s3par}), we obtain
the right formulae for NOR/M$_\parallel$ in transmission of
circularly and linearly polarized radiations; they correspond to
the mentioned mistaken one (3.9)\footnote{In the numerator of this
formula there is $2C_{\mathrm{a}}^2$ instead of
$C_{\mathrm{a}}^2/2$.} and (3.10) from \cite{SSSh85}:
\begin{equation}
 s^{(3)}_{\mathrm{circ}}(B^{(0)})
=\frac{2}{3} \left( 1 -\frac{C_{\mathrm{a}}^2/2}{\nu^2
+C_{\mathrm{a}}^2 +\varDelta_{IJ}^2} \right), \label{circ_}
\end{equation}
and
\begin{equation}
 s^{(3)}_{\mathrm{line}}(B^{(0)})
 =\frac{1}{3} \left[1 -\frac{C_{\mathrm{a}}^2/2}{\nu^2
 +C_{\mathrm{a}}^2 +\varDelta_{IJ}^2}\,
 \frac{(\varDelta_I +\varDelta_J)^2}{\nu^2 +(\varDelta_I +\varDelta_J)^2}
\right].
\end{equation}
We have here put $\nu_j=\nu$ for simplicity. The adduced formulae
well illustrate the classification described on
p.~\pageref{4leveldescript}.

The simplified account of ballast structure can be demonstrated in
case, when $\nu\ll C_{\mathrm{a}}$. Here it is enough to take into
account pair of optically connected magnetic sublevels
${[}^{m_1}_n$. Let, when the magnetic field is varied, one of
them, \emph {e.~g}.\ $m_1$, is crossed with a sublevel $m_2$,
optically not connected with sublevel $n$. Let now sublevels $m_1$
and $m_2$ mix up by hyperfine interaction. If we go over to basis
diagonalizing both magnetic and hyperfine interactions, we obtain
two (anticrossing) sublevels, $\tilde{m}_1$ and $\tilde{m}_2$,
optically interacting with sublevel $n$ under the diagram
${}^{\tilde{m}_1}\underline{\vee}^{\tilde{m}_2}_{n}$. Introducing
an angle of mixing $\beta$, it is possible to write out a rule of
sums for probabilities of the transition: $\cos^2\beta
+\sin^2\beta =1$. Taking into account that linear-optical
resonance is calculated exactly this way, it is possible to assert
that it does not depend on mixing. The nonlinear-optical resonance
\begin{equation}
\propto \cos^4\beta + \sin^4\beta = 1-\frac{\sin^2 2\beta}{2},
\label{SimpleModel}
\end{equation}
and the dependence on mixing remains. Comparing this expression
with (\ref{circ_}), we obtain\footnote{It is supposed that
$\beta\rightarrow0$ when $|\varDelta_{IJ}|\gg |C_{\mathrm{a}}|$.}
\[\sin^2 2\beta
=\frac{C_{\mathrm{a}}^2}{C_{\mathrm{a}}^2 +\varDelta_{IJ}^2},
\]
\ie\ how the mixing (or angle of mixing) depends on magnetic
field. The similar reasonings in case $J\gg1$ can give the formula
(\ref{dipsign}).

\newcommand{\noopsort}[1]{} \newcommand{\printfirst}[2]{#1}
  \newcommand{\singleletter}[1]{#1} \newcommand{\switchargs}[2]{#2#1}

\end{document}